\RequirePackage{fix-cm}
\documentclass{svjour3}
\usepackage{graphicx,url,hyperref,microtype,subcaption,amsmath,graphbox}
\usepackage{latexsym,courier,upgreek,algorithmic,xcolor}
\usepackage[square,sort&compress,numbers]{natbib}
\usepackage[displaymath,mathlines]{lineno}
\hypersetup{colorlinks,citecolor=blue,linkcolor=red,urlcolor=green}
\smartqed

\chardef\_=`_
\journalname{Transport in Porous Media}

\begin{document}
\title{A Dynamic Network Simulator for Immiscible Two-Phase Flow in
  Porous Media}

\titlerunning{Dynamic Network Simulator}  
\author{Santanu Sinha \and Magnus Aa. Gjennestad \and Morten Vassvik \and Alex Hansen}

\institute{
  Santanu Sinha \at
  Beijing Computational Science Research Center,
  10 East Xibeiwang Road, Haidian District, Beijing 100193, China.\\
  \email{santanu@csrc.ac.cn}\\
  \and
  Magnus Aa Gjennestad
  \and
   Morten Vassvik
  \and
  Alex Hansen \at
  PoreLab, Department of Physics, Norwegian University of Science and Technology, NTNU,
  N-7491 Trondheim, Norway.\\
  \email{alex.hansen@ntnu.no}\\
}

\date{Received: date / Accepted: date}
\maketitle

\begin{abstract}
We present in detail a set of algorithms to carry out fluid
displacements in a dynamic pore-network model of immiscible two-phase
flow in porous media. The algorithms are general and applicable to
regular and irregular pore networks in two and three dimensions with
different boundary conditions. Implementing these sets of algorithms,
we describe a dynamic pore-network model and reproduce some of the
fundamental properties of both the transient and steady-state
two-phase flow. During drainage displacements, we show that the model
can reproduce the flow patterns corresponding to viscous fingering,
capillary fingering and stable displacement by altering the capillary
number and the viscosity ratio. In steady-state flow, the model
verifies the linear to non-linear transition of the effective
rheological properties and satisfy the relations between the seepage
velocities of two-phase flow in porous media.
\end{abstract}

\section{Introduction}
Flow of multiple immiscible fluids inside a porous medium shows a
range of complex characteristics during transient as well as in steady
state \cite{d92, b72}. A number of factors, such as the capillary
forces at the interfaces, viscosity contrast between the fluids,
wettability and geometry of the system, make the multiphase flow
properties very different compared to single phase flow. When a
non-wetting fluid displaces a wetting fluid the flow is called as
drainage whereas the opposite is called imbibition. During drainage, a
less-viscous fluid displacing a more-viscous fluid in a porous medium
creates a variety of fingering patterns, whereas a more viscous fluid
displacing a less-viscous one shows a stable displacement front
\cite{lmtsm04, vhj19}. The fingering patterns show different
properties depending on whether the flow is dominated by capillary or
viscous forces and correspondingly they are named as the capillary and
viscous fingerings respectively \cite{ltz88, mfj85}. The capillary
fingering patterns appear during slow displacement process and are
well described by invasion percolation \cite{ww83, lz85}, whereas the
viscous fingering patterns appear during fast displacement and can be
modeled by diffusion limited aggregation (DLA) model \cite{ws81,
  p84}. If both the fluids are continuously fed into the porous
medium, the initial transients will eventually die out and the system
enters a steady state. It has been discovered that under steady-state
conditions, the two-phase flow rate do not obey the linear Darcy law
in the capillary dominated regime. Rather, it was found to have a
power law dependence on the total pressure drop \cite{tkrlmtf09,
  tlkrfm09, rcs11, sh12, sbdk17}.

The problem of two-phase flow in porous media has extensively been
studied using laboratory experiments \cite{lmtsm04, vhj19, ltz88,
  mfj85, cw85, st58}, statistical models \cite{ws81,ww83,bkl86} and
numerical simulations \cite{vhj19, kl85}. There have been significant
theoretical developments in describing the steady-state properties
\cite{hg93, h98, gm14, hsbkgv18}. Due to the vast advancement in the
computer power and development in scanning instruments for pore space
reconstruction, the computational modeling techniques play a leading
role in the study of two-phase flow problems. There are different
modeling approaches. The direct numerical simulations, such as the
volume-of-fluid method \cite{rb12} and the level-set method
\cite{jh13, gm15}, perform discretization of the pore space into
smaller cells and solve the Navier-Stokes equation. A popular
voxel-based simulation technique is the lattice Boltzmann method (LBM)
which solves the Boltzmann transport equations for different species
of lattice gases at the discretized pore space \cite{gr91, rob10,
  alk11}. These techniques all provide detailed information on the
flow propagation at pore scale and are useful where the actual shape
of the pores matter. However, due to the discretization of the pores,
these models become computationally expensive when studying systems
with large number of pores, such as for the scale-up problems.

Pore-network modeling \cite{b01, jh12}, in this regard, proves a
computationally efficient method that can treat much larger
systems. In pore-network modeling, a porous matrix is modeled as a
network of links (pore throats) that are connected at nodes (pore
bodies). The actual shapes of the pore space are replaced with
simplified geometries and the average flow properties for each pore is
considered to model the flow of the system. Unlike the voxel-based
methods that discretize the pore space in smaller grids, here one pore
is the smallest computational unit in the pore-network
simulations. This simplification looses the detailed evolution of the
fluid arrangements inside a pore, however it allows the pore-network
models to apply for large systems and thereby study their statistical
properties. There are two main ingredients for transporting fluids in
a pore-network model: (1) the local pressure drops in the pores and
(2) the resulting displacements of the fluids. Based on these two
ingredients, there are two major groups of pore-network models,
quasi-static models and dynamic models. The quasi-static models are
intended for the flow that is dominated by capillary forces so that
the viscous forces can be neglected. The displacement of fluids in the
quasi-static models are performed by filling a whole pore at a time
with invasion percolation-type algorithms \cite{ww83, ltz88, ck82,
  b98} where the filling of pores are decided by capillary entry
pressures or by determining the stability of an interface for a given
contact angle \cite{cr88, cr90}. Quasi-static models can successfully
describe the equilibrium properties of two-phase flow at capillary
dominated regime \cite{ob98, bj02, pt18}, however they are unable to
capture the dynamic effects from the interaction between viscous and
capillary forces at higher flow rates. This interaction between the
viscous and capillary forces at the pore scale are taken into account
in the dynamic pore-network models where the fluids inside pores are
displaced under both the viscous and capillary pressure drops
\cite{jh10, hu12, ak98}. The viscous pressure drops are calculated by
solving flow equations for fully developed viscous flow inside the
pores and the capillary pressure drops are obtained from the local
fluid configuration inside a pore. There are many factors that make
the dynamic models computationally more complex to implement compared
to the quasi-static models and efficient algorithms are therefore
necessary. One such factor is the mixing of fluids at the nodes and
distributing them to the neighboring links while conserving the
volumes of each fluid.

In this article, we present in detail a set of algorithms to implement
the displacements of two fluids in the links of a pore network and to
distribute them to the neighboring links after they pass the network
nodes. Together with these algorithms, we describe an {\it
  interface-tracking\/} dynamic pore network model in which the
location of two fluids inside the pores are marked explicitly by the
positions of all the interfaces at any time step. These interfaces are
displaced in small steps under the instantaneous viscous and capillary
pressure drops at the pores. We therefore call the algorithms as {\it
  interface-dynamics} algorithms. The displacements of all the
interfaces with time describe the transport of the fluids through the
network. This is different from other pore-network models, such as in
\cite{jh10}, where the interface positions are only available
implicitly from the volumetric saturation of the fluid elements inside
a pore, or from the model in \cite{hu12} where an interface is moved
through a whole volume element at each time step. Explicit positions
of all the interfaces in this interface-tracking model provides a
detailed picture of the fluid configuration at any time. Calculation
of the capillary forces thus become straightforward from the interface
positions and different dynamical events, such as the retraction of
invasion fronts after a Haines jump \cite{h90, bo13, ab13, mf92}, can
easily be resolved.

The interface-tracking model we describe here was first introduced by
Aker et al.\ in late nineties \cite{ak98} to study transient two-phase
flow -- i.e. drainage -- in a pore network. The interface algorithms
in that model were developed for the drainage in a regular pore
network with open boundaries, where a non-wetting fluid invades the
network filled with wetting fluid. There, the algorithms were made in
such a way that the fluid-volumes were not conserved as they pass
across the nodes. Later, the model was extended for steady-state flow
by implementing periodic boundary conditions in the flow direction
\cite{ka02}. There the interface-dynamics algorithms of the previous
model were updated to maintain the volume conservation. However, the
rules to transport the interfaces through nodes described in the model
were based on different events at the nodes of a regular square
network in two dimension and were not straightforward to implement in
an irregular network topology, such as the networks reconstructed from
real samples. Since then, the model has been updated over the years
and efforts have been made to generalize the rules for interface
dynamics that can be applied for any type of network topology and
spatial dimension, as well as for both the transient and steady-state
flow \cite{sbdk17}. However any detailed description of the new
algorithms were absent. Therefore, one of the main goals of this
article is to describe these new interface algorithms with sufficient
details so that the reader may reproduce the model. The interface
algorithms we present here are universal in terms of the network
connectivity and topology and can be used for different networks
without any further modification. Though these algorithms are simple
and straight-forward to implement, they are capable to capture the
essential physical properties of both the transient as well as of
steady-state immiscible two-phase flow. We will show this by
presenting a few fundamental results of transient and steady-state
two-phase flow in porous media using this model. When the statistical
properties are concerned, fine details of dynamics are generally
dropped out. This makes it possible to model the fluid displacements
with simplified interface dynamics rules, while still preserving the
fundamental statistical properties of the flow.

We like to point out that, complex flow mechanisms such as the wetting
films along the pore walls can also be included in the model as
presented in \cite{toh12}. However, we do not consider them in this
article and restrict ourselves only to the simplest case of Darcy-like
creep flow. Moreover, here we will use an explicit Euler integration
method to do the time steps, whereas semi-implicit methods can be used
for computational efficiency at very low flow rates as illustrated in
\cite{gvkh18}. A Monte-Carlo algorithm was also developed recently for
this model to achieve the steady state in less computational time
\cite{sshbkv16}. We also note that important phenomena such as wetting
angle hysteresis is straight-forward to implement but it has not been
included so far.  However, wetting angle alteration due to changes in
the composition of the immiscible fluids has been considered in the
past \cite{sgosh11,fsh15}.

The article is structured as follows. In section \ref{secEqn}, we will
present the governing equations describing the two-phase flow at the
pore level and will describe how we solve the equations to find the
local pressures at the nodes at any time step. Then we present in
detail in section \ref{secInt} the interface-dynamics algorithms which
transport the fluids inside the links and distribute them to the next
connected links after they pass a node. The algorithms are divided
into three functions, which will be presented in the subsections. In
section \ref{secBoundary}, we will described how different boundary
conditions can be implemented in this model. In order to validate the
model, we will present a few examples in section \ref{secRes} where
the model can successfully reproduce some fundamental results of
two-phase flow. Both the transient and steady-state flow will be
simulated and the corresponding results will be presented in the
subsections \ref{secTrn} and \ref{secStd} respectively. Finally, we
will draw our conclusions in section \ref{secCon}.

\begin{figure}[t]%=================================================
  \centerline{\hfill
    \includegraphics[height=0.2\textwidth,clip]{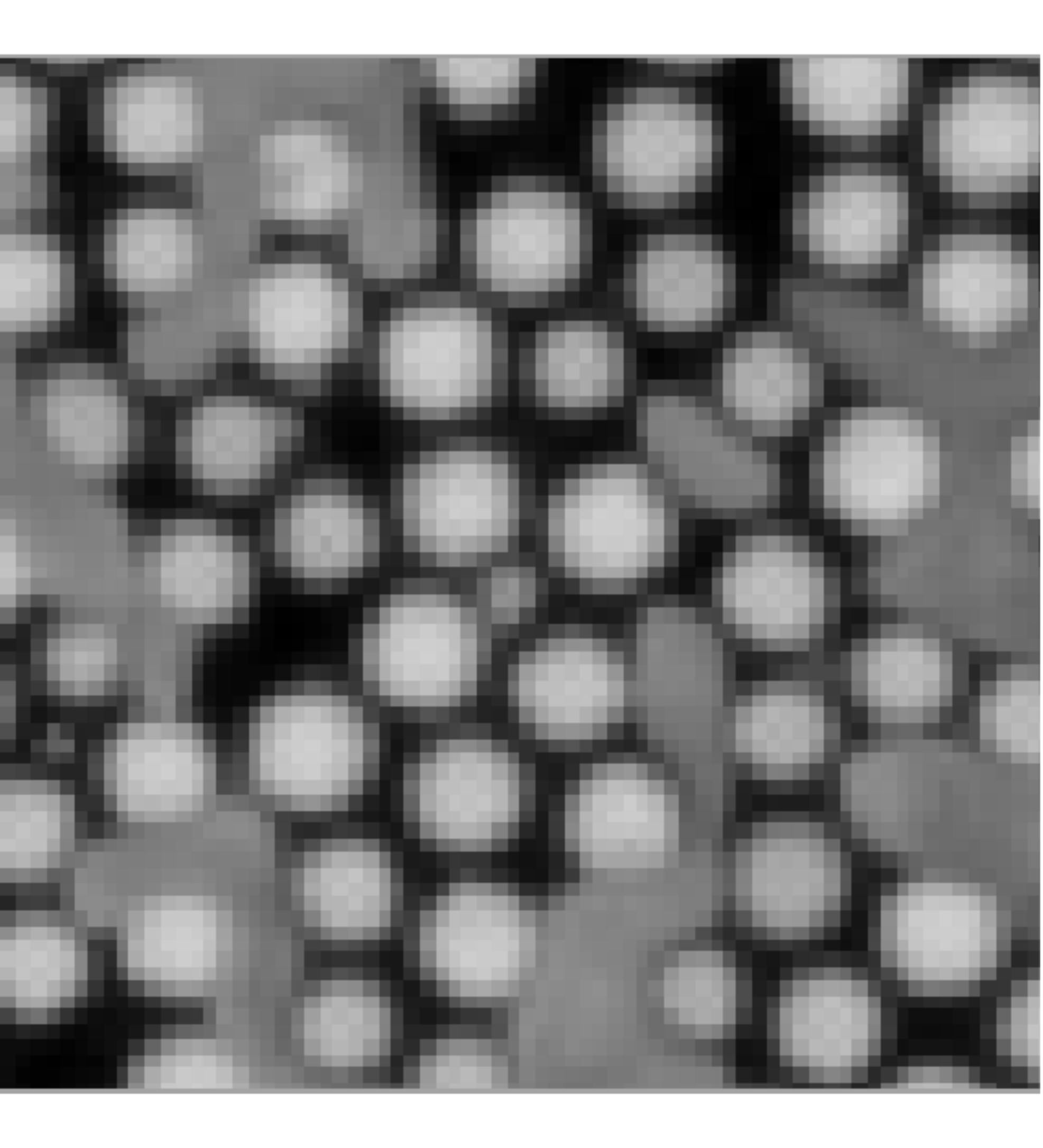}
    \includegraphics[height=0.2\textwidth,clip]{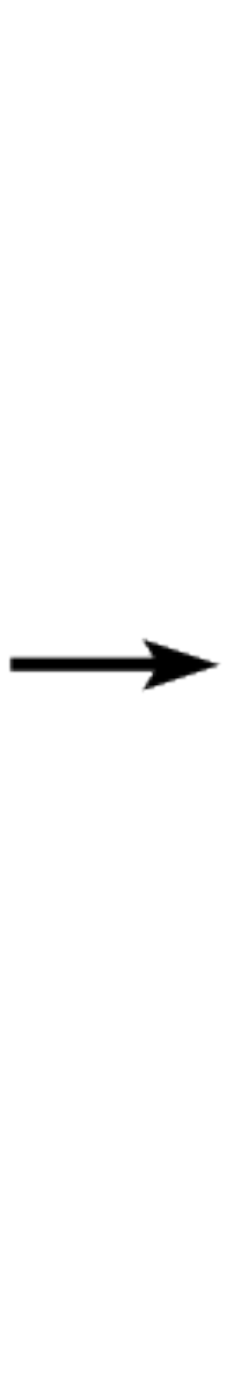}
    \includegraphics[height=0.2\textwidth,clip]{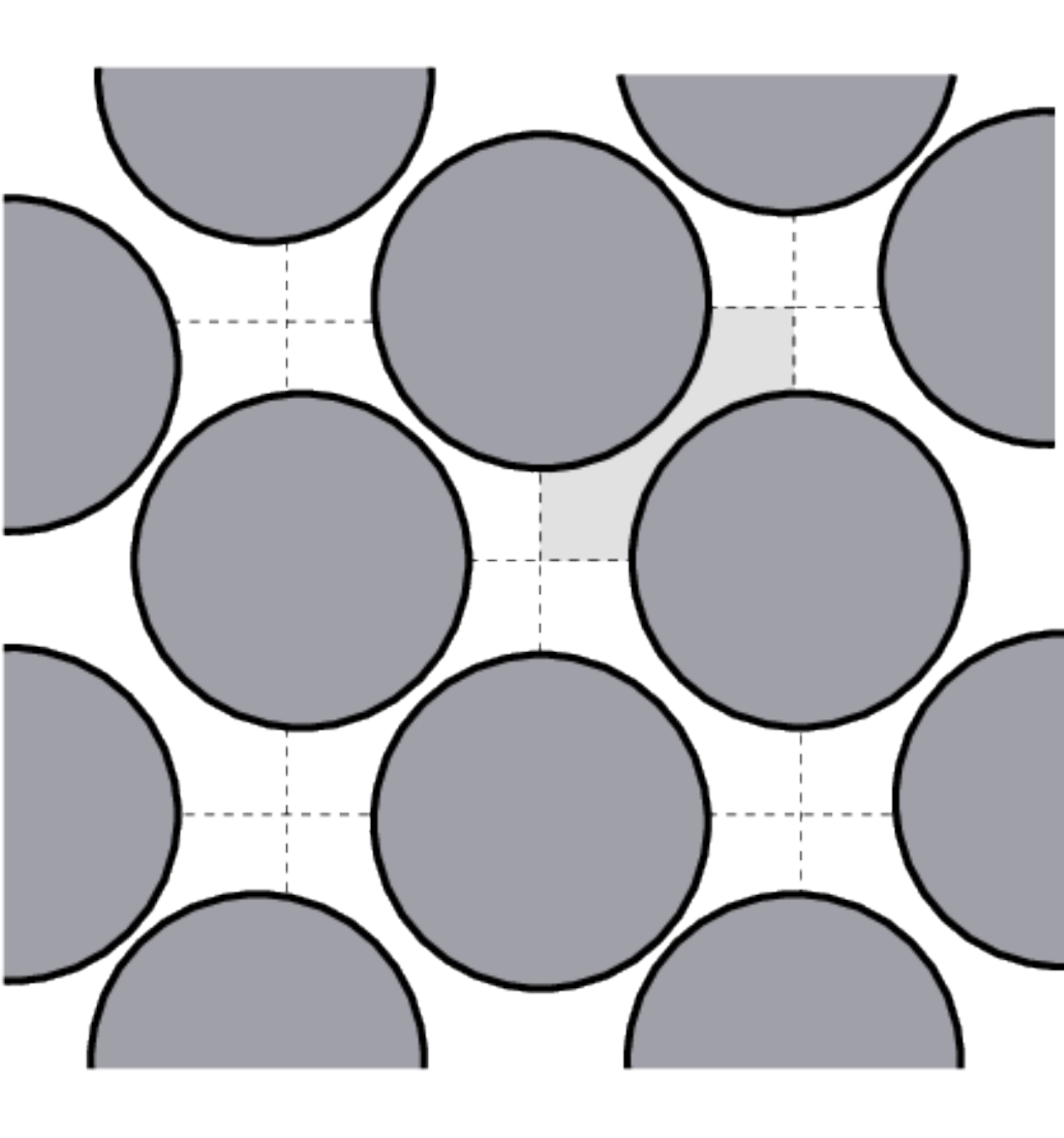}\hfill\hfill
    \includegraphics[height=0.2\textwidth,clip]{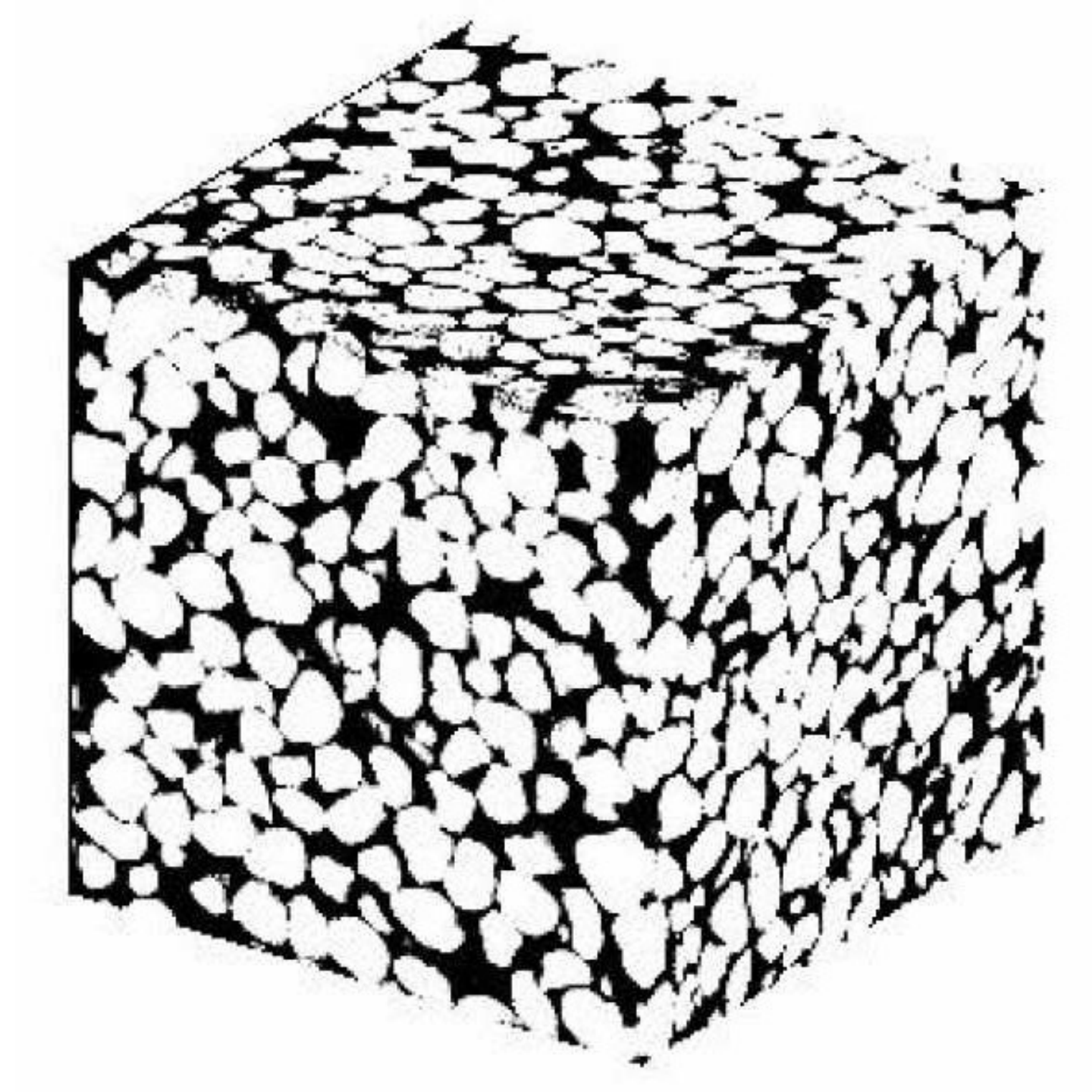}
    \includegraphics[height=0.2\textwidth,clip]{fig_Arrow_a.pdf}
    \includegraphics[height=0.2\textwidth,clip]{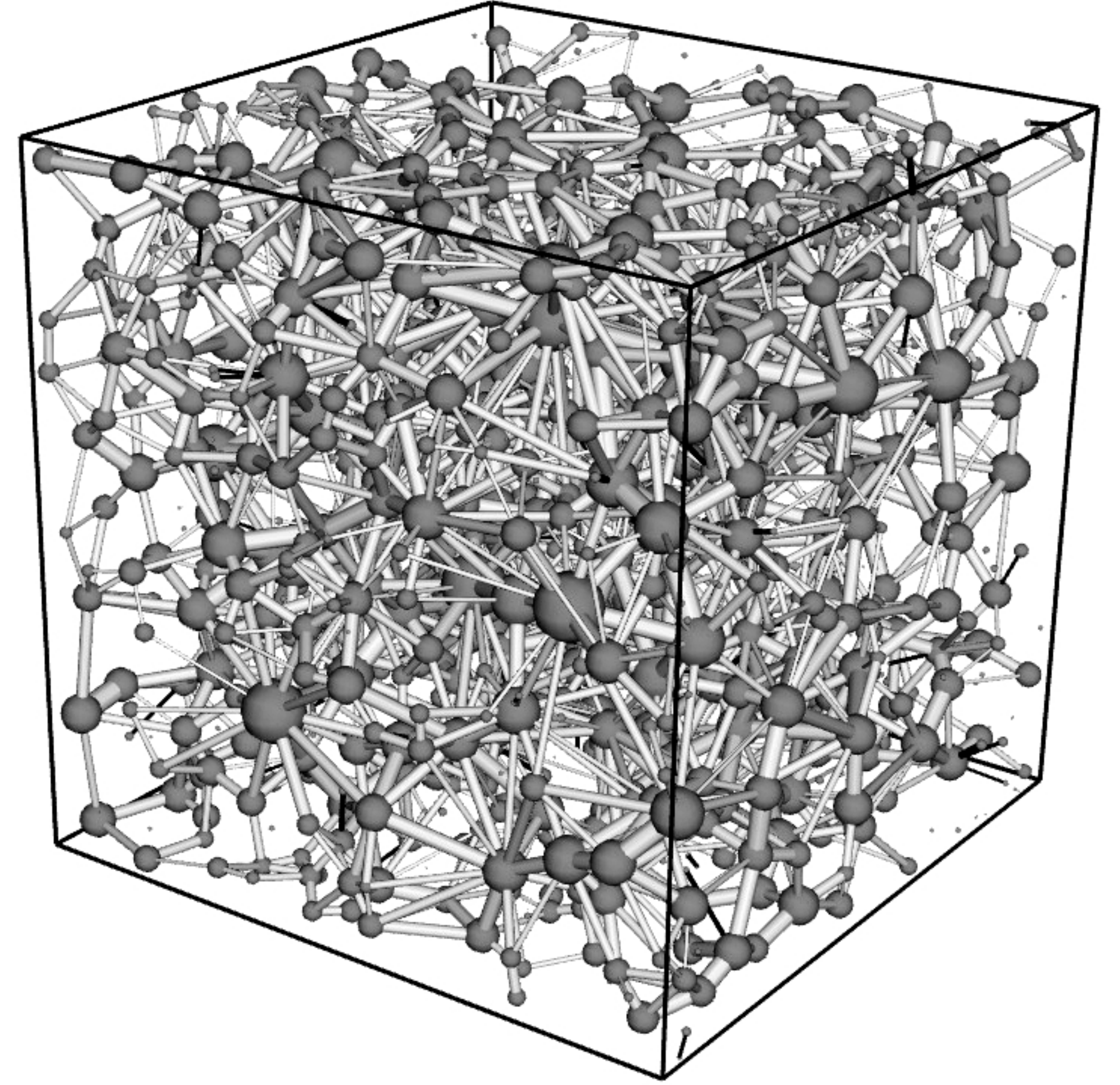}\hfill}
  \centerline{\hfill(a) \hfill \hfill (b) \hfill \hfill (c) \hfill \hfill (d) \hfill}
\caption{\label{figSys} An overview of the network representation of
  two and three dimensional porous media. Figure (a) shows a $10\,{\rm
    mm} \times 10\,{\rm mm}$ crop of a $42\,{\rm cm} \times 85\,{\rm
    cm}$ Hele-Shaw cell, filled randomly with a monolayer of $1\,{\rm
    cm}$-diameter glass beads, indicated by the white circles
  \cite{esth13}. The black and gray colors in the image show the
  wetting and non-wetting fluids. Such a system can be modeled by a
  two-dimensional network of hour-glass shaped links as shown in (b),
  where the distribution in link radii and node positions can be tuned
  according to the system properties. In (b), the dark gray circles
  represent the beads and white represents the pores. The links are
  separated by dotted lines and the intersection of two such lines
  defines the position of a node. One link is colored by light
  gray. We like to point out that, (b) is not an exact reconstruction
  of the image shown in (a), rather it is a simplified
  illustration. In (c), we show a three dimensional Micro CT image of
  sand-pack (sample F42A in \cite{mbweb}) and the corresponding
  reconstructed pore network is shown in (d).}
\end{figure}%====================================================

\section{\label{secEqn}Flow equations}
We consider immiscible flow of two incompressible Newtonian fluids
through a network of pores, where one of the fluids is more wetting
than the other with respect to the pore walls. We denote the less and
more wetting fluids as the non-wetting (n) and wetting (w),
respectively. For a given network, two dimensionless macroscopic
parameters that characterize the dynamics of two-phase flow are the
capillary number (Ca) and the viscosity ratio ($M$). The capillary
number is a measure of the ratio of viscous to capillary forces in the
system. These parameters are defined as,
\begin{equation}
\label{eqnCaM}
\displaystyle
\textrm{Ca} = \frac{\mu_{\rm e}Q}{\gamma A}
\quad \textrm{and} \quad
M = \frac{\mu_{\rm n}}{\mu_{\rm w}}
\end{equation}
where $Q$ is the total flow rate, $\gamma$ is the interfacial tension
between two fluids and $A$ is the cross-sectional pore area of the
network. $\mu_{\rm n}$ and $\mu_{\rm w}$ are the viscosities of the
two phases. In case of transient studies $\mu_{\rm e}$ is the
viscosity of the invading phase, whereas for steady-state flow
$\mu_{\rm e}$ is considered as the saturation-weighted effective
viscosity of the two phases. Hydraulic properties of a pore-network
depend on the geometrical shape of the individual pores, as well as on
the network topology, that is, the connectivity and spatial
organization of the nodes and links of the network \cite{vr01}. With
our model, we can consider pore networks in two (2D) or three (3D)
dimensions with different topologies as illustrated in figure
\ref{figSys}. In (a), we show a crop of a Hele-Shaw cell filled with
monolayer of glass beads \cite{esth13} that is widely used as a
two-dimensional model porous medium in laboratory experiments. Such a
porous medium may modeled as a two-dimensional network of disordered
pores as shown in (b). A porous medium in 3D, a sample of sand-pack
\cite{rh09,mbweb}, is shown in (c) and a network that is reconstructed
from the sample is shown in (d). There are different techniques for
reconstructing the pore-network from scanned images of the sample,
which can be found in a wide range of literature \cite{ob05, ob02,
  ob03, d07, db09}.

\begin{figure}%=================================================
  \begin{minipage}[c][][t]{0.3\textwidth}
    \centerline{\includegraphics[width=0.8\textwidth,clip]{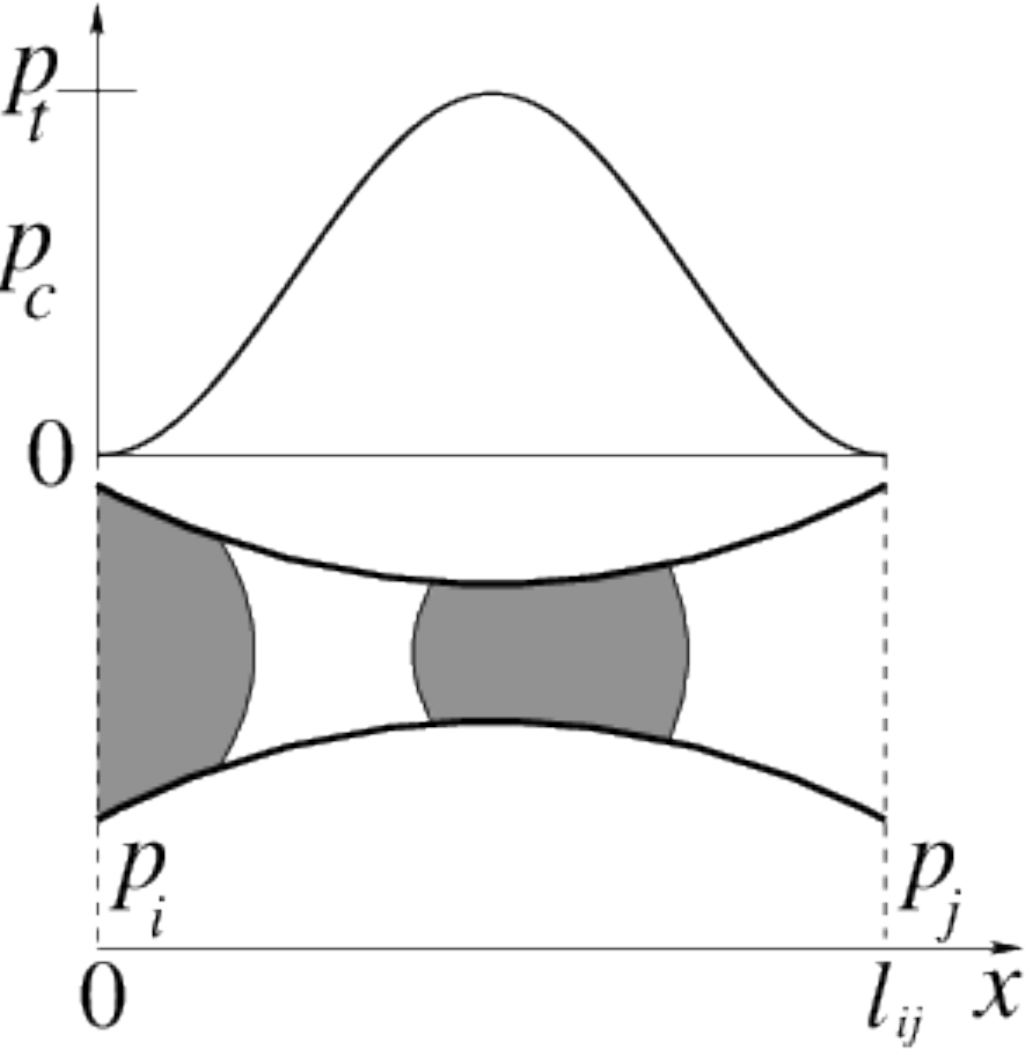}}
  \end{minipage}%
  \begin{minipage}{0.7\textwidth}
  \caption{\label{figEqn} Schematics of a link in the network, where
    the narrow throat and wide pore body is modeled by an hour-glass
    shape in terms of the capillary forces ($p_c$). The variation of
    $p_c$ at an interface with its position inside the link is shown
    above. The wetting and non-wetting fluids inside the link are
    colored as white and gray respectively. Here, $p_c$ is modeled
    with a cosine function, however one may consider any other
    function with a minimum in the middle.}
  \end{minipage}
\end{figure}%===================================================

\begin{figure}
  \begin{minipage}[c][][t]{0.4\textwidth}
    \centerline{\includegraphics[width=0.9\textwidth,clip]{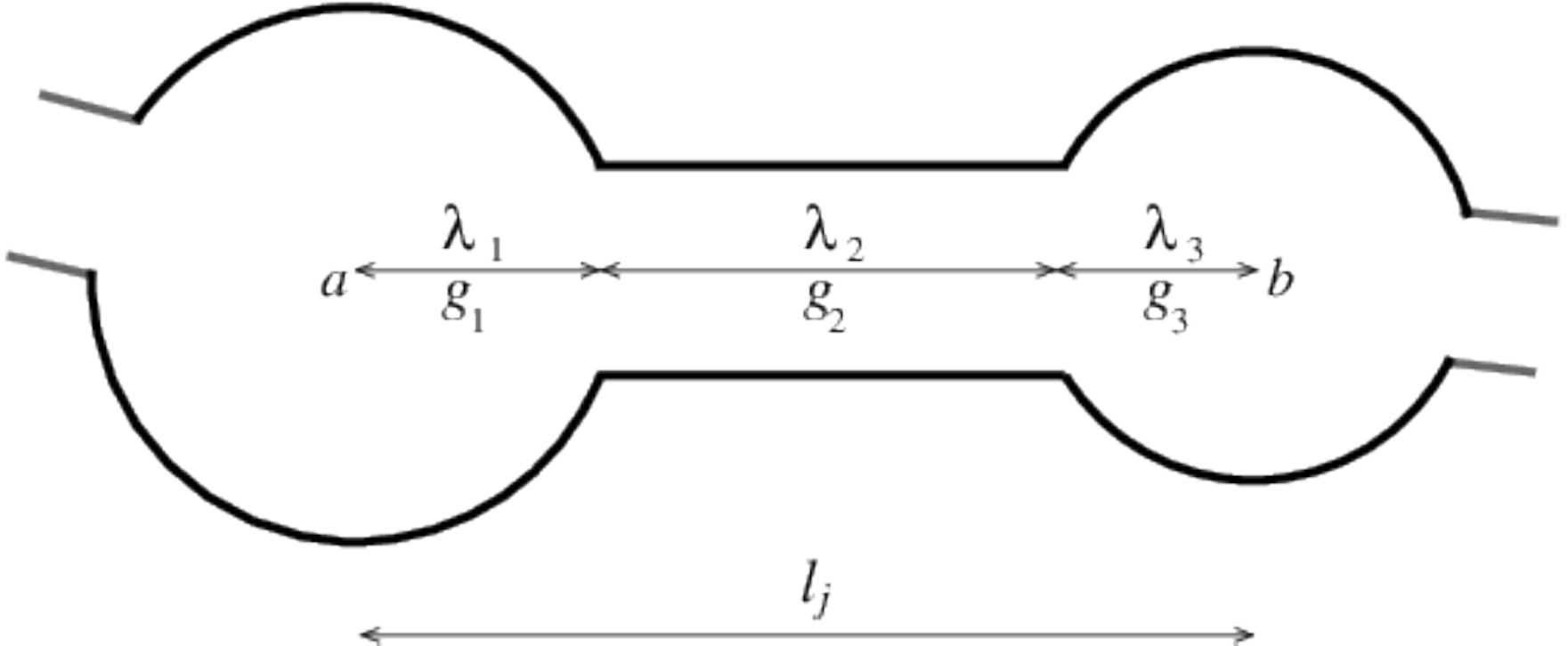}}
  \end{minipage}%
  \begin{minipage}{0.6\textwidth}
  \caption{\label{figPorepart} Schematics of one pore in the
    reconstructed network which consists of three pore parts, two pore
    bodies at the end and a narrow pore throat in the
    middle. $\lambda_{1,2,3}$ and $g_{1,2,3}$ are the lengths and
    mobility contributions. The total length of the link
    $l_j=\lambda_1+\lambda_2+\lambda_3$ and the total mobility of the
    link is obtained from equation \ref{eqngav}.}
  \end{minipage}
\end{figure}

A pore typically consists two wider pore bodies that are connected by
a narrower pore throat. In our network representation, the centers of
the pore bodies provide the positions of nodes which are connected by
links. In this model, we assign all the volume of the pore space to
the links of the network and the nodes do not contain any pore
volume. This introduces a variation in the cross-sectional area along
the length of the link due to the wider pore bodies at the ends and
the narrow pore throat in between. We model the cross section of the
link by a simplified hourglass shape as shown in figure
\ref{figEqn}. The interfacial pressure ($p_c$) at an interface
therefore depends on the position of the interface as it moves along
the link. The functional dependence of $p_c$ on the position, obtained
from Young-Laplace equation, takes the form \cite{shbk13},
\begin{equation}
  \displaystyle
  |p_c\left(x_k\right)| = \frac{2\gamma\cos\theta}{r_j}
  \left[1-\cos\left(\frac{2\pi x_k}{l_j}\right)\right]
  \label{eqnpc}
\end{equation}
where $r_j$ is the average radius of a link $j$ and $x_k \in [0,l_j]$
is the position of the $k$th interface inside the link. Here $\gamma$
is the surface tension and $\theta$ is the contact angle between the
interface and the pore wall. If $p_a$ and $p_b$ are the local
pressures at the two nodes $a$ and $b$ across the link, the
instantaneous local flow rate $q_j$ inside the link from node $a$ to
$b$ is proportional to the difference between the viscous pressure
drop $\Delta p$ ($=p_b-p_a$) across the link and the total capillary
pressure drop due to all the interfaces inside the link. This can be
calculated by \cite{w21},
\begin{equation}
  \displaystyle
  q_j = -\frac{g_j}{l_j\mu_j}\left[\Delta p_j - \sum_k p_c(x_k)\right]
  \label{eqnWB}
\end{equation}
where the sum is over all the interfaces inside the link $j$, taking
into account the direction of the capillary forces. $\mu_j$ is the
saturation-weighted viscosity of the fluids present inside the link at
that instant, given by $\mu_j=s_{j,{\rm n}}\mu_{\rm n}+s_{j,{\rm
    w}}\mu_{\rm w}$. Here $s_{j,{\rm n}}=l_{j,{\rm n}}/l_j$ and
$s_{j,{\rm w}}=l_{j,{\rm w}}/l_j$ are the fractions of the link length
occupied by the non-wetting and wetting fluids respectively, so that
$s_{j,{\rm n}}+s_{j,{\rm w}} = 1$. The term $g_j$ is the mobility of
the link, which depends on the cross section of the link. This model
deals with piston-like creep flow at low Reynolds number without any
corner flow or film flow. In that case, the effect on $q_j$ due to a
different cross-sectional shape can be taken into account by the link
mobility term $g_j$. For the regular network in 2D we chose the links
to be cylindrical with circular cross section for which
$g_j=a_jr_j^2/8$ for Hagen-Poiseuille flow \cite{d92}, where $a_j=\pi
r_j^2$ is the cross sectional area. In the reconstructed 3D network,
the pores are triangular in shape and characterized by a shape factor
($G$), defined as the ratio between the effective cross-sectional area
of the pore and the square of its circumference. The effective cross
sectional areas of each of the three pore parts, the two wider pore
bodies at the ends and the narrow throat in the middle, are calculated
from the relation $\alpha=\rho^2/(4G)$, where $\rho$ is the radius of
the inscribed circle in that pore part \cite{mm91}. The mobility
contribution from each part of the pore is then obtained from the
relation $g = 3\rho^2\alpha/20$ \cite{l64,jddy08} and the mobility
term $g_j$ for the total link is then calculated from the harmonic
average of the contribution from the three individual terms given by,
\begin{equation}
  \displaystyle
  \frac{l_j}{g_j} = \frac{\lambda_1}{g_1} +
  \frac{\lambda_2}{g_2} + \frac{\lambda_3}{g_3}
  \label{eqngav}
\end{equation}
where $\lambda_{1,2,3}$ and $g_{1,2,3}$ are the lengths and mobility
of the each pore parts respectively as shown in figure
\ref{figPorepart}.

In order to move the fluids through the pores, the local pressures at
nodes are needed to be solved at each time step. The net volume flux
($f_i$) through every node at each time step will be zero with the
Kirchhoff equations, therefore for any node $i$,
\begin{equation}
  \displaystyle
  f_i = \sum_h q_h = 0,
  \label{eqnKirch}
\end{equation}
where the subscript $h$ runs for the links connected to the node
$i$. This sum, along with the equations \ref{eqnpc} and \ref{eqnWB},
constructs a set of linear equations. We solve these equations with
conjugate gradient solver \cite{bh88} or the Cholesky factorization
method \cite{numrec} and the solutions of which at any time step
provides the local node pressures $p_i$ at that step. The system can
be driven by a constant global pressure drop $\Delta P$ or a total
flow rate $Q$ with open or periodic boundary conditions. More details
about implementing different boundary conditions will be presented in
section \ref{secBoundary}.

\section{\label{secInt} Interface-dynamics algorithms}
\begin{figure}[t]
  \begin{minipage}[c][][t]{0.7\textwidth}
    \centerline{\hfill\hfill
      \includegraphics[width=0.4\textwidth,clip]{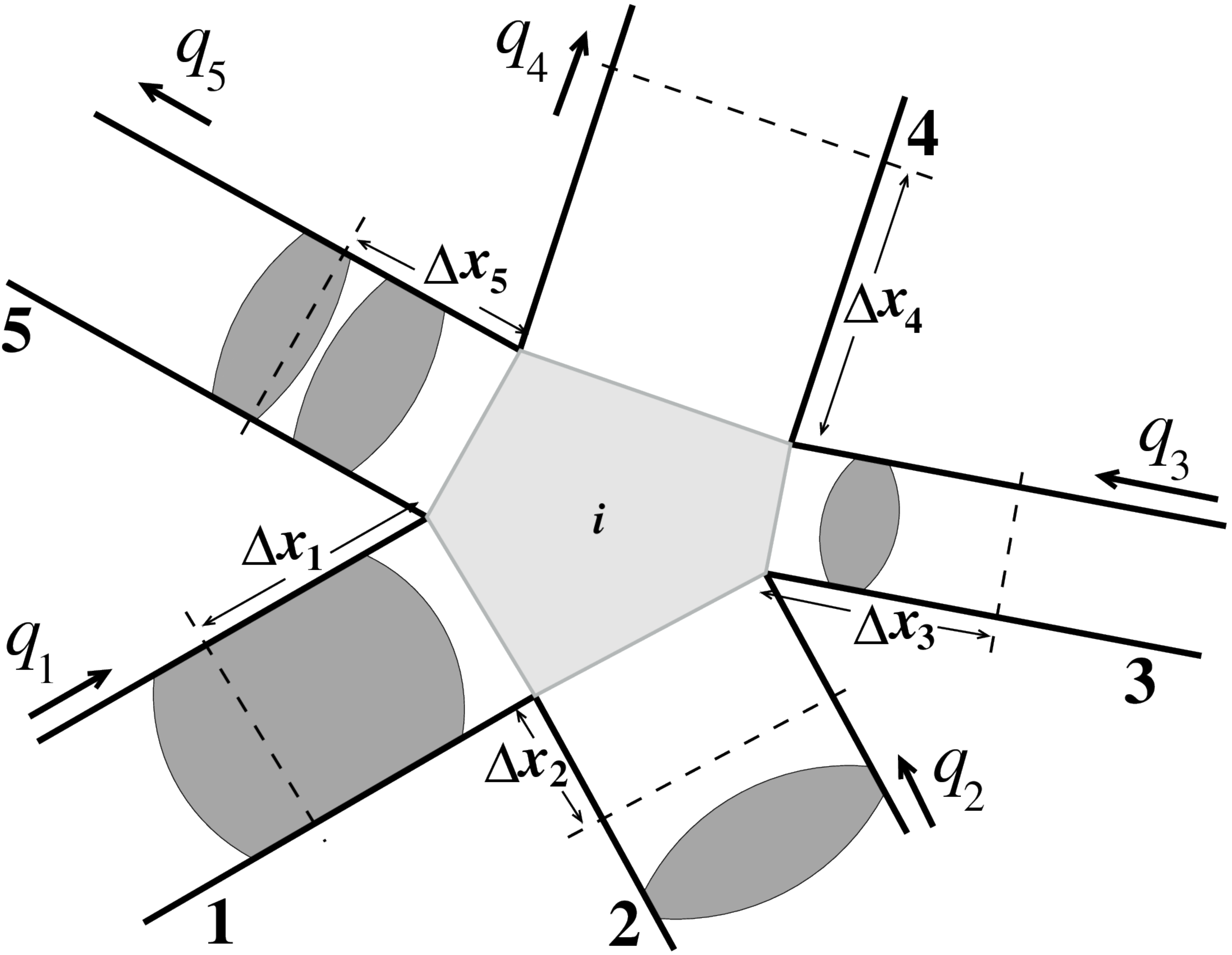}\hfill
      \includegraphics[width=0.4\textwidth,clip]{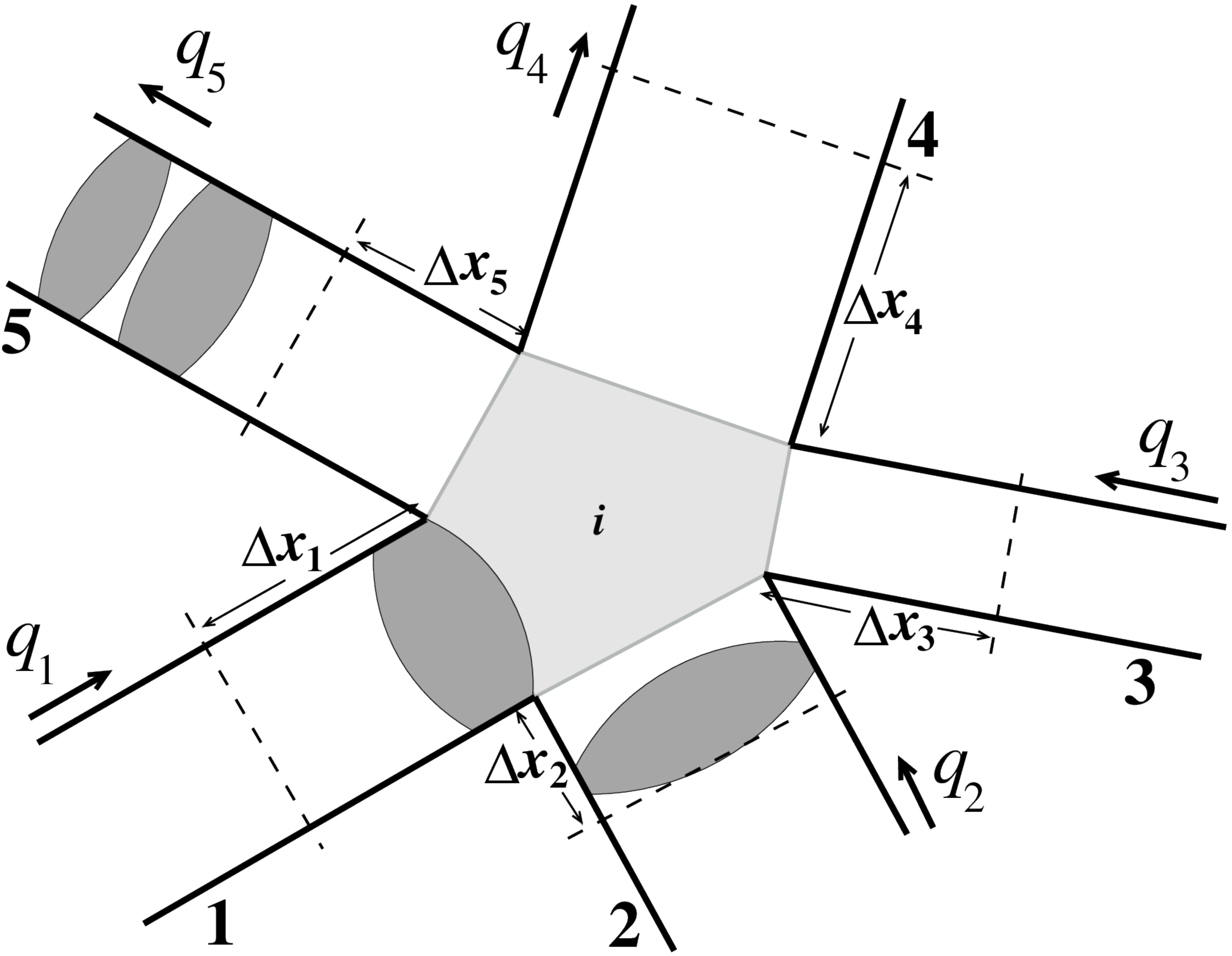}\hfill
      \hfill\includegraphics[width=0.06\textwidth,clip]{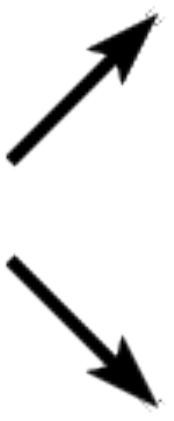}\hfill}
    \centerline{\hfill\hfill(a)\hfill \hfill (b)\hfill\hfill}
    \bigskip
    \bigskip
    \centerline{\hfill\includegraphics[width=0.5\textwidth,clip]{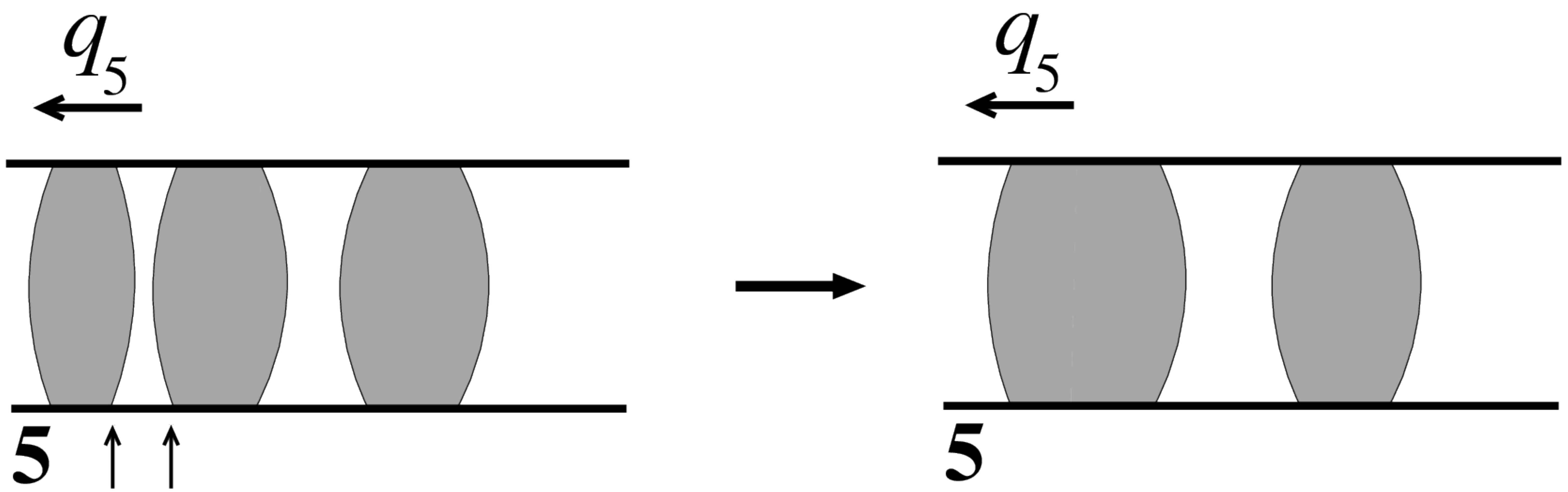}\hfill}
    \centerline{\hfill(e)\hfill}
  \end{minipage}%
  \begin{minipage}{0.3\textwidth}
    \centerline{\hfill\includegraphics[width=0.9\textwidth,clip]{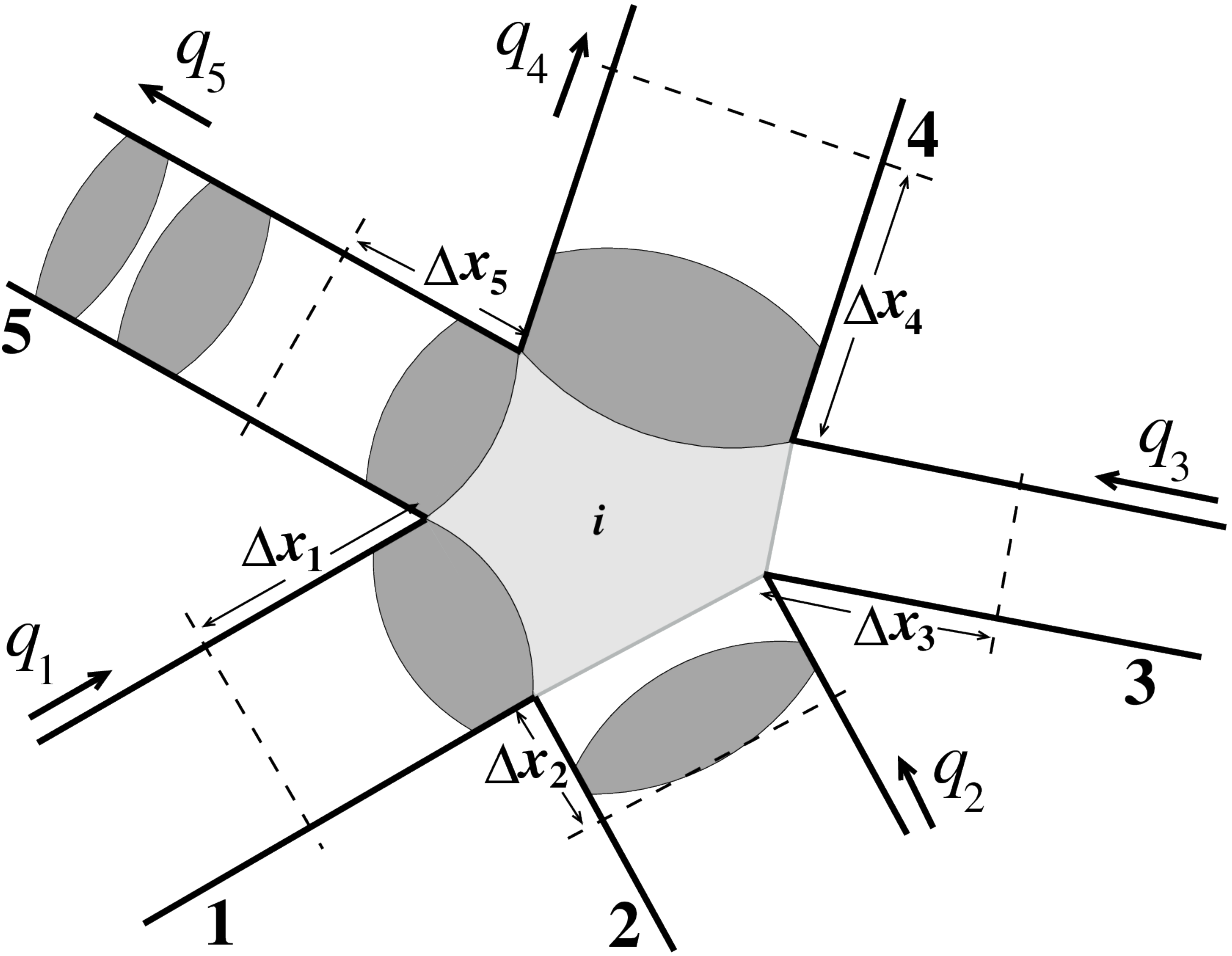}\hfill\hfill}
    \centerline{\hfill(c)\hfill}
    \medskip
    \centerline{\hfill\includegraphics[width=0.9\textwidth,clip]{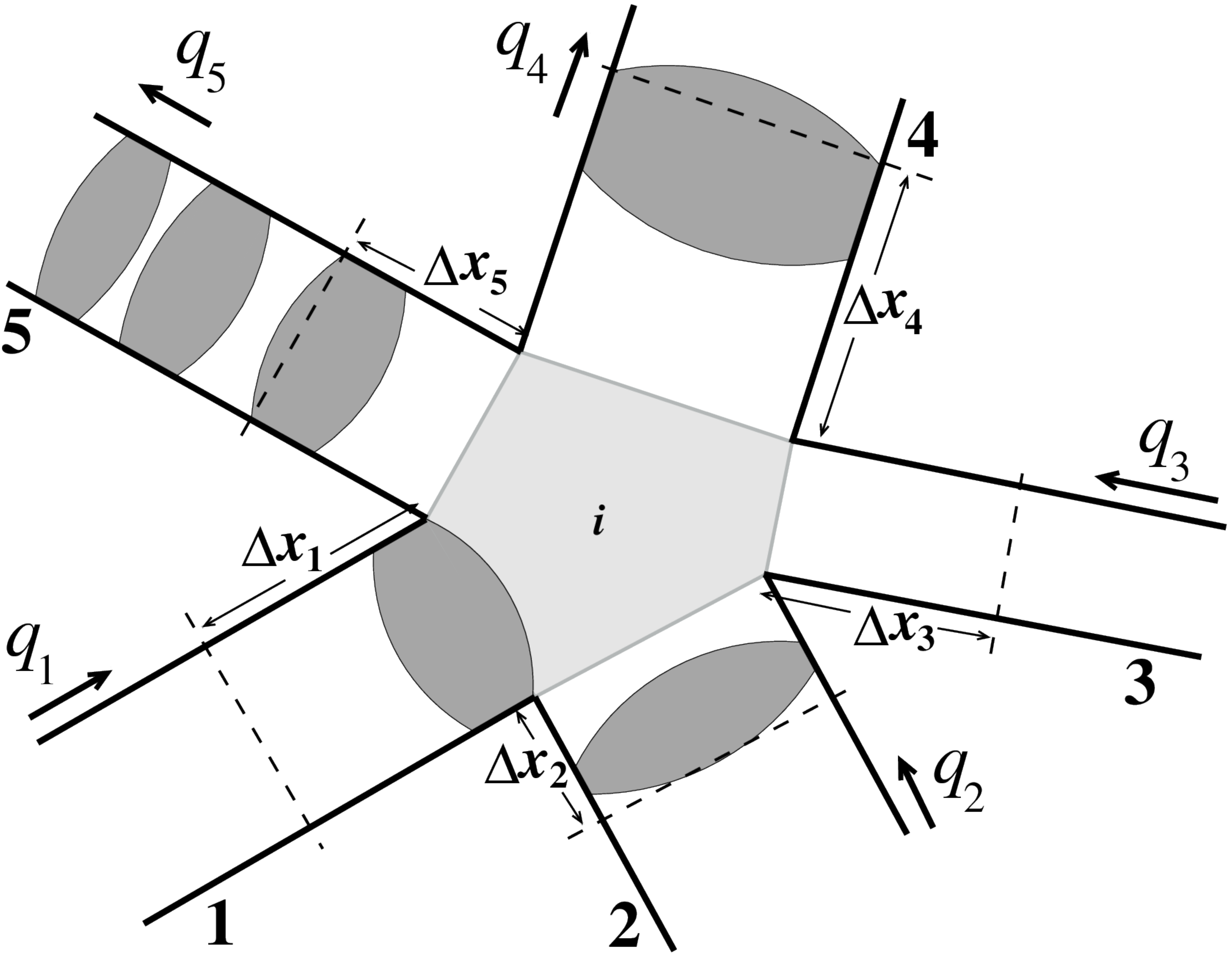}\hfill\hfill}
    \centerline{\hfill(d)\hfill}
  \end{minipage}
\medskip
\caption{\label{figBrule} The three intermediate steps for the
  interface dynamics, the \texttt{interface\_move},
  \texttt{interface\_create} and \texttt{interface\_merge}, at every
  time step are illustrated here. Here five links, numbered as $j=1$,
  $2$, $3$, $4$ and $5$, are connected to the node $i$. However the
  above illustrations are also valid for any number of links. Wetting
  and non-wetting fluids inside the links are colored by white and
  dark gray respectively. All of the pore space are distributed to the
  links in this model and the area around the node $i$, colored by
  light gray, does not contain any real volume.  (a) shows an example
  configuration of two fluids at a time step. Solution of the
  Kirchhoff equations as described in section \ref{secEqn} provides
  the local flow rates $q_j$. Based on the directions of $q_j$ at this
  time step, say the links $1$ and $2$ are identified as the incoming
  links whereas $3$, $4$ and $5$ are identified as the outgoing links
  here. Values of the displacements $\Delta x_j$ for each link,
  calculated from $\Delta x_j=q_j\Delta t/a_j$ (see text) are shown by
  dashed lines. Every interface in a link $j$ are moved towards (in
  the incoming links) or away from (in the outgoing links) the node by
  a distance $\Delta x_j$ which is shown in (b). This is performed by
  the function \texttt{interface\_move}. The total volume of the
  fluids ($a_1\Delta x_1+a_2\Delta x_2$) which left the links $1$ and
  $2$ into the node $i$ is now to be placed at the beginning of the
  links $3$, $4$ and $5$, which will fill the volume $a_3\Delta
  x_3+a_4\Delta x_4+a_5\Delta x_5$ and will create new interfaces in
  the outgoing links as described in \texttt{interface\_create}. There
  are two different ways of creating the new interfaces, one is to
  place the wetting fluid first and the non-wetting next as shown in
  (c), and the other is to place the non-wetting fluid first and the
  wetting next as shown in (d). We adopt (c) and (d) alternatively at
  each consecutive time steps. (e) shows the \texttt{interface\_merge}
  function for the link $5$ when there are four allowed
  interfaces. There link $5$ has total $6$ interfaces before merging
  and the two bubbles around the two nearest interfaces marked by the
  arrows are merged. The merging process are illustrated in more
  detail in figure \ref{figBmrg}.}
\end{figure}

We specify the locations and displacements of the two fluids inside
the pore space by the positions of all the interfaces. Transport of
the fluid bubbles through the links, the coalescence and the snap off
of the bubbles -- all are treated by the interface-dynamics algorithms
by the displacement, creation and deletion of the interfaces. The
underlying idea of these algorithms are very simple, at every time
step we calculate the volumes of fluids arriving at each node from
incoming links and then distribute them to the outgoing links with the
volumes proportional to the flow rates in them. In our implementation,
the full algorithm for the interface dynamics is sub-divided into
three intermediate functions, we name them (a)
\texttt{interface\_move}, (b) \texttt{interface\_create} and (c)
\texttt{interface\_merge}. Each of these functions is applied at every
time step. The purpose of \texttt{interface\_move} is to move every
interface in each link according to the local link flow rate and to
measure the amount of fluids that enter to a node from all the
connected links with incoming flow. Next, the
\texttt{interface\_create} function displaces all the accumulated
fluids from each node to the connected links that have the outgoing
flow and creates new interfaces at the entrance of those
links. Finally the \texttt{interface\_merge} function controls the
maximum number of interfaces in any link, which can be viewed as an
equivalent of the merging and mixing of fluids at the nodes. These
three intermediate steps are applied at every links and/or nodes at
each time step which together manage the whole process of fluid
displacements. We describe them in the following and illustrate them
in Fig \ref{figBrule}. When describing the algorithms, we will show
simulation results at limiting values of parameters to make sure that
the rules do not produce any unphysical results. This also helps to
select the best possible rule for an algorithm when there can be a few
different choices.

%-----------------------------------------------------------------
%-----------------------------------------------------------------

\subsection{\texttt{interface\_move}:}
The pore network consists of $N_{\rm L}$ number of links that are
connected to each other by $N_{\rm N}$ number of nodes. We will use
the subscript $i$ for a node ($i=1,...,N_{\rm N}$) and the subscript
$j$ for a link ($j=1,...,N_{\rm L}$). The interface number inside a
link is denoted by $k$. Inside the code, one also needs to mark the
type of the interfaces, for example, whether it is an interface at the
beginning of a non-wetting bubble or at the end. This is necessary in
order to calculate the direction of the capillary forces and also to
measure the amount of two individual fluids that enter from a link to
node. Whether it is a regular network with constant coordination
number for each node or an irregular network where the nodes have
different coordination numbers, such as a reconstructed network from a
real sample, it does not change any of the rules described in the
following. In figure \ref{figBrule}, we illustrate them for an example
of five links connected to a node $i$. At a time step, when all the
pressures ($p_i$) at the nodes for an existing fluid configuration are
known from the solutions of Kirchhoff equations, the flow rates $q_j$
inside the links are obtained from equation \ref{eqnWB}. A time
interval ($\Delta t$) is then decided in such a way that an interface
inside any link does not move more than $10\%$ of corresponding link
length at that time step. In order to do so, the link-velocities $c_j$
are calculated from local flow rates $q_j$ by $c_j=q_j/a_j$, where
$a_j$ is the cross sectional area of the link $j$. The time step is
then calculated as
\begin{equation}
\label{tstep}
\displaystyle
\Delta t = \frac{0.1l_j}{c_{j,\rm max}},
\end{equation}
where $c_{j,\rm max}$ is the largest fluid velocity among all the
links and $l_j$ is the length of that link. For the time evolution, we
use an explicit Euler-type procedure. This works well for a large
range of capillary numbers with only the time step criterion mentioned
in equation \ref{tstep}. In order to ensure numerical stability at low
capillary numbers, however, the sensitivity of interfacial pressure
jumps to perturbations in interface positions must also be taken into
account when choosing the time step, see \cite{gvkh18}. This puts a
limit on the maximum time step size that is independent of flow rate
and thus becomes a severe criterion when flow rates are low. In
\cite{gvkh18}, computational efficiency was improved at low capillary
numbers by a semi-implicit method. Here, however, we consider only
capillary numbers that are large enough for equation \ref{tstep} to be
sufficient to ensure numerical stability.

\begin{figure}[t]
  \centerline{\hfill
    \includegraphics[width=0.25\textwidth,clip]{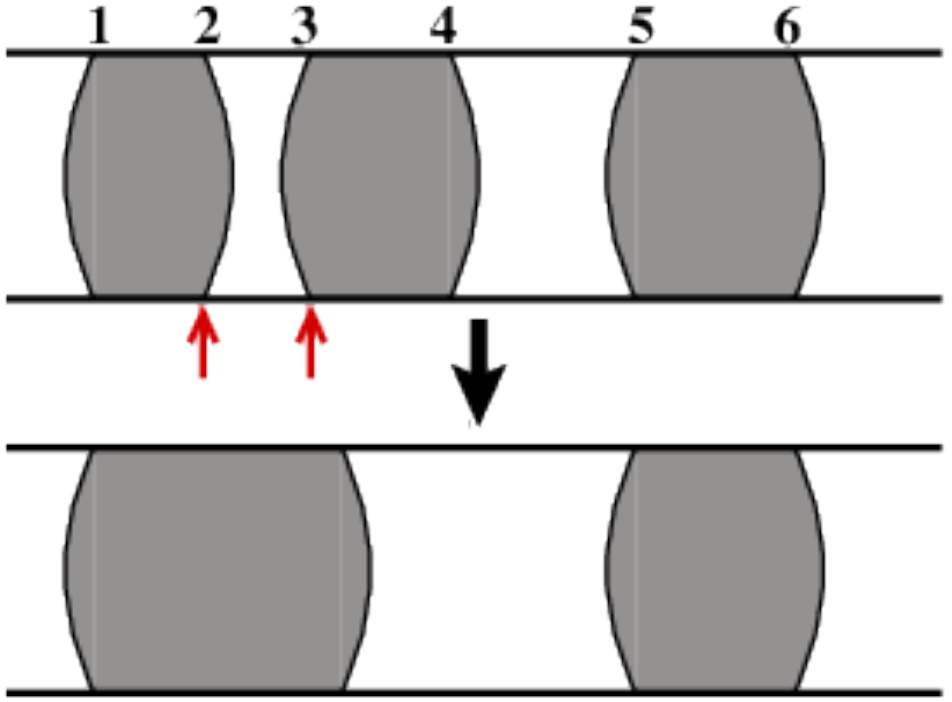}\hfill
    \includegraphics[width=0.25\textwidth,clip]{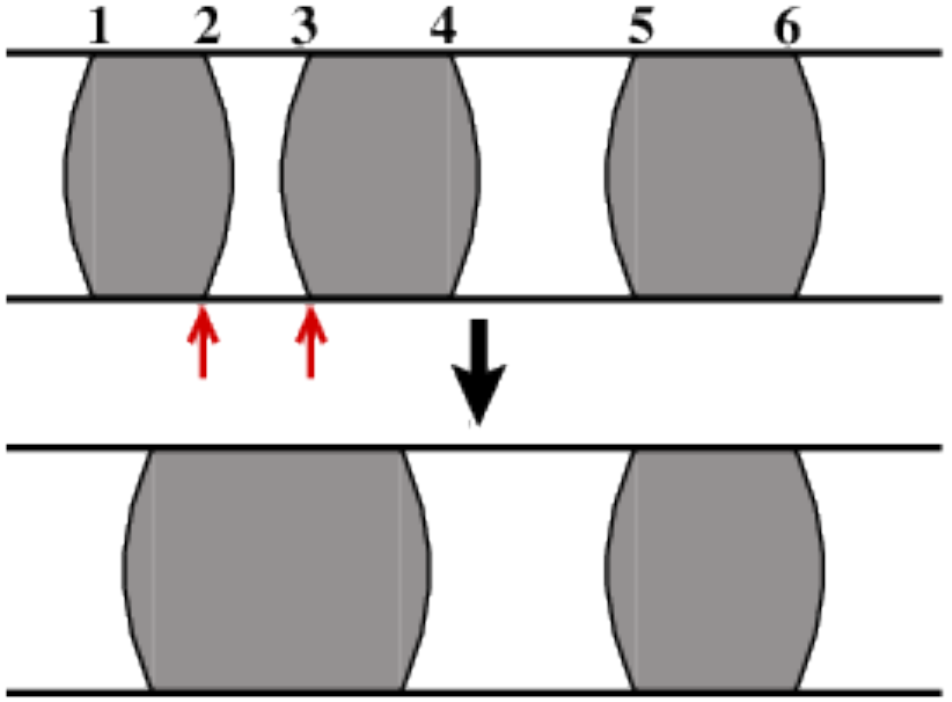}\hfill
    \includegraphics[width=0.25\textwidth,clip]{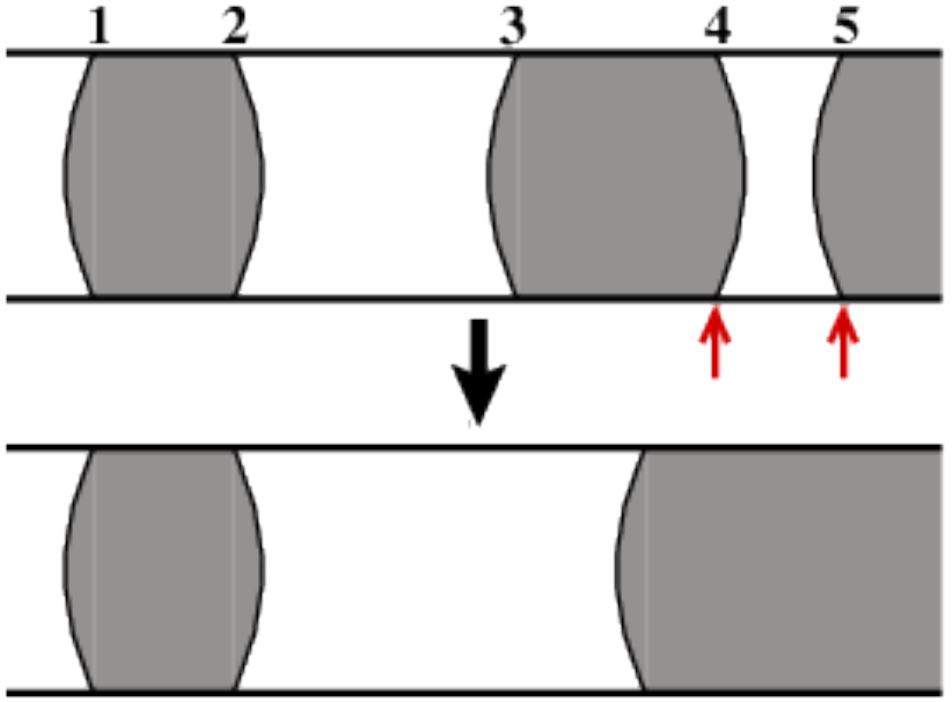}\hfill}
  \centerline{\hfill (a) \hfill \hfill (b) \hfill \hfill (c) \hfill}
\caption{\label{figBmrg}Illustration of the \texttt{interface\_merge}
  function for maximum interface count $N_{\rm max}=4$. The top row
  shows the interface positions before merging and the bottom row
  shows after merging. White and gray represent the two fluids inside
  the links. The nearest two interfaces are indicated by red
  arrows. In (a) we show the simplest possible try, namely the
  \texttt{merge\_back} scheme. The link had $6$ interfaces before
  merging among which the interfaces $2$ and $3$ are the nearest. We
  merge them by moving only one bubble towards the another, so the
  interface $1$ does not change its position, only the interface $4$
  is moved towards $1$ by a distance that is equal to the distance
  between $2$ and $3$. In (b) we show the next possible try, namely
  the \texttt{merge\_cm}. Here instead of moving the interface $4$
  towards $1$, we move both $1$ and $4$ towards each other by such
  distances so that the center of mass of the two bubbles do not
  change after merging. The third and final interface merging scheme
  is \texttt{merge\_cmnn}, which is same as (b) with an additional
  criteria as illustrated in (c). In the situation when one of the two
  nearest bubbles is connected to a node, such as the non-wetting
  bubble connected to the interface $5$ in (c), it will get
  disconnected from the node according to the merging scheme (b). We
  avoid this disconnection of any bubble from a node by moving only
  the other bubble in such a situation as shown in (c). So the to
  merge the nearest interfaces $4$ and $5$, we only move the interface
  $3$ towards the node. The \texttt{merge\_cmnn} scheme is therefore a
  combination of (b) and (c) which we adopt here.}
\end{figure}

After deciding the time step, interfaces inside each link $j$ are to
be moved in the direction of $q_j$ by a distance,
\begin{equation}
\label{xstep}
\displaystyle
\Delta x_j = \frac{q_j}{a_j}\Delta t
\end{equation}
as shown in figure \ref{figBrule}(b). Here, the links have uniform
cross-sectional area in terms of the mobility, so all the interfaces
inside the same link move by the same distance. The positions of all
the interfaces inside any link ($x_{j,k}$) are then updated,
$x_{j,k}=x_{j,k}+\Delta x_j$. By doing this, if any of the interfaces
moves outside any link ($x_{j,k}>l_j$), it is deleted from the list of
interfaces. The set of $p_i$s at the nodes decide which links, among
all the links connected to a node, have fluids that flow towards the
node and have fluids that flow away from the node at that time
step. We name these links respectively as the {\it incoming links} and
the {\it outgoing links} for that node. A typical example is shown in
figure \ref{figBrule}(a) for five links connected to a node $i$. Due
to the displacement of the interfaces by an amount $\Delta x_j$, a
node $i$ receives a certain volume ($\phi_{i,j}$) of wetting and/or
non-wetting fluids from an incoming link $j$, given by
$\phi_{i,j}=a_j\Delta x_j=q_j\Delta t$ that leaves from the end of the
incoming link $j$. The total volume of fluids ($V_i$) received by a
node $i$ from all of its incoming links is therefore,
\begin{equation}
\label{vflux}
\displaystyle
V_i = \sum_{j\in \rm I} \phi_{i,j} = V_i^{\rm w} + V_i^{\rm n},
\end{equation}
where $V_i^{\rm w}$ and $V_i^{\rm n}$ respectively are the total
volume of wetting and non-wetting fluids arrived from all the incoming
links to the node $i$. Here $j\in \rm I$ denotes the set of all the
incoming links connected to $i$. We point out that, in order to apply
this \texttt{interface\_move} function for the whole network, one can
loop through all links $j$, move the interfaces by the distance
$\Delta x_j$ and add up the amount of fluids that exit from the link
to the connected node in the direction of the link-flow. This will
produce the arrays of $V_i^{\rm w}$ and $V_i^{\rm n}$ the end of the
loop which contain the incoming volume flux at each the nodes at that
time step. The marking of the interface types, as mentioned before,
will allow here to calculate the individual terms $V_i^{\rm w}$ and
$V_i^{\rm n}$. This is illustrated in figure \ref{figBrule}.

\subsection{\texttt{interface\_create}:}
After the function \texttt{interface\_move} is performed over all the
links, the list of volumes $V_i^w$ and $V_i^n$ of incoming fluids for
all the nodes are generated for that time step. As all of the pore
space in this model is assigned to the links of the network and the
nodes do not contain any physical volume, the total volume of incoming
fluids ($V_i$) to a node are to be injected at the beginning of the
outgoing links at the same time step. This will create new bubbles and
interfaces at the beginning of the outgoing links. The total volume of
fluids that will enter from a node $i$ to an outgoing link $j$ is
given by, $\phi_{i,j}=q_j\Delta t$, and due to the balancing of the
Kirchhoff equations it ensures that $\sum_{j\in \rm I}\phi_{i,j} =
\sum_{j\in \rm O}\phi_{i,j}$ where $j\in \rm I$ and $j\in \rm O$
respectively denote the set of all incoming and outgoing links for the
node $i$. However, we still need to find out the volumes of each
individual fluid for any outgoing link, that is, how much of the
wetting ($\phi_{i,j}^{\rm w}$) and the non-wetting ($\phi_{i,j}^{\rm
  n}$) volumes will enter into $j$. We adopt a ``democratic" rule to
calculate this, that means both the wetting and non-wetting fluids get
the same preference and the volumes depend only on the flow rate $q_j$
of the respective outgoing link. For any outgoing link $j$,
$\phi_{i,j}^{\rm w}$ and $\phi_{i,j}^{\rm n}$ are therefore calculated
as,
\begin{equation}
\label{eqnvout}
\displaystyle
\phi_{i,j}^{\rm w} = q_jV_i^{\rm w}\Delta t/V_i
\quad \textrm{and} \quad
\phi_{i,j}^{\rm n} = q_jV_i^{\rm n}\Delta t/V_i
\end{equation}
which imply that the ratio of the volumes of a fluid among all the
outgoing links are the same as the ratio between flow rates among
those links. Distributing the fluids in this way also preserves the
volume conservation of {\it each individual fluid}, that is,
\begin{equation}
\label{eqnvcon}
\displaystyle
\sum_{j\in \rm O}\phi_{i,j}^{\rm w} = V_i^{\rm w}
\quad \textrm{and} \quad
\sum_{j\in \rm O}\phi_{i,j}^{\rm n} = V_i^{\rm n}.
\end{equation}
In each link, the wetting and non-wetting bubbles can be placed in two
different ways, the non-wetting fluid at the beginning and the wetting
fluid at the next, as shown in figure \ref{figBrule}(c) or in the
other way as shown in figure \ref{figBrule}(d). Here we adopt (c) or
(d) alternatively at every consecutive time steps. One can also chose
(c) or (d) randomly at every time step. This is equivalent of assuming
that at some time step the wetting fluid coming from the incoming
nodes pass the node before the non-wetting fluid enters the node, and
the situation is the opposite in the other case.

\begin{figure}
  \centerline{\hfill
    \includegraphics[width=0.3\textwidth,clip]{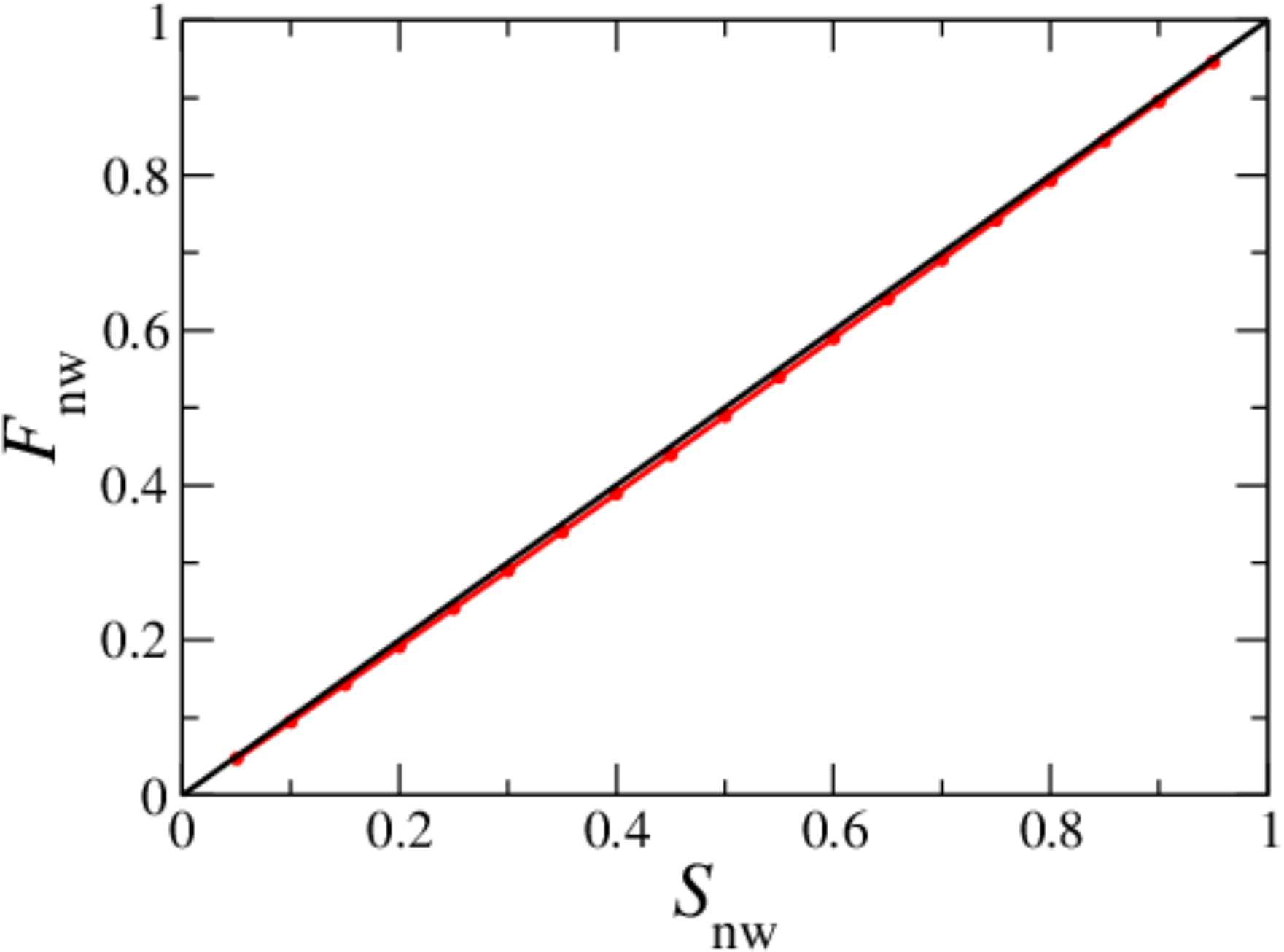}\hfill
    \includegraphics[width=0.3\textwidth,clip]{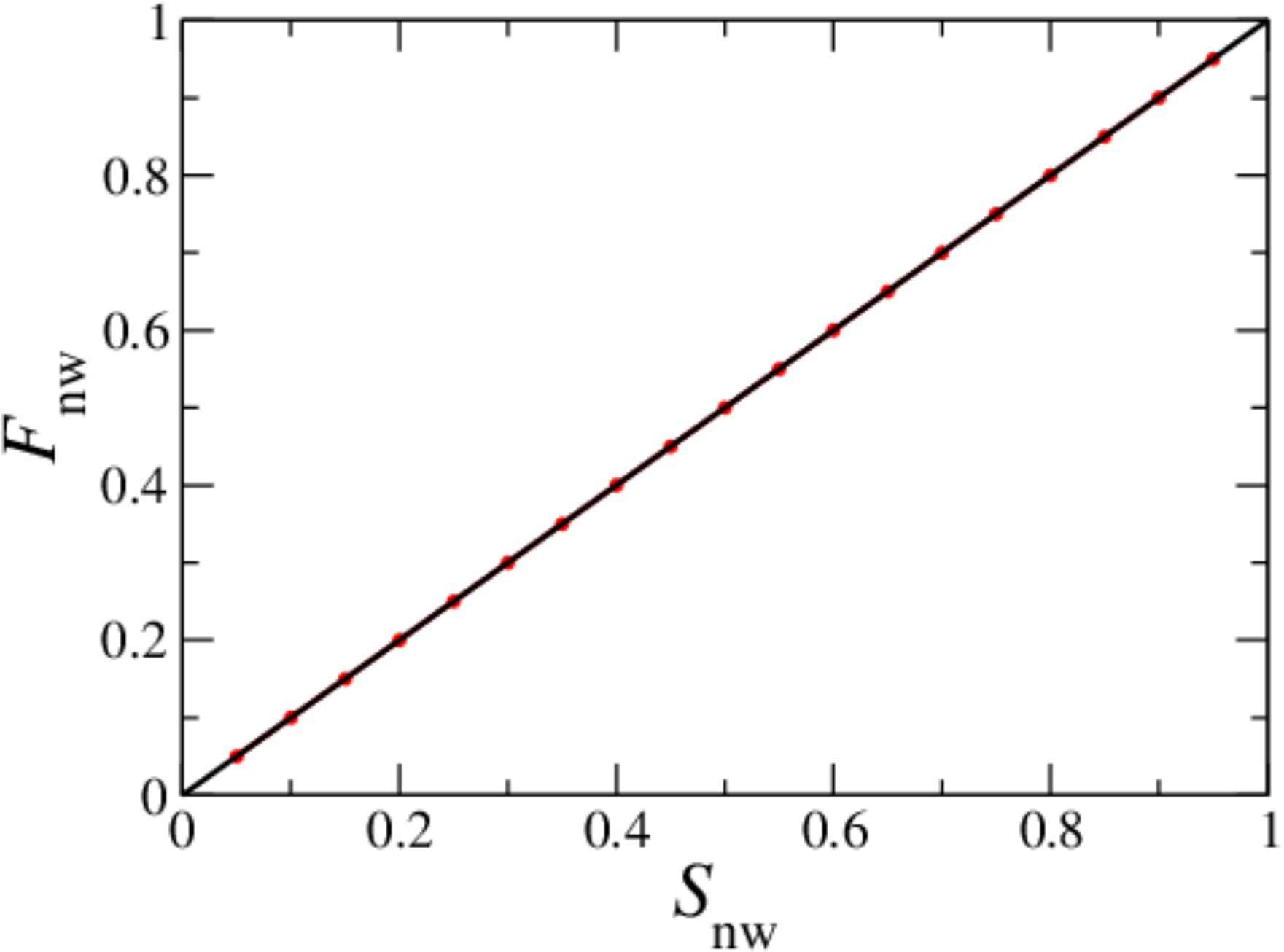}\hfill
    \includegraphics[width=0.3\textwidth,clip]{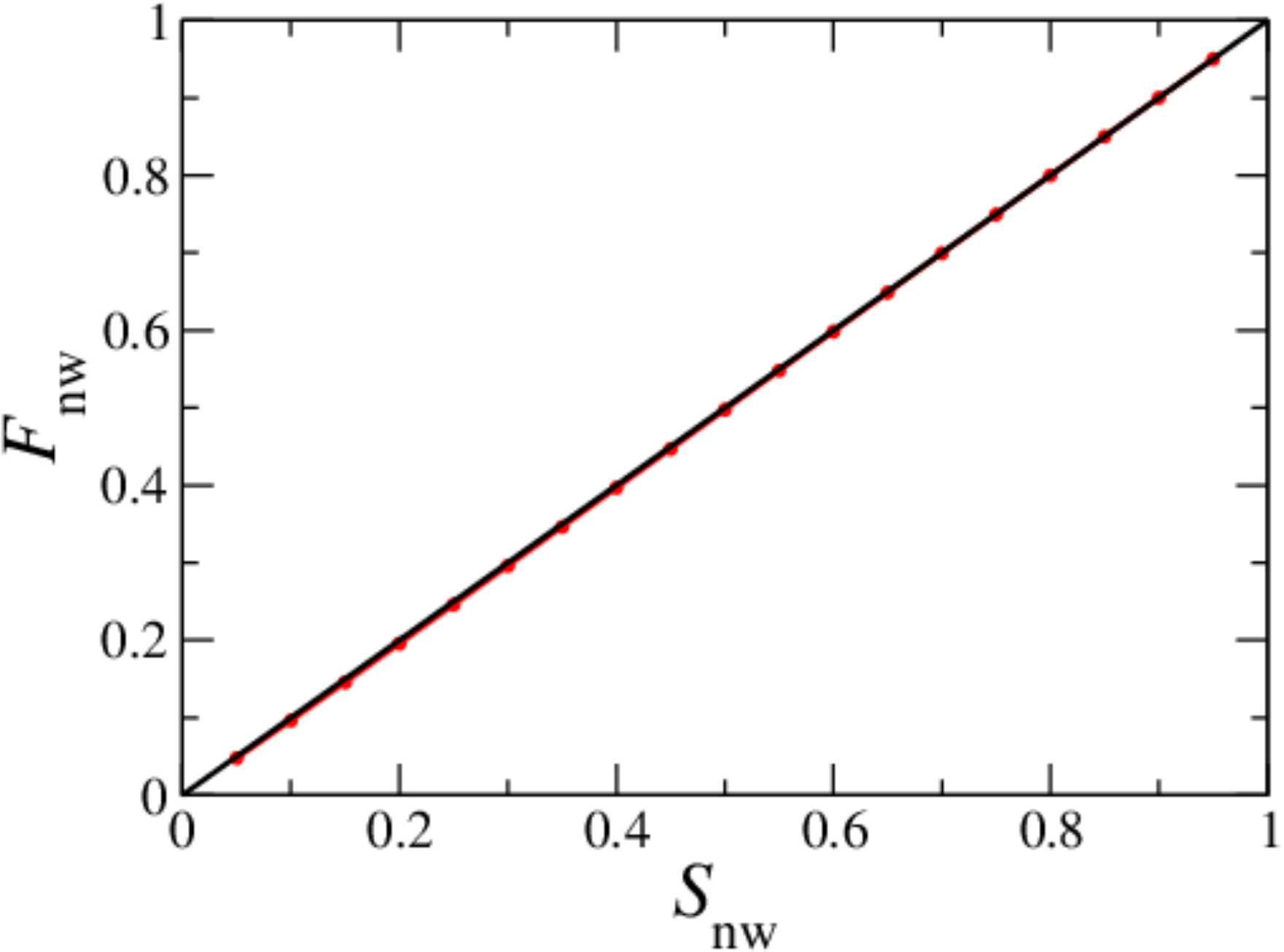}\hfill}
  \centerline{\hfill{\bf A} \hfill \hfill {\bf B} \hfill \hfill {\bf C} \hfill}
\caption{\label{figMrgZero}Plot of steady-state non-wetting fractional
  flow $F_{\rm n}$ as a function of non-wetting saturation $S_{\rm n}$
  for zero capillary pressure at the interfaces. The three plots
  correspond to the three merging schemes ({\bf A})
  \texttt{merge\_back}, ({\bf B}) \texttt{merge\_cm} and ({\bf C})
  \texttt{merge\_cmnn} as described in the section \ref{secmrg}. A
  small but systematic deviation from the diagonal $F_{\rm n}=S_{\rm
    n}$ line can be observed for {\bf A}.}
\end{figure}

These ``democratic" rules are symmetric in terms of the wetting and
the non-wetting fluids. Therefore when the surface tension is set to
zero, the capillary forces will disappear and one should obtain
\cite{sg19},
\begin{equation}
  Q_{\rm w}(S_{\rm w},M) = Q_{\rm n}(1-S_{\rm w},1/M)
  \label{eqnQsym}
\end{equation}
where $Q_{\rm n}$ and $S_{\rm n}$ are the total non-wetting flow rate
and the non-wetting saturation. Moreover, if we further set $\mu_{\rm
  w}=\mu_{\rm n}$ when the surface tension is zero, we should obtain,
\begin{equation}
  F_{\rm w} = S_{\rm w}
  \label{eqnFSEq}
\end{equation}
Where, $F_{\rm n} = Q_{\rm n}/Q$, the non-wetting fractional flow in
the steady state. These conditions can be used as preliminary tests
for the interface functions and to verify their implementations in the
code. The precise way to measure the fractional flow and other
measurable quantities will be presented in section \ref{secRes}.

\subsection{\label{secmrg}\texttt{interface\_merge}:}
Creating new interfaces in the outgoing links using the
\texttt{interface\_create} function can increase the number of bubbles
inside a link indefinitely with time. In a real system, coalescence of
bubbles occur inside the pores which prevents the number of interfaces
from increasing indefinitely. The limit depends on the pore geometry
such as the aspect ratio as well as on the flow parameters such as the
capillary number and surface tension \cite{gf06}. To limit the number
of interfaces inside a link in our model, we use an
\texttt{interface\_merge} function which merges two bubbles when the
number of interfaces inside a link exceeds a maximum number ($N_{\rm
  max}$). After the two functions \texttt{interface\_move} and
\texttt{interface\_create} are executed for each link, we look for any
link in which the number of interfaces exceeds the maximum number
$N_{\rm max}$. Notice that, limiting the number of interfaces inside a
link does not necessarily impose any restriction on the minimum or
maximum size of a bubble. The merging of interfaces can be viewed as
the coalescence of fluids during they pass the nodes. However, it is a
crucial step to formulate how exactly we merge bubbles inside a link
so that the volume of the two fluids remain conserved and it does not
introduce any artificial effect on the flow properties.

\begin{figure}[t]
  \centerline{\hfill
    \includegraphics[width=0.24\textwidth,clip]{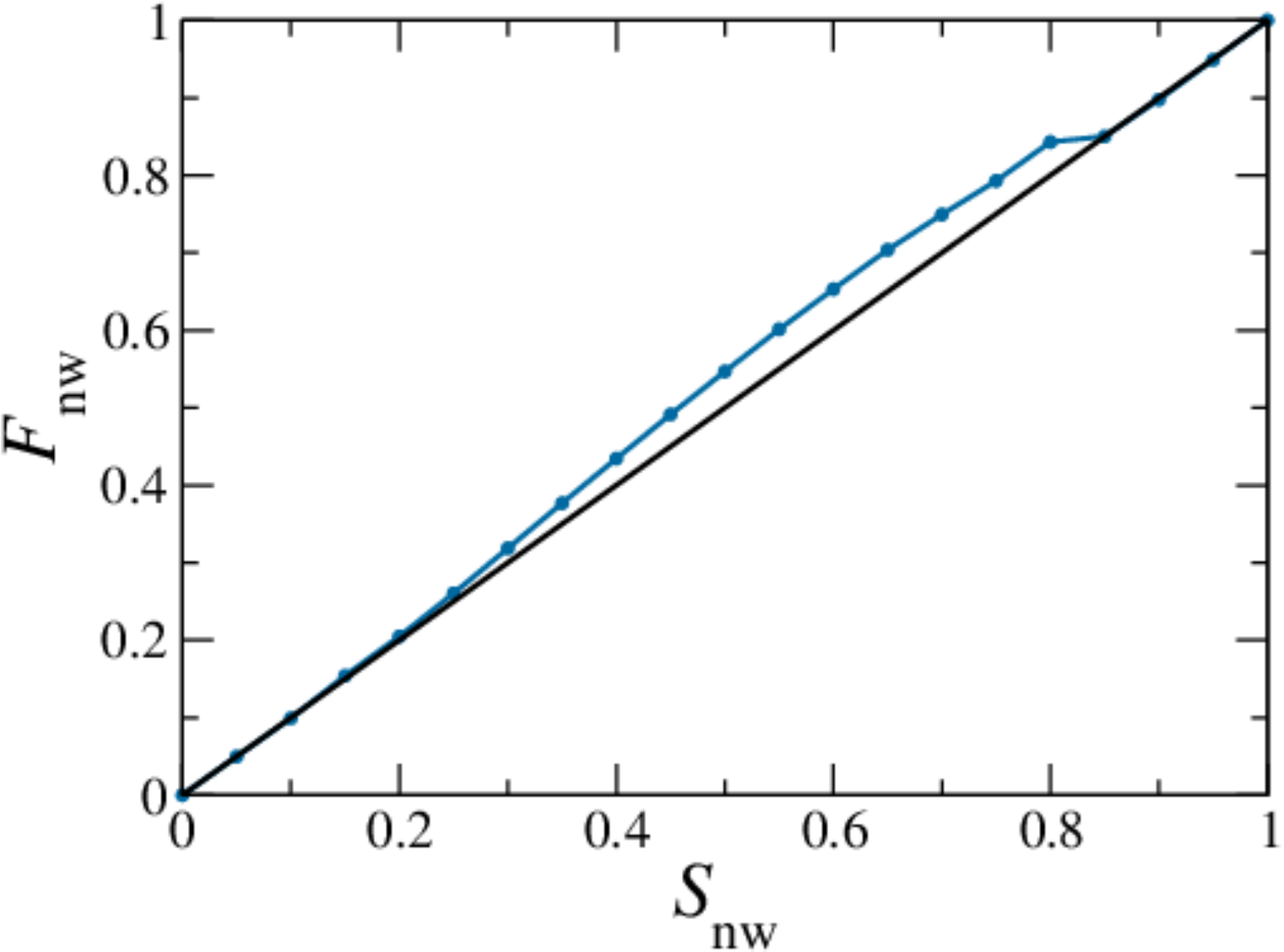}\hfill
    \includegraphics[width=0.24\textwidth,clip]{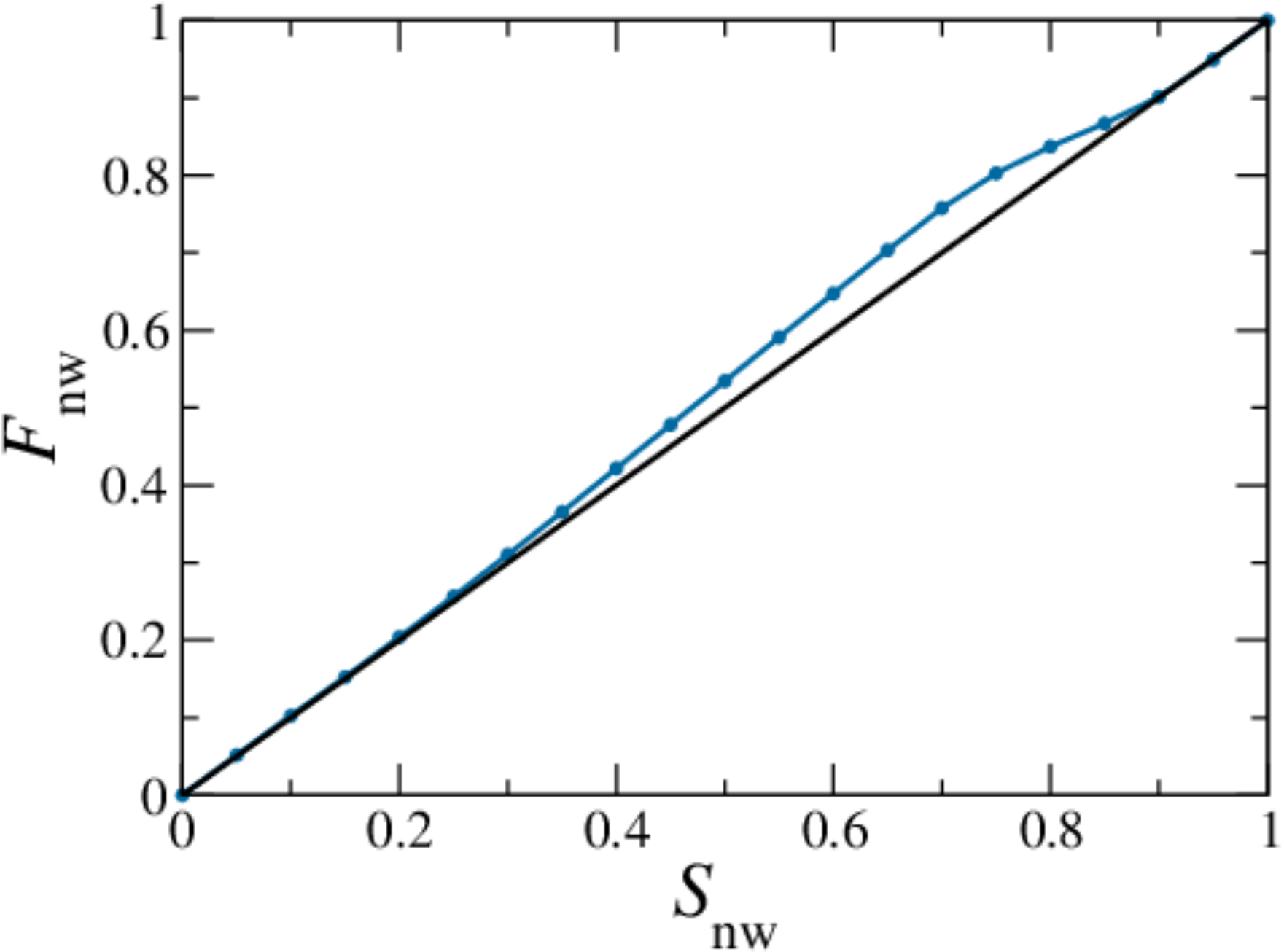}\hfill
    \includegraphics[width=0.24\textwidth,clip]{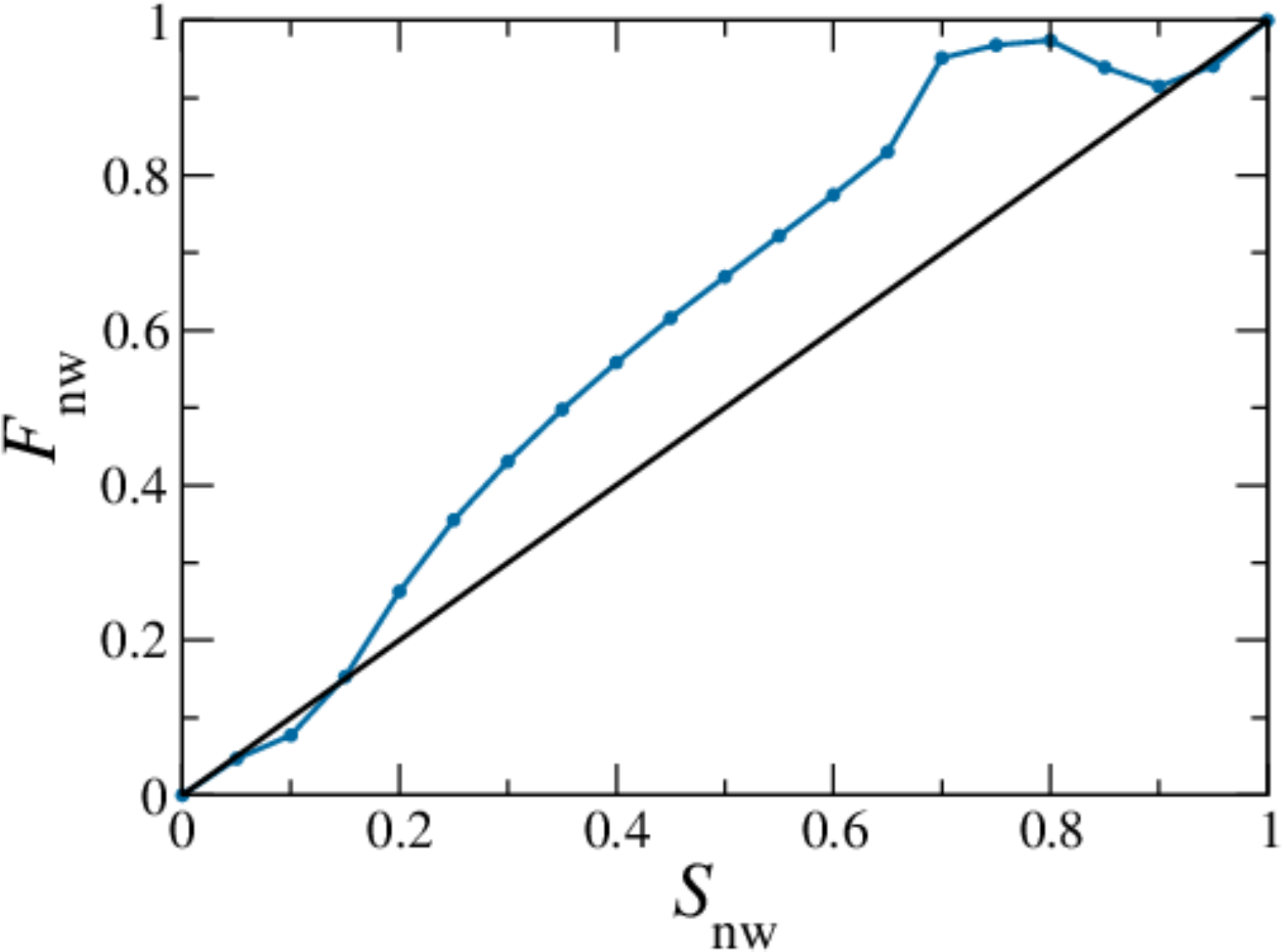}\hfill
    \includegraphics[width=0.24\textwidth,clip]{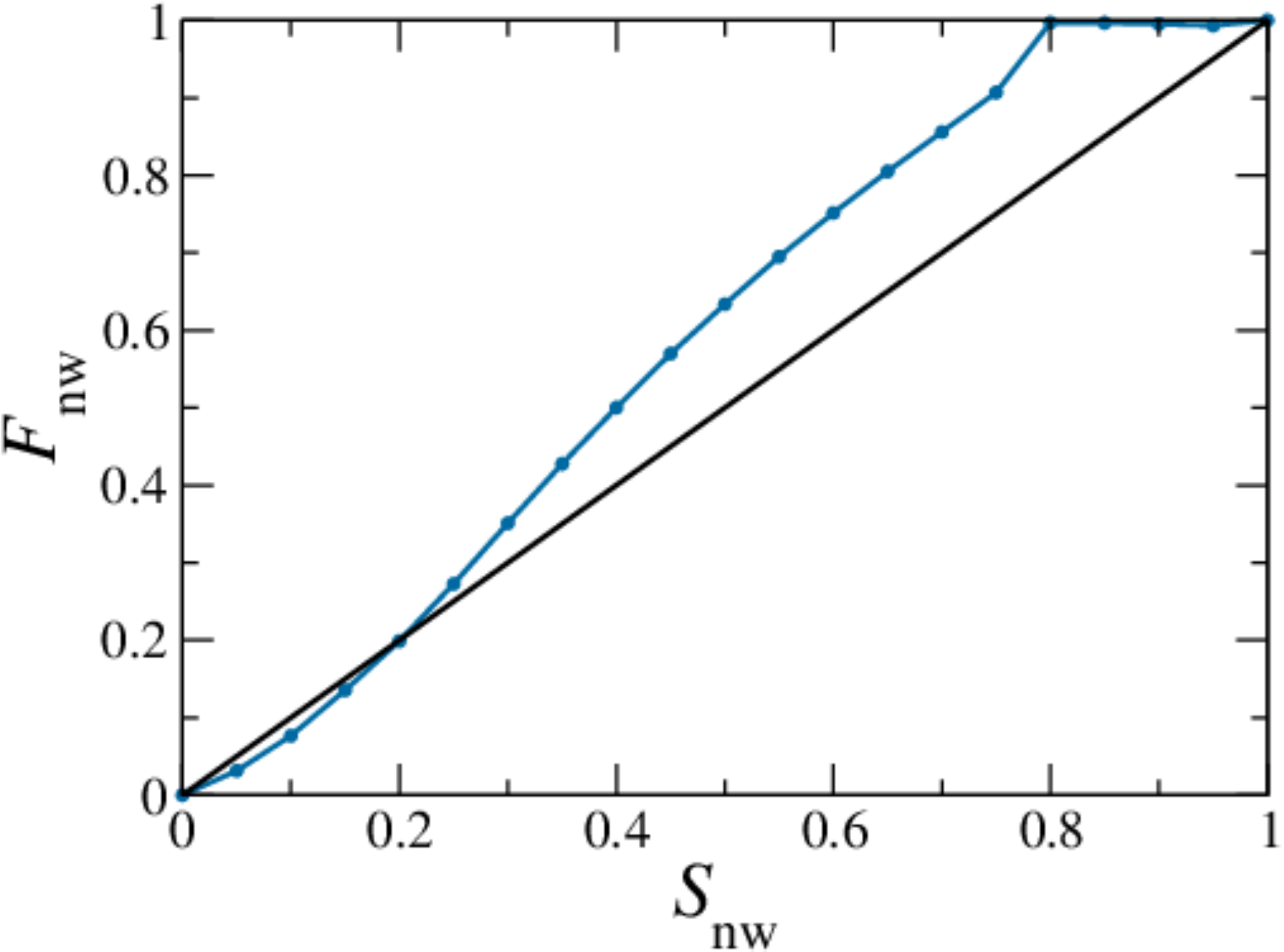}\hfill}
  \centerline{\hfill
    \includegraphics[width=0.24\textwidth,clip]{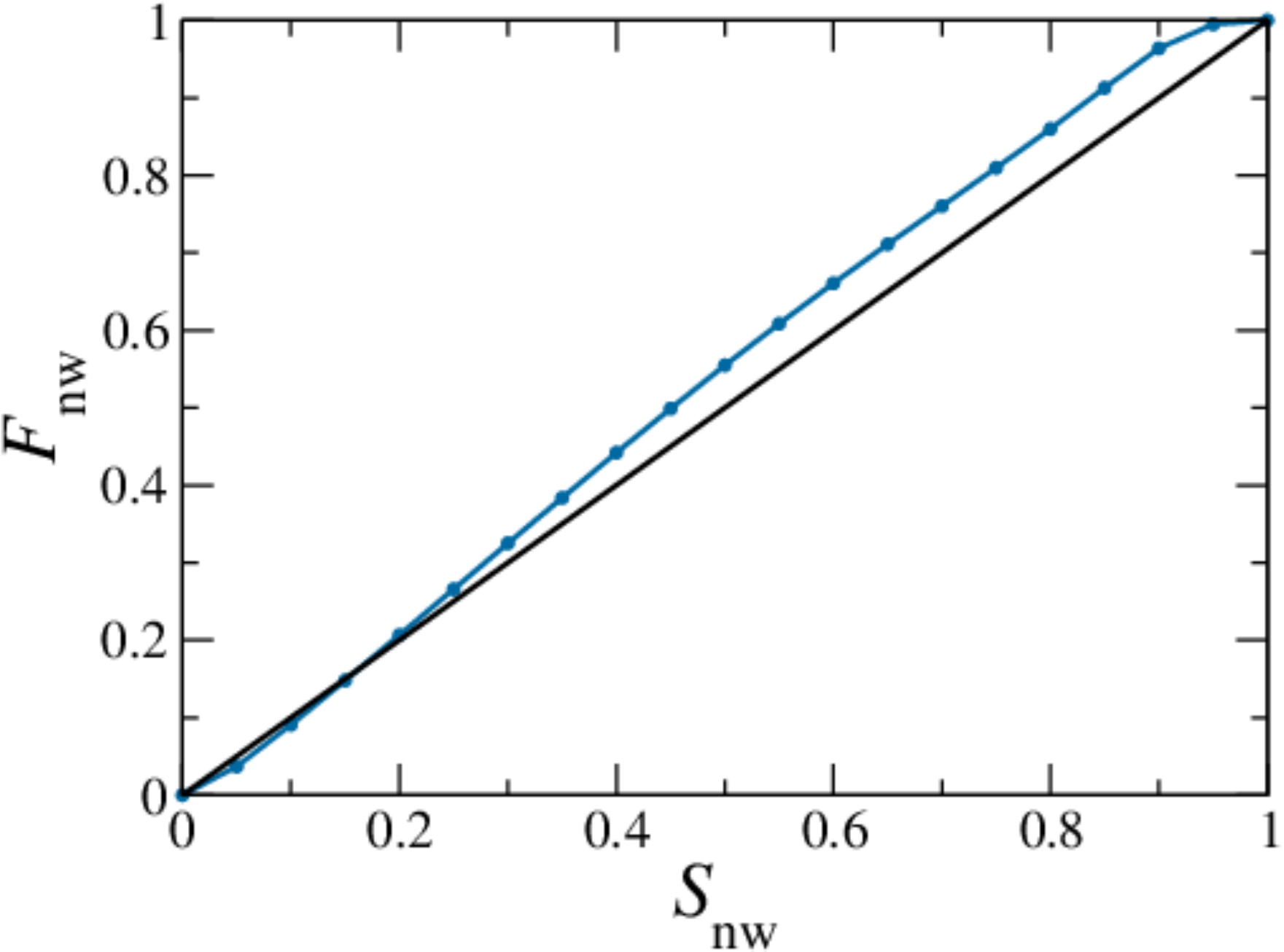}\hfill
    \includegraphics[width=0.24\textwidth,clip]{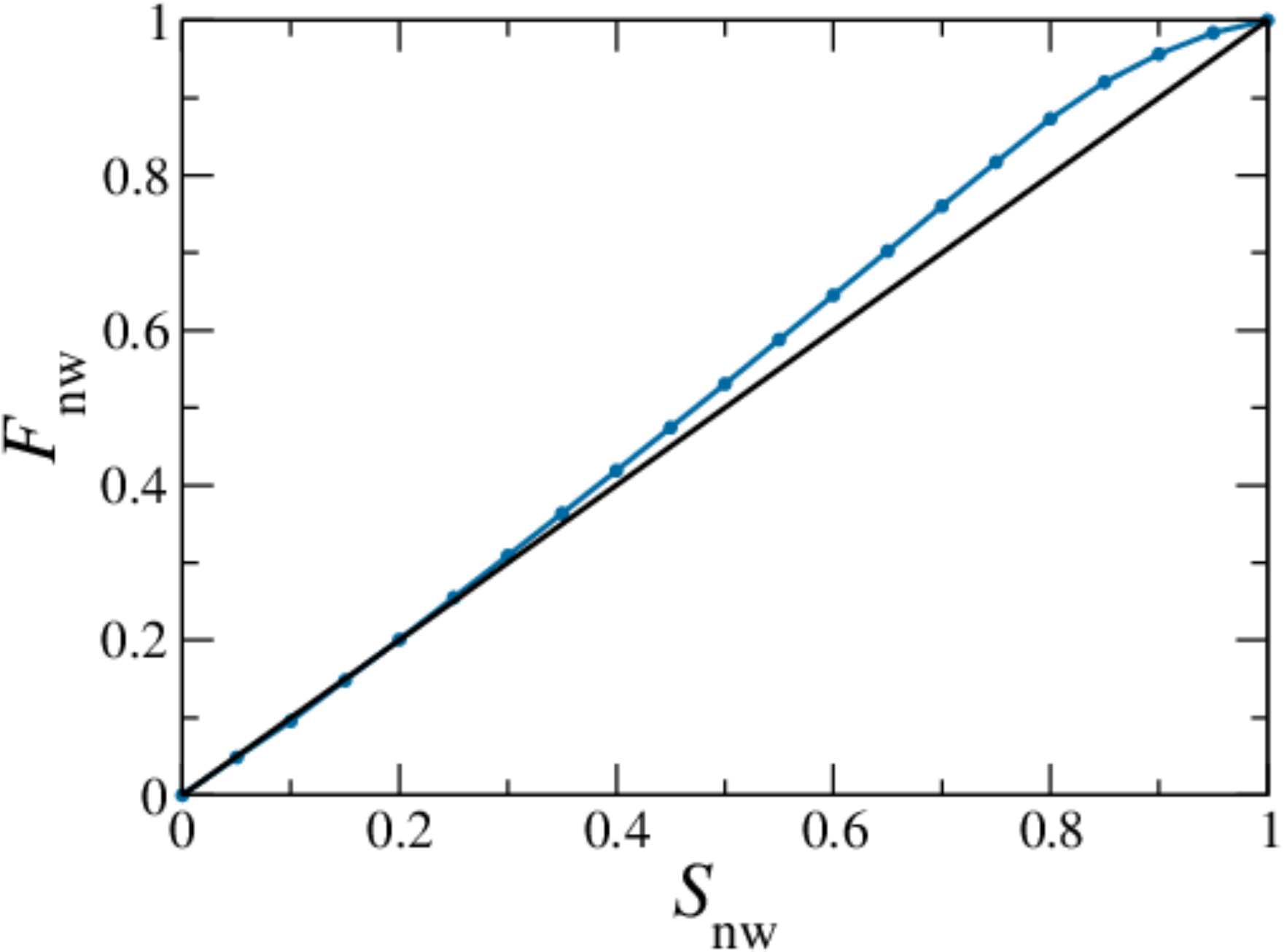}\hfill
    \includegraphics[width=0.24\textwidth,clip]{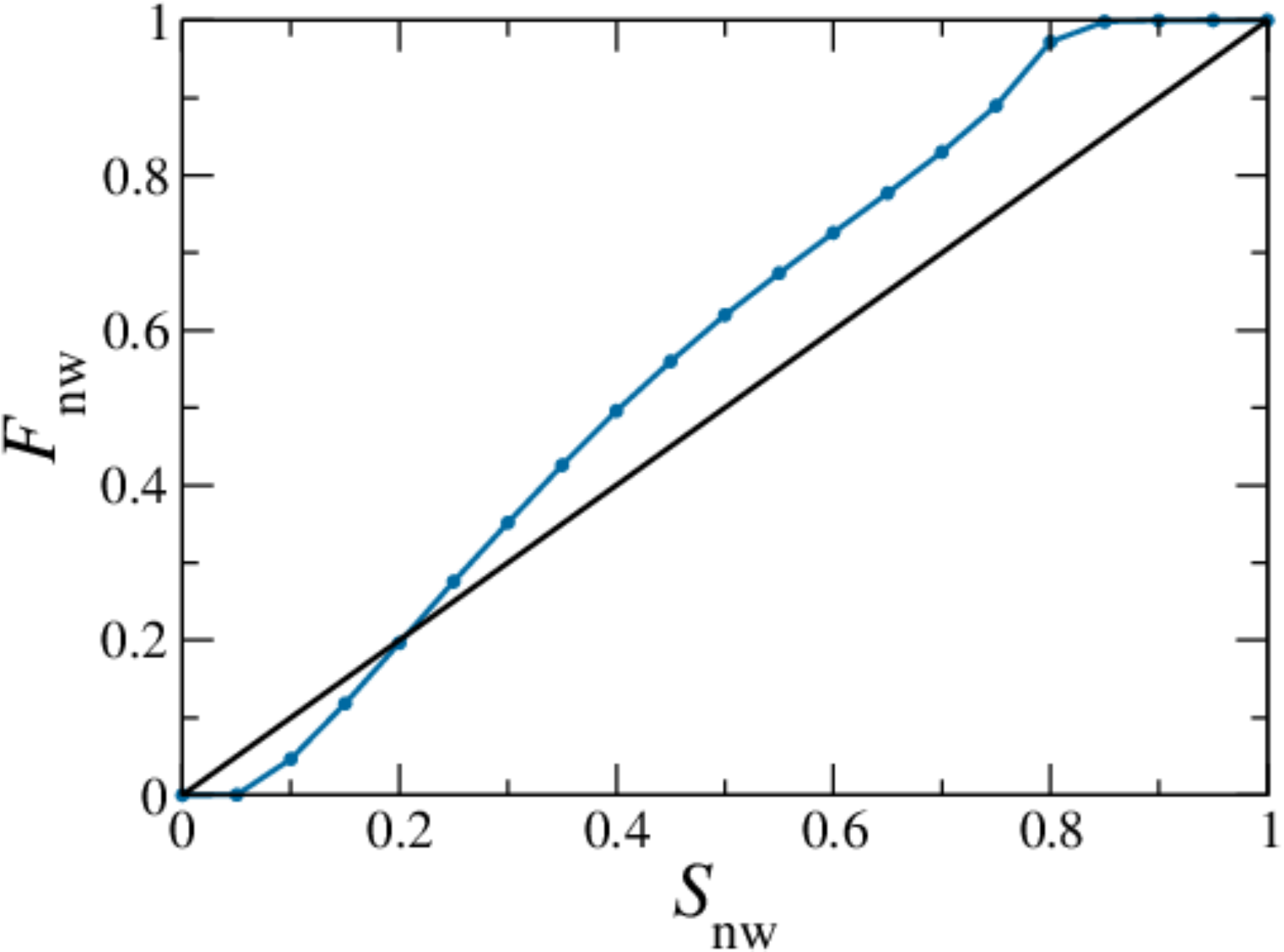}\hfill
    \includegraphics[width=0.24\textwidth,clip]{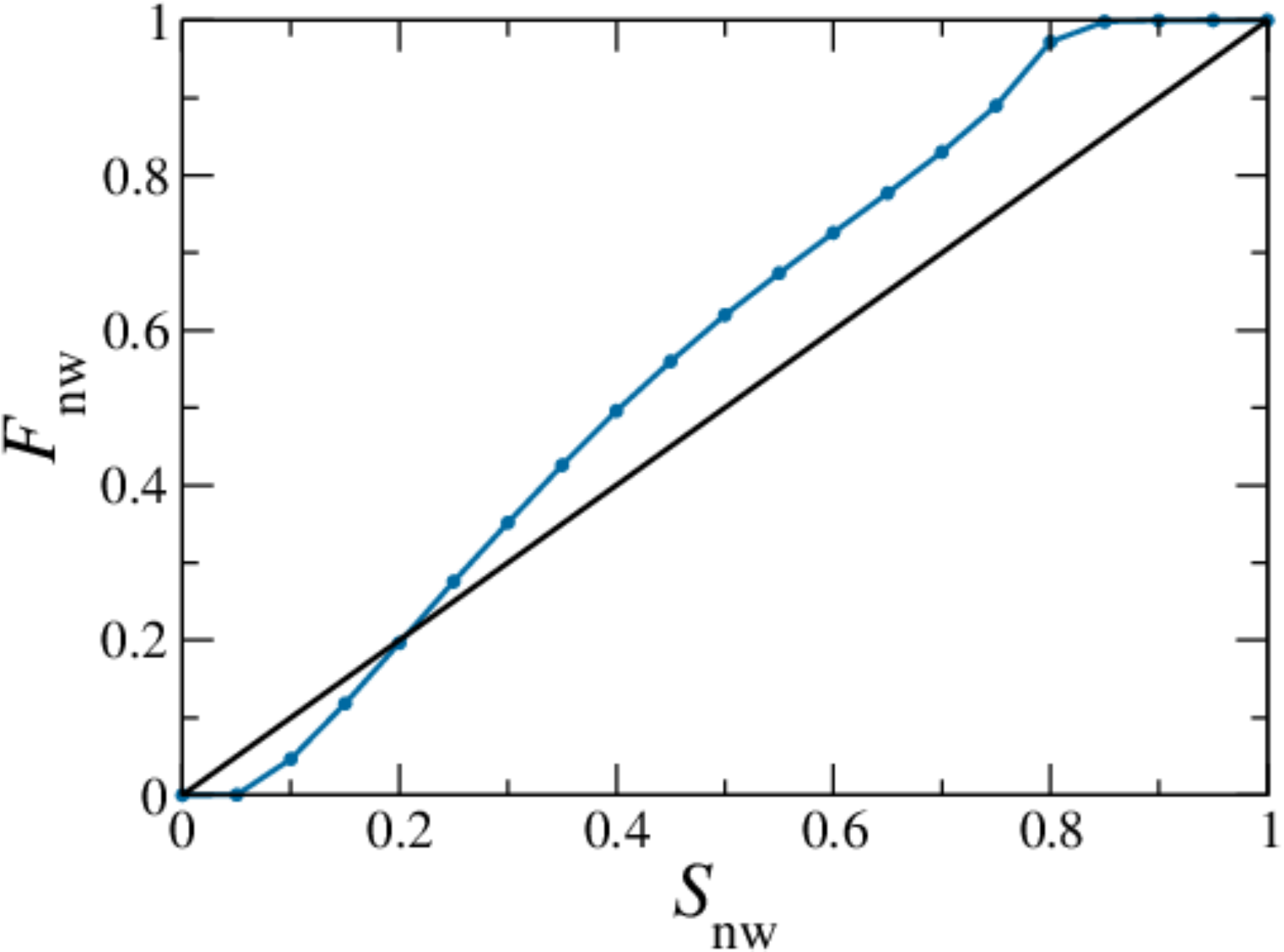}\hfill}
  \centerline{\hfill ${\rm Ca}=0.1$ \hfill \hfill ${\rm Ca}=0.01$ \hfill}
\caption{\label{figMrgCa}Steady-state fractional flow at finite
  capillary numbers ${\rm Ca}=0.1$ and $0.01$ with the merging rules
  {\bf B}: \texttt{merge\_cm} (top row) and {\bf C}:
  \texttt{merge\_cmnn} (bottom row). The left and right plots for each
  capillary number correspond to $N_{\rm max}=2$ and $3$ respectively.
  With {\bf B}, $F_{\rm n}$ approaches towards the diagonal line at
  higher saturation $S_{\rm n}>0.8$ for the first three plots.}
\end{figure}

\begin{figure}[t]
  \centerline{\hfill
    \includegraphics[width=0.24\textwidth,clip]{fig_F-S_cmnn_Ca1e-02_2m.pdf}\hfill
    \includegraphics[width=0.24\textwidth,clip]{fig_F-S_cmnn_Ca1e-02_3m.pdf}\hfill
    \includegraphics[width=0.24\textwidth,clip]{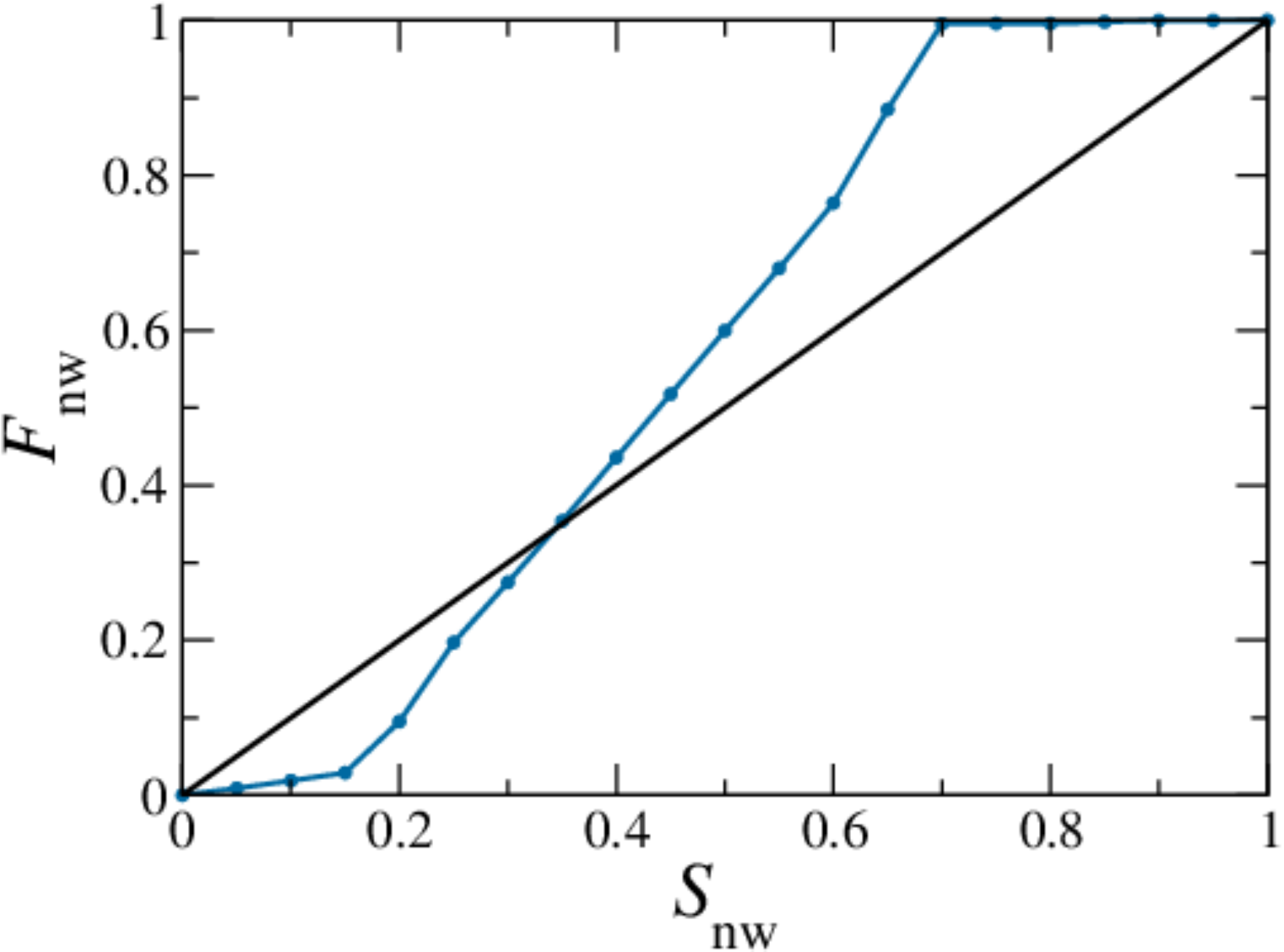}\hfill
    \includegraphics[width=0.24\textwidth,clip]{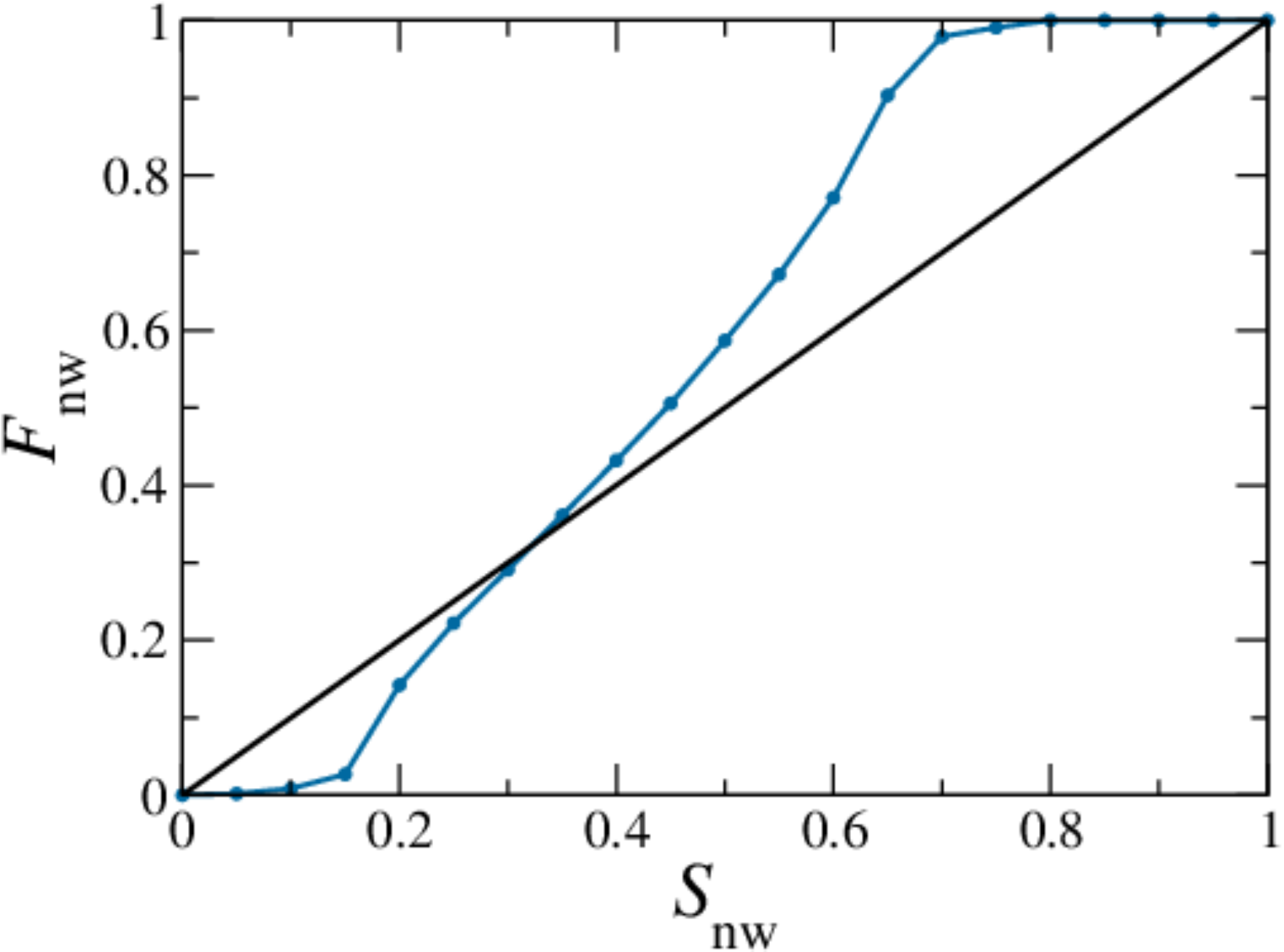}\hfill}
  \centerline{\hfill$N_{\rm max}=2$\hfill\hfill$N_{\rm max}=3$
    \hfill\hfill$N_{\rm max}=4$\hfill\hfill$N_{\rm max}=5$\hfill}
\caption{\label{figNbubl} We show the effect of changing the maximum
  number of interface ($N_{\rm max}$) with the merging scheme {\bf C},
  the \texttt{merge\_cmnn}. The steady-state non-wetting fractional
  flow is plotted as a function of non-wetting saturation for $N_{\rm
    max}=2$, $3$, $4$ and $5$. With such a change in the interface
  count, no noticeable difference is observed in the qualitative
  behavior in $F_{\rm n}$.}
\end{figure}

We start with a simple rule {\bf A} (\texttt{merge\_back}), where we
identified two nearest non-wetting bubbles inside a link. Among these
two non-wetting bubbles, only one, say the one in the front towards
the flow direction, is moved back towards the other bubble and then
merged. The length of this merged non-wetting bubble is then sum of
the two non-wetting bubbles before merging. This reduces the interface
count by two inside the link and the other interface positions are
updated accordingly. This is illustrated in figure \ref{figBmrg}
(a). However, when we measure the non-wetting fractional flow for zero
surface tension with equal viscosities of the fluids, we find a
deviation from equation \ref{eqnFSEq}. This is shown in figure
\ref{figMrgZero} (a) where a small but systematic deviation from the
diagonal $F_{\rm n}=S_{\rm n}$ line can be observed. This effect
appears due to displacing the non-wetting bubble opposite to the flow
which introduces a decrease in the non-wetting fractional flow. We
then tried the next rule {\bf B} (\texttt{merge\_cm}), where instead
of merging two nearest non-wetting bubbles, we identified any two
nearest interfaces and merge the two bubbles across them. With this
process, both the non-wetting or wetting bubbles can be merged
whichever are the nearest. Moreover, instead of moving only one bubble
towards the other, here we moved both the bubbles towards each other
by such a distance so that the center of mass of these two bubbles
does not change after merging. We illustrate this in figure
\ref{figBmrg} (b). This rule shows satisfactory $F_{\rm n}=S_{\rm n}$
behavior at the zero capillary pressure as shown in figure
\ref{figMrgZero} (b).  However, when we measured $F_{\rm n}$ at finite
capillary numbers ${\rm Ca} = 0.1$ and $0.01$, we see some discrepancy
in the qualitative behavior. The results are shown in the top row of
figure \ref{figMrgCa} with different maximum bubble counts $N_{\rm
  max}=2$ and $3$. The $F_{\rm n}$ versus $S_{\rm n}$ plot generally
shows an S-shaped curve for finite Ca. There the fluids need to
overcome the capillary barriers at the narrow pore throats and they
flow with different velocities. The phase with larger volume fraction
wins and as a result the non-wetting fractional flow curve stays under
the diagonal for lower non-wetting saturation and stays above the
diagonal for higher non-wetting saturation. Due to the asymmetry
between the wetting and non-wetting phases, the curve does not cross
the diagonal at the middle. This makes fractional flow curve S-shaped
when plotted against the saturation. When the saturation of a fluid is
very high, it percolates through the system by trapping the other
fluid in small clusters and the fractional flow of the percolating
fluid become close to $1$. However, for the merging scheme
\texttt{merge\_cm} ({\bf B}), we see that at higher saturation values
($S_{\rm n}>0.8$), the curve again approaches to the diagonal line. We
do not see any physical reason for this behavior other than an
artificial effect introduced by the merging scheme {\bf B}. It seems
that, as we moved both the bubbles towards each other while merging,
small bubbles connected to different links at the nodes got
disconnected and started flowing. We therefore updated the merging
scheme further on. In the third and final rule {\bf C}
(\texttt{merge\_cmnn}), we added one additional criteria compared to
{\bf B}. There we made sure that any fluid bubble that is in contact
to a node, does not get disconnected from the node during the merging
process. In order to do so, if one of the two nearest bubbles are
connected to a node, we did not move that bubble during the merging
process and only moved the other one. Everything else in this rule
{\bf C} are the same as rule {\bf B}. With this merging scheme, we
found exact match of fractional flow with equation \ref{eqnFSEq} at
zero surface tension and also obtained the expected qualitative
behavior for non-zero capillary pressure. These are shown in figure
\ref{figMrgZero} (c) and in the bottom row of figure \ref{figMrgCa}
respectively. Moreover, while changing the maximum bubble count with
this merging scheme, we find no noticeable difference in the
qualitative behavior of $F_{\rm n}$ with the change in $N_{\rm
  max}$. This is shown in figure \ref{figNbubl} for $N_{\rm max}=2$,
$3$, $4$ and $5$. We therefore finally adopt the \texttt{merge\_cmnn}
as the merging scheme.

\section{\label{secBoundary} Boundary conditions}
Simulations of different types of flow need proper boundary conditions
to be implemented. The drainage displacement simulations can be
performed with open boundary conditions (OB) where two opposite edges
can be used as the {\it inlets} and {\it outlets} and the other edges
are kept closed. Depending on the setup, all or some of the nodes and
links at the inlet edge can be considered to inject a fluid. Depending
on whether the system is driven under constant pressure drop or
constant flow rate, either the node pressures ($p_i$) or the link flow
rates ($q_j$) at the inlets are to be set externally. We also need to
set the node fluxes ($V^{\rm w}_i$ and $V^{\rm n}_i$) at the inlets
depending upon the type of fluid injected. For the injection of
non-wetting fluid, all the $V^{\rm n}_i$s are set to be one and the
$V^{\rm w}_i$s are set to be zero for all the inlet nodes. With these,
we solve the equations for pressures and flow rates for all the other
nodes and links inside the network.

The open boundary conditions can also be used for steady-state flow
for the setup that is generally used in laboratory experiments
\cite{tlkrfm09}. There, instead of injecting one fluid, two fluids are
injected simultaneously through alternate inlets. For this setup, all
we need to do is to set the inlet node fluxes accordingly, all the
$V^{\rm n}_i$s are set to one for the non-wetting inlets and all the
$V^{\rm w}_i$s are set to one for the wetting inlets. The inlet flow
rates define the total flow rate ($Q$) and the fractional flow $F_{\rm
  n}=Q_{\rm n}/(Q_{\rm n}+Q_{\rm w})$. Here, $Q_{\rm n}=\sum'_{\rm
  n}q_j$ and $Q_{\rm w}=\sum'_{\rm w}q_j$ where $\sum'_{\rm n}$ and
$\sum'_{\rm w}$ indicate the sum over all non-wetting and wetting
inlets respectively. The fractional flow $F_{\rm n}$ is an external
parameter in this setup and the saturation $S_{\rm n}$ is decided by
the system.

\begin{figure}[t]
  \centerline{\hfill
    \includegraphics[height=0.16\textwidth,clip]{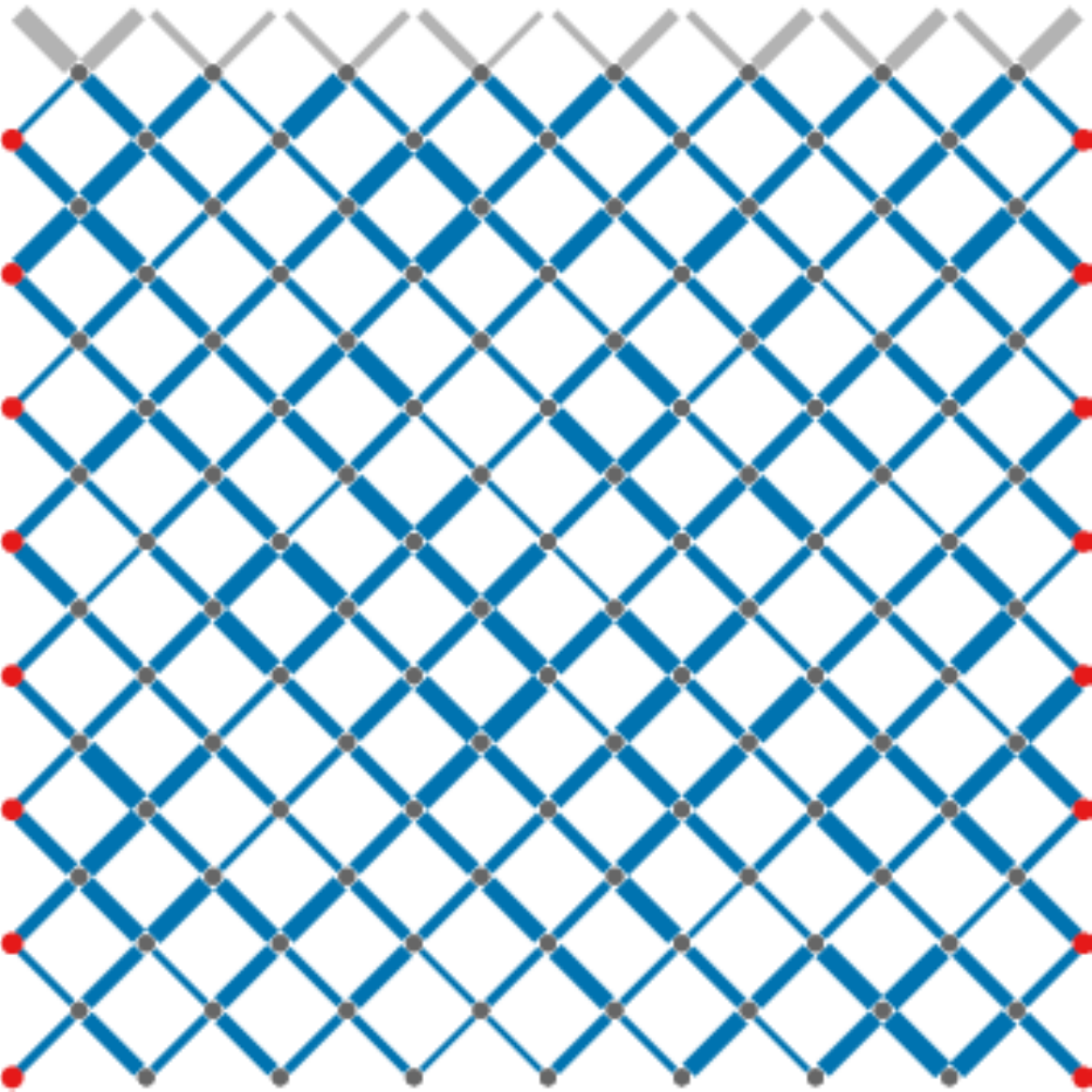}\hfill
    \includegraphics[height=0.16\textwidth,clip]{fig_Arrow_a.pdf}\hfill
    \includegraphics[height=0.16\textwidth,clip]{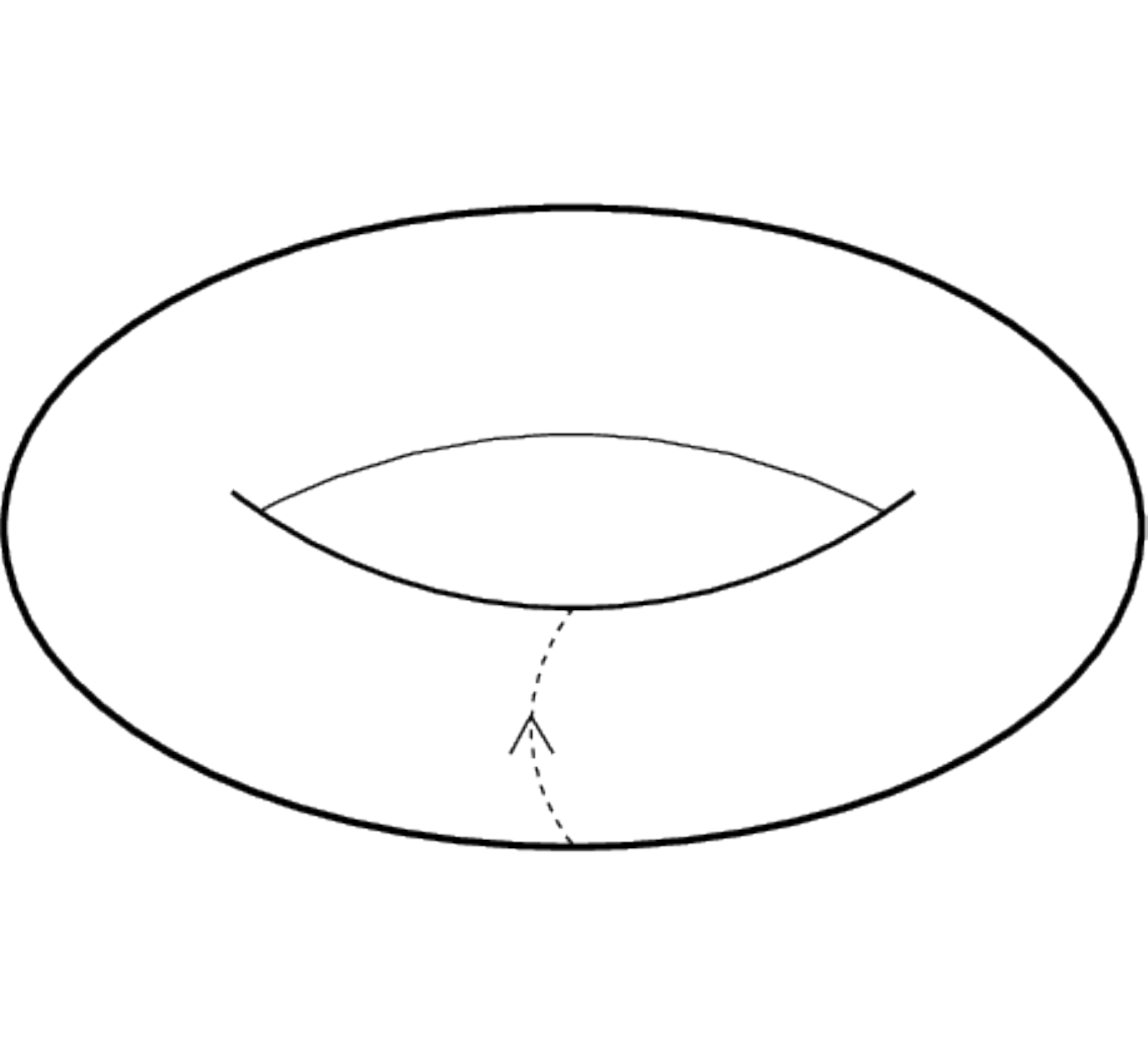}\hfill\hfill\hfill
    \includegraphics[height=0.16\textwidth,clip]{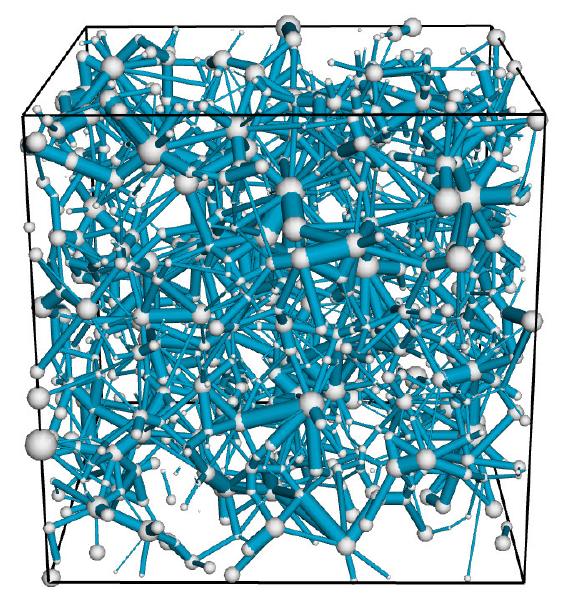}\hfill
    \includegraphics[height=0.16\textwidth,clip]{fig_Arrow_a.pdf}\hfill
    \includegraphics[height=0.16\textwidth,clip]{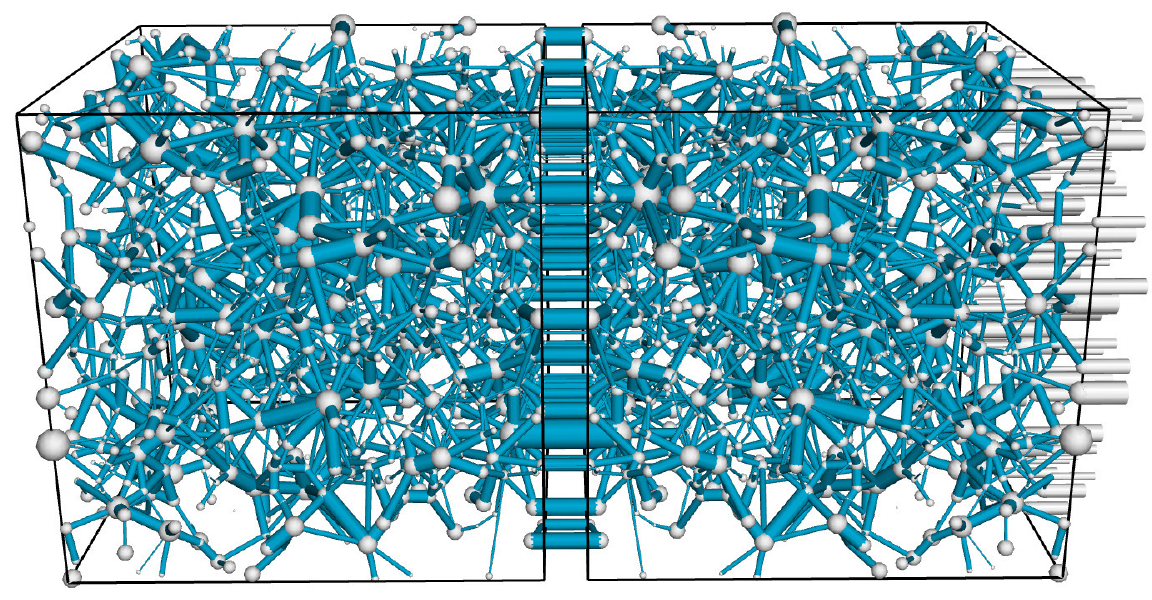}\hfill
  }
  {\hspace{0.08\textwidth} (a) \hspace{0.19\textwidth} (b) \hspace{0.18\textwidth} (c) \hspace{0.25\textwidth} (d)}
\caption{\label{figPBDemo} Implementation of periodic boundary
  conditions for 2D and 3D networks. A 2D regular network is shown in
  (a) where periodic boundary conditions are applied in both the
  directions. Here the overall flow is in the upward direction and the
  gray links at the top row represent the ghost links. The links are
  hourglass shaped in terms of the capillary pressure, but indicated
  here by uniform tubes for the simplicity of drawing. The nodes are
  marked by small dots where the red dots at the left and right edges
  represent identical nodes. This makes the flow essentially on the
  surface of a torus as shown in (b) where the arrow represents the
  effective flow direction. For 3D, a network reconstructed from Beria
  sandstone is shown in (c) where the overall flow is from left to
  right. Periodic boundary conditions are applied in the direction of
  flow by adding a mirror copy of the the network as shown in (d). The
  ghost links at the right are colored by gray.}
\end{figure}

There is a better way to run steady-state flow in the simulations
which is not possible in case of the experiments. This is by
implementing the periodic boundary (PB) conditions in the direction of
overall flow. For this, we connect the inlet and outlet edges so that
the fluids leaving at the outlets, enter the system again through the
inlets. The links that connect the inlets and outlets, we call {\it
  ghost links}. With this, the network becomes a closed system and the
two fluids keep flowing through the system. The saturation $S_{\rm n}$
therefore is an external parameter here and the fractional flow is
decided by the system. This is illustrated in figure \ref{figPBDemo}
for 2D and 3D networks. In case of 2D, periodic boundary conditions
are applied in both directions which makes the flow equivalent to the
flow on the surface of a torus as shown in (b). For the reconstructed
network in 3D, the opposite inlet and outlet edges are not
identical. Therefore, to apply to periodic boundary conditions to this
network we double the system by a mirror copy of the network in the
direction of the flow, as shown by the two cuboids in (d). This makes
the inlets and outlets at the opposite edges identical, which we then
connect with the ghost links to make the system periodic. With the
periodic boundary, the inlets are seen as the neighbors of the outlets
for the interface-dynamics algorithms. Here, the global pressure drop
$\Delta P$ to drive the flow needs to be added with the node pressure
drops across the ghost links while solving the equations \ref{eqnWB}
and \ref{eqnKirch}.

The structure of the complete simulation with all the interface
dynamics functions is the following:
\begin{algorithmic}[1]
  \STATE Network:    construct or read
  \STATE Define:     boundary conditions
  \STATE Initialize: random or sequential fluid distribution
  \FOR {$t = 1$ to $time steps$}
      \STATE Solve the pressure field
      \FOR {$j = 1$ to $total links$}
           \STATE \texttt{interface\_move}($j$)
      \ENDFOR
      \FOR {$j = 1$ to $total links$}
           \STATE \texttt{interface\_create}($j$)
      %\ENDFOR
      %\FOR {$l = 1$ to $total links$}
           \STATE \texttt{interface\_merge}($j$)
      \ENDFOR
      \STATE Calculate measurable quantities at $t$
  \ENDFOR    
\end{algorithmic}
During initialization, the initial positions of all the interfaces in
each link will be defined depending on the saturation and how we want
to start the simulation. Then after solving the pressure field, the
\texttt{interface\_move} function is performed on all the links and
that will generate the array of the node fluxes $V^w_i$ and $V^n_i$
for each node $i$. These arrays will then be used as input to the
\texttt{interface\_create} function that will create the new
interfaces in the links.

\section{\label{secRes} Applications and validation}
As stated earlier, the interface algorithms in this pore-network model
has the flexibility to be applied for different network topologies and
boundary conditions. Moreover, we can study both the transient and the
steady-state properties. In our simulations, we consider a network of
links forming a tilted square lattice at an angle $45^\circ$ with the
direction of applied pressure drop in 2D. The length of each link
($l_j$) is $1\,{\rm mm}$ and their radii ($r_j$) are taken from a
uniform distribution of random numbers in the range from $0.1\,{\rm
  mm}$ to $0.4\,{\rm mm}$. In 3D, we consider a network that is
reconstructed from a real sample of Berea sandstone by using micro-CT
scanning \cite{rh09,mbweb}. The reconstructed network contains $2274$
links that are connected via $1163$ nodes and has a physical dimension
of $1.8^3\,{\rm mm}^3$. We doubled this network by adding a mirror
image of itself in the direction of the applied pressure drop in order
to apply periodic boundary conditions. In the following, we present
some of the fundamental results of transient and steady-state of
two-phase flow in porous media by using this pore-network model. All
the following results are obtained only by changing the parameters
values and the boundary conditions as necessary, while using the
exactly same code for the interface dynamics algorithms.

\begin{figure}
  \centerline{\hfill
    \rotatebox{90}{(a) ${\rm Ca}=10^{-2}$, $M=10^{-3}$}\hfill
    \includegraphics[width=0.26\textwidth,clip]{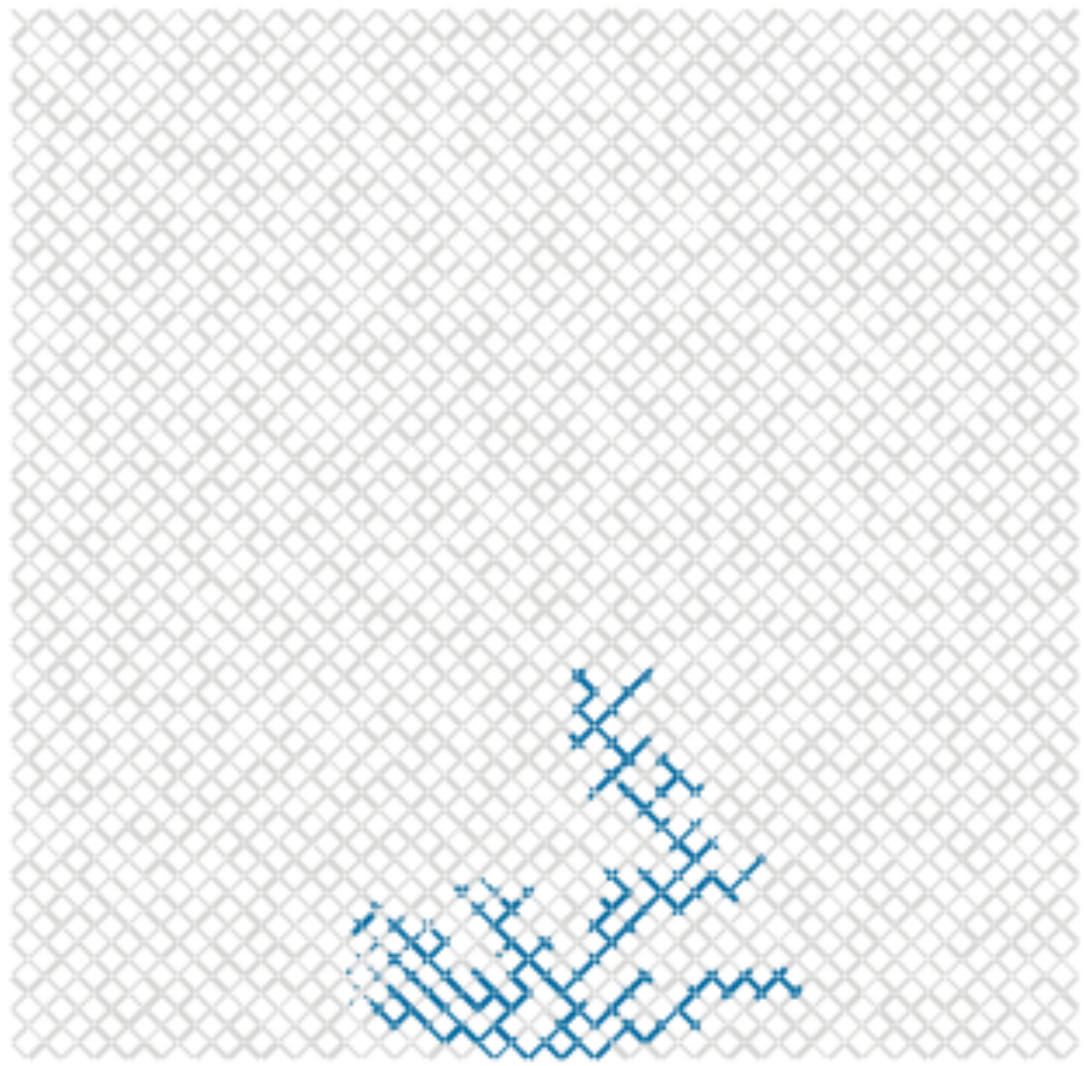}\hfill
    \includegraphics[width=0.26\textwidth,clip]{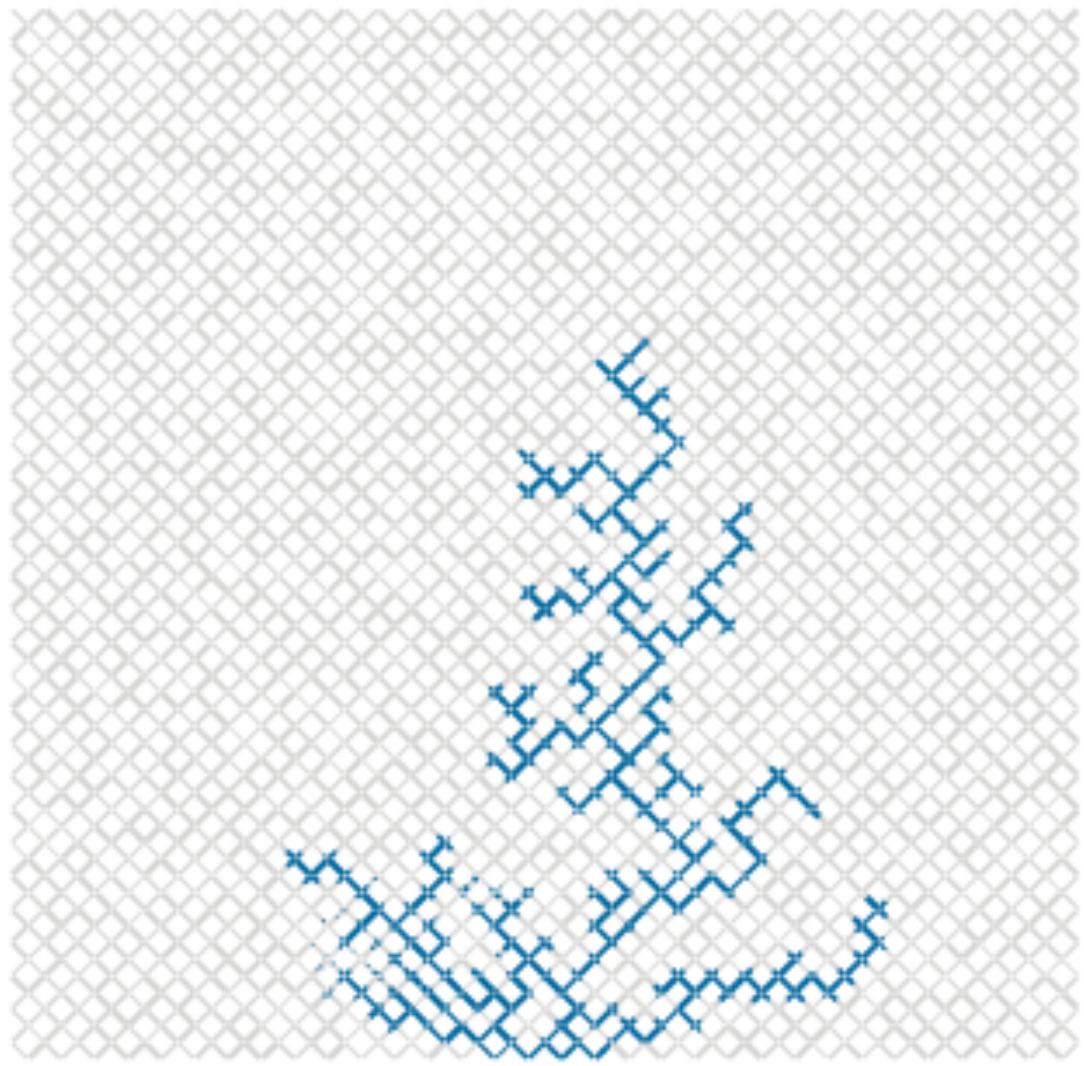}\hfill
    \includegraphics[width=0.26\textwidth,clip]{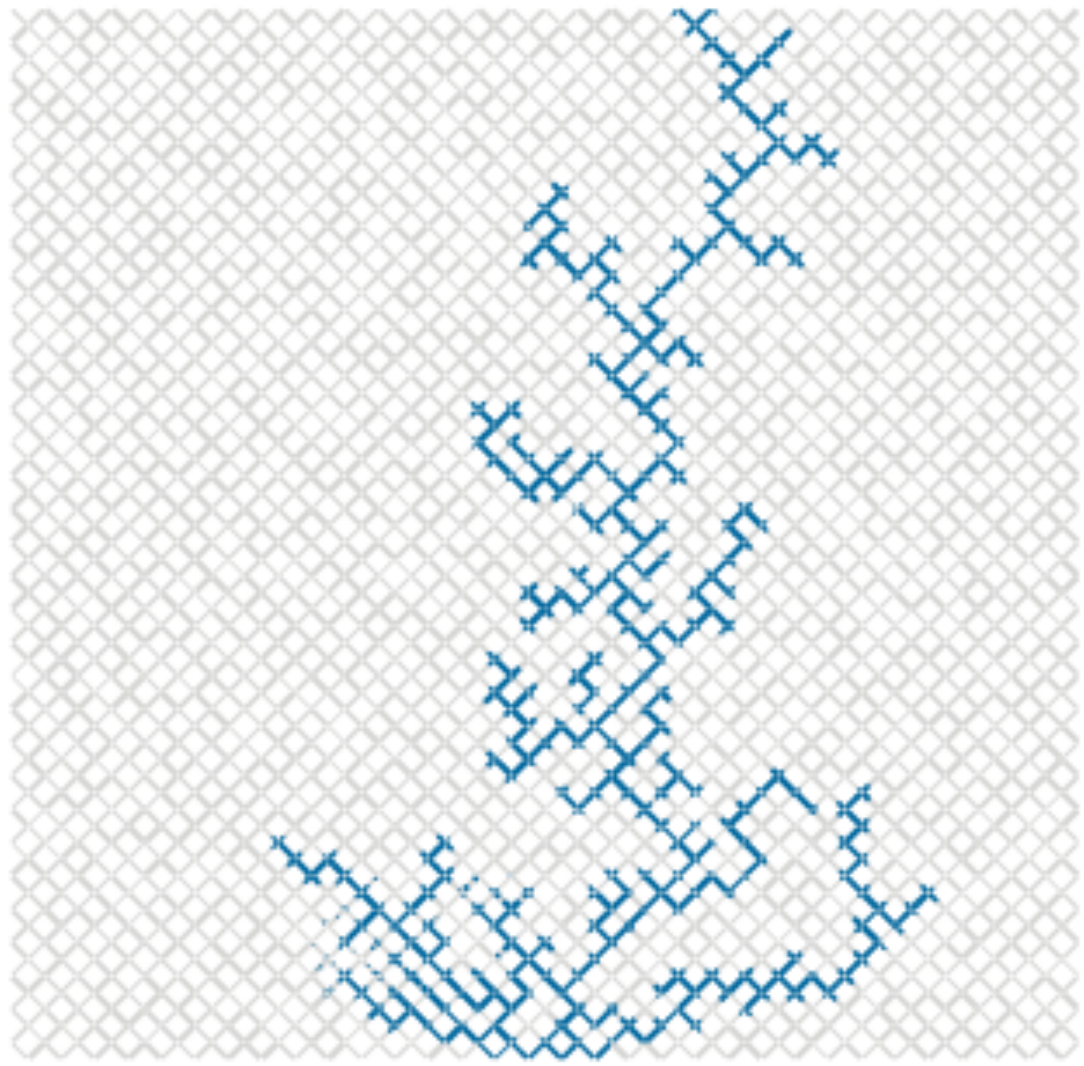}\hfill}
  \centerline{\hspace{0.06\textwidth}\hfill
    \includegraphics[width=0.26\textwidth,clip]{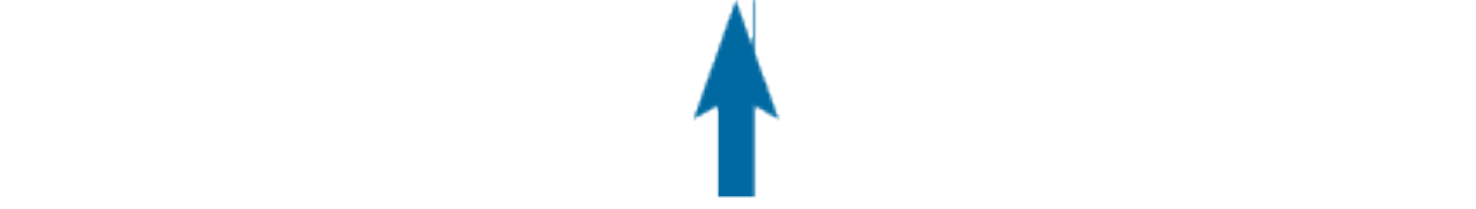}\hfill
    \includegraphics[width=0.26\textwidth,clip]{fig_Arrow_up.pdf}\hfill
    \includegraphics[width=0.26\textwidth,clip]{fig_Arrow_up.pdf}\hfill}
  \medskip
  \centerline{\hfill
    \rotatebox{90}{(b) ${\rm Ca}=10^{-2}$, $M=10^2$}\hfill
    \includegraphics[width=0.26\textwidth,clip]{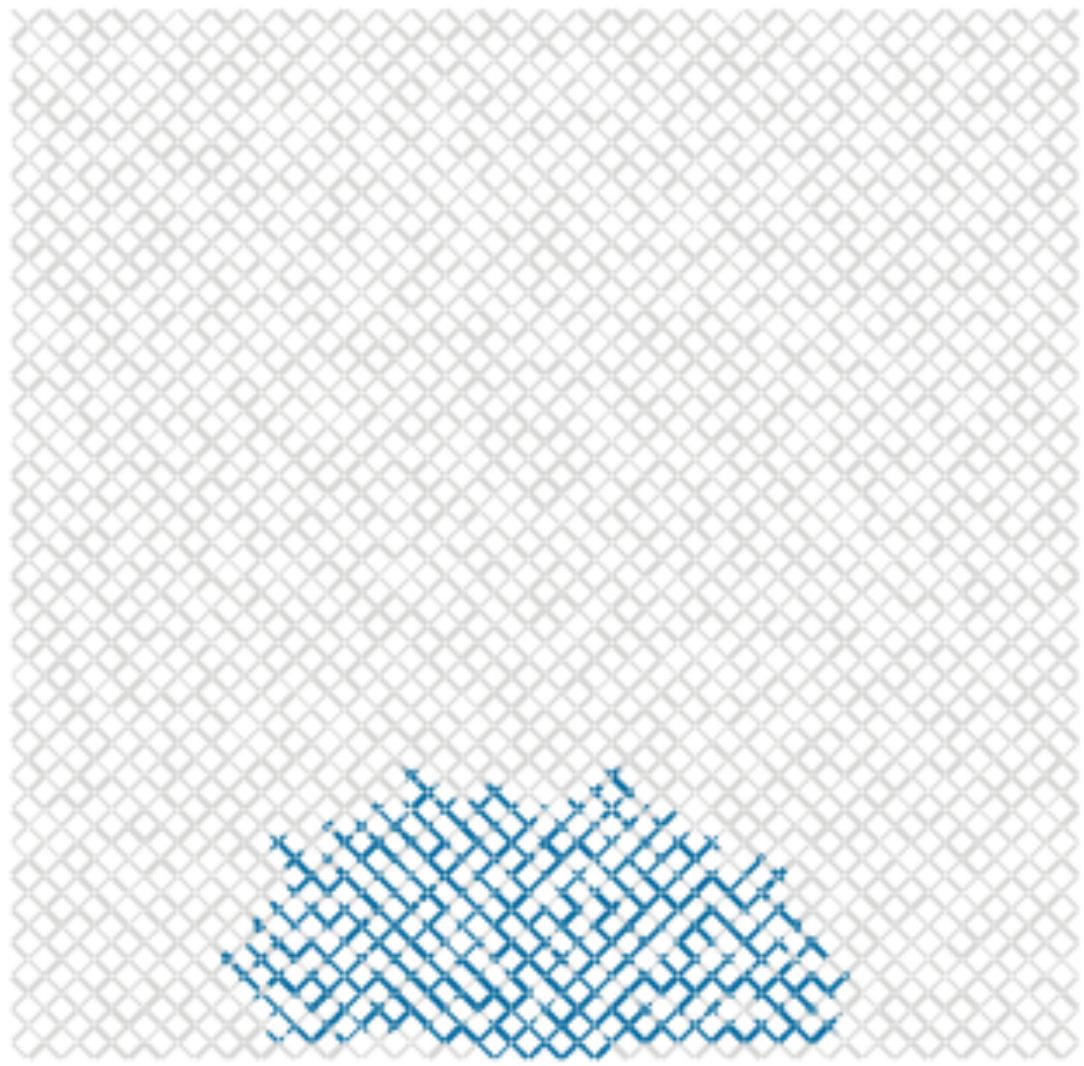}\hfill
    \includegraphics[width=0.26\textwidth,clip]{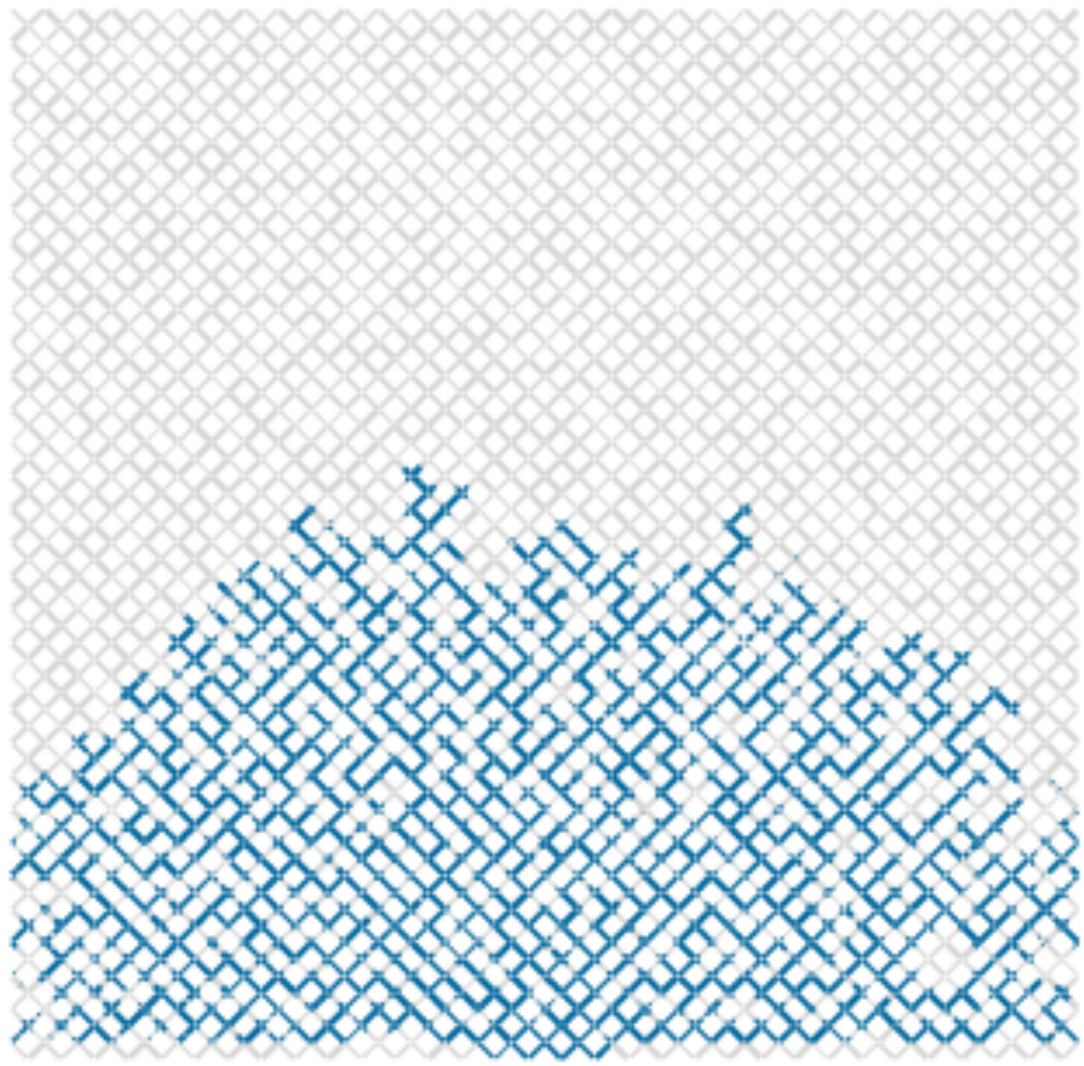}\hfill
    \includegraphics[width=0.26\textwidth,clip]{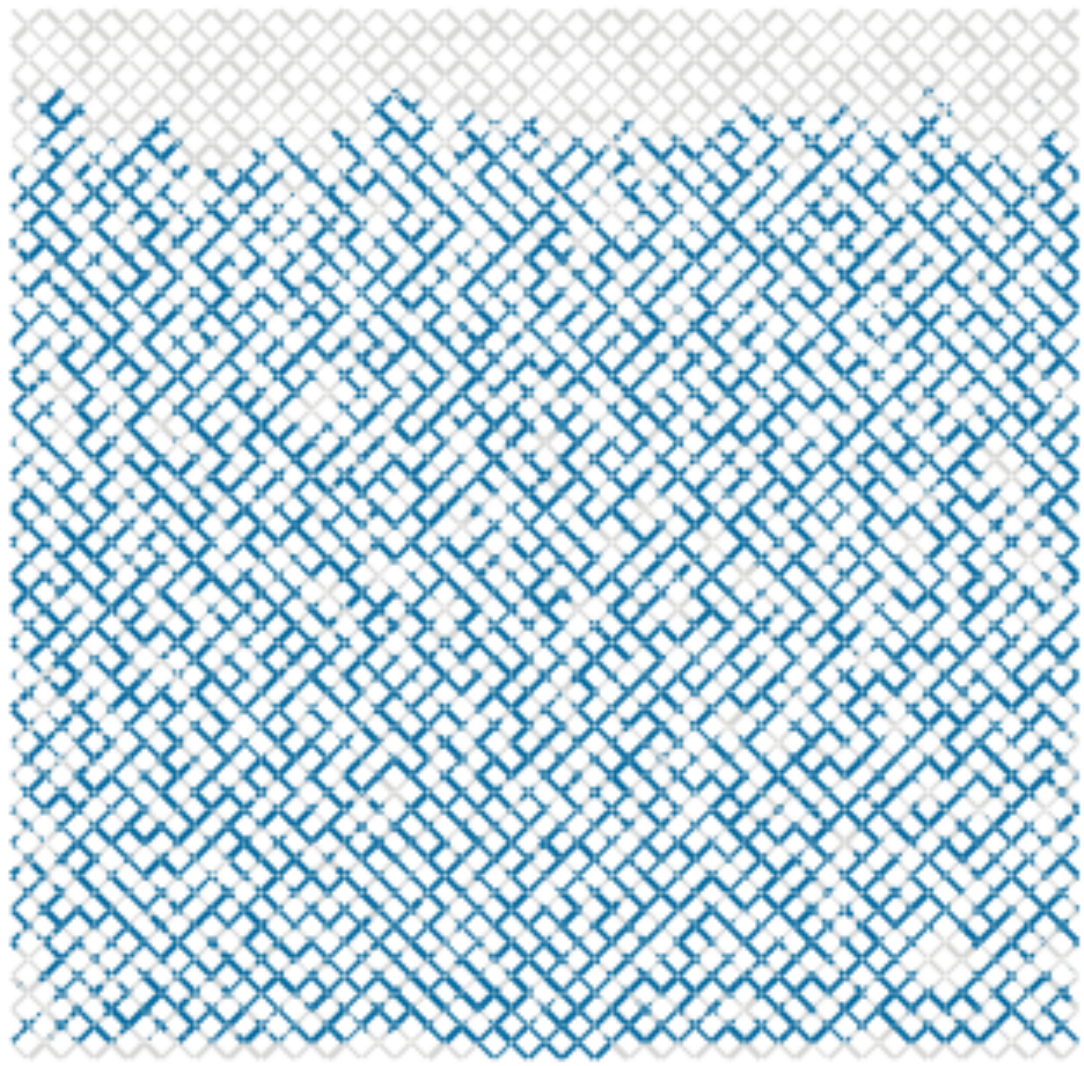}\hfill}
  \centerline{\hspace{0.06\textwidth}\hfill
    \includegraphics[width=0.26\textwidth,clip]{fig_Arrow_up.pdf}\hfill
    \includegraphics[width=0.26\textwidth,clip]{fig_Arrow_up.pdf}\hfill
    \includegraphics[width=0.26\textwidth,clip]{fig_Arrow_up.pdf}\hfill}
  \medskip
  \centerline{\hfill
    \rotatebox{90}{(c) ${\rm Ca}=10^{-5}$, $M=1$}\hfill
    \includegraphics[width=0.26\textwidth,clip]{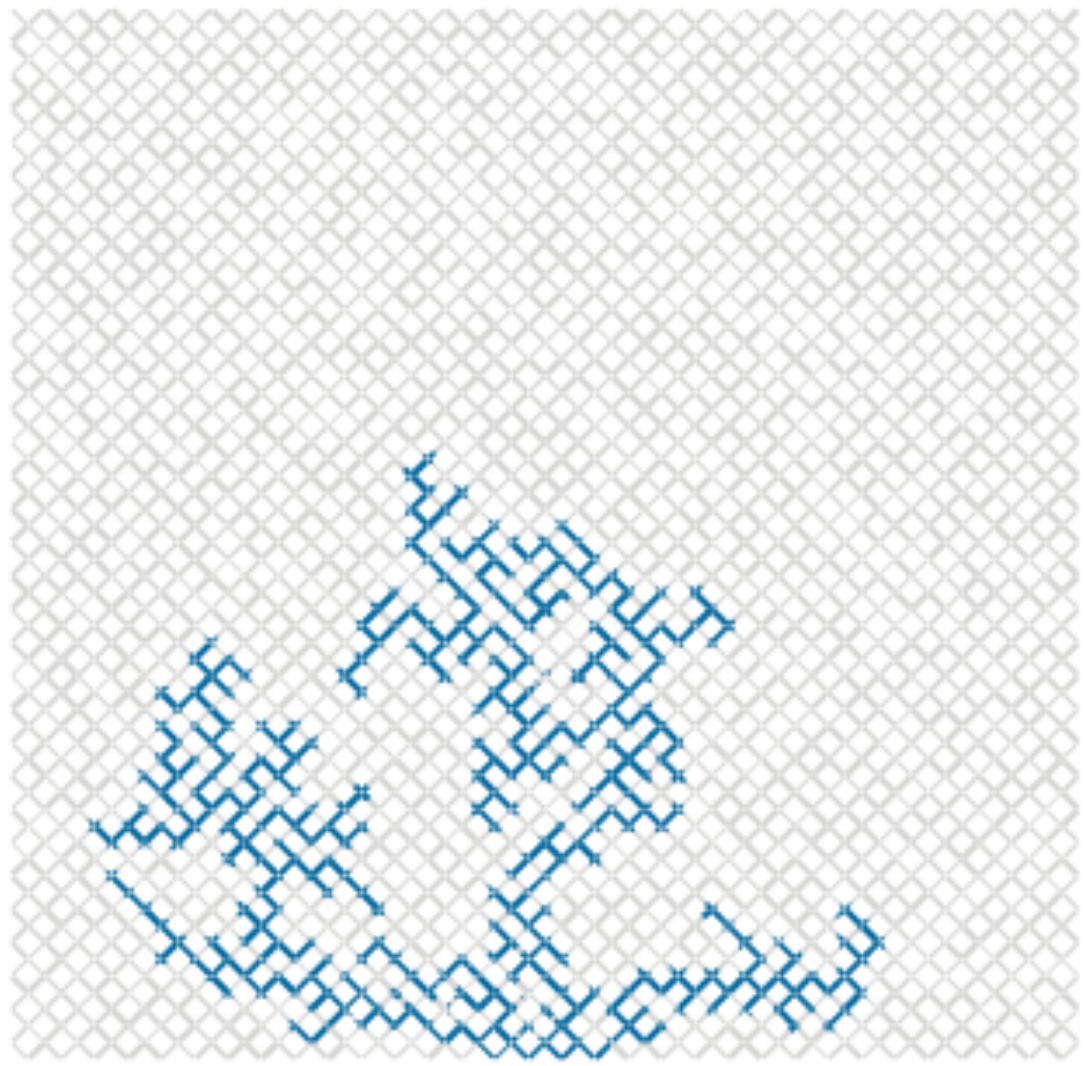}\hfill
    \includegraphics[width=0.26\textwidth,clip]{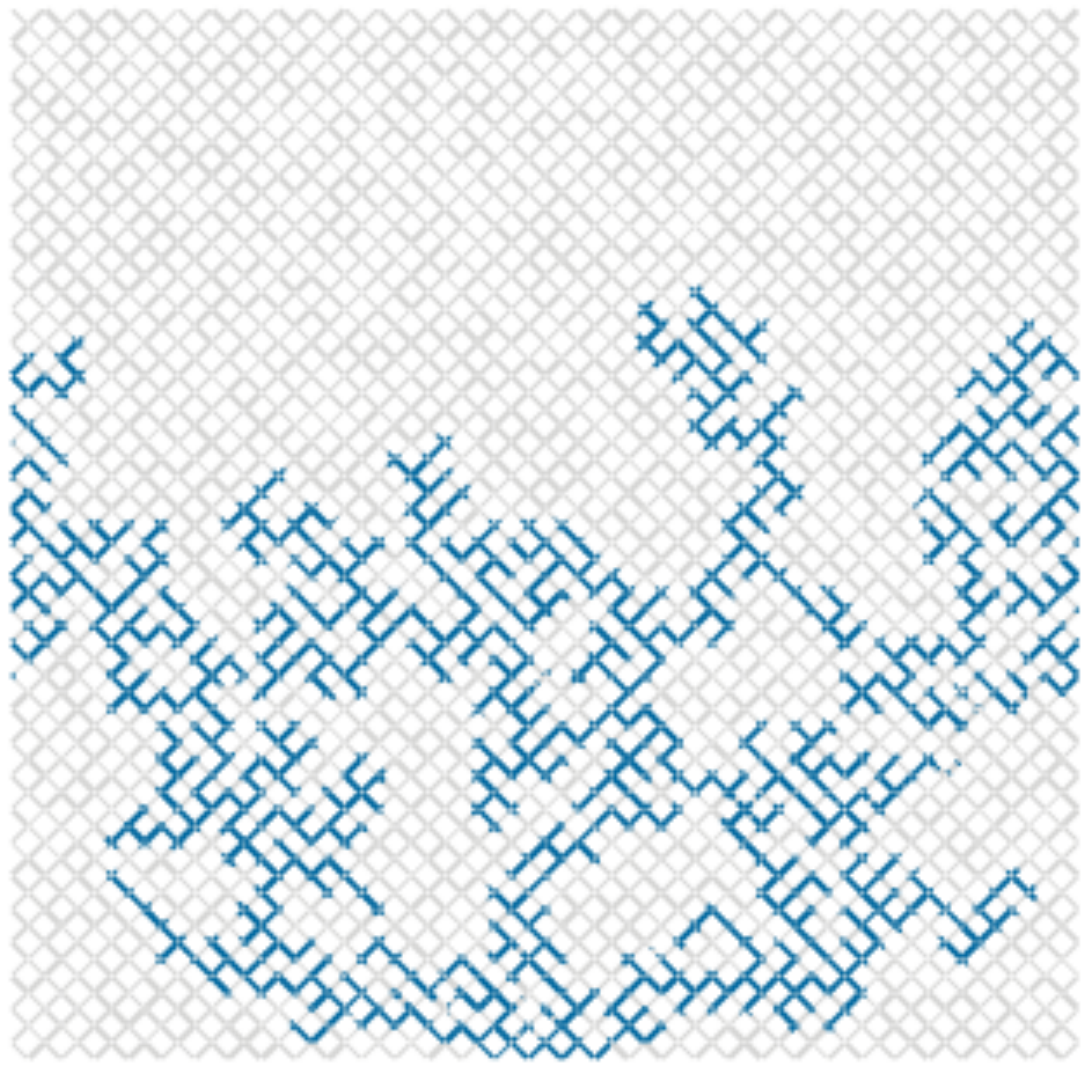}\hfill
    \includegraphics[width=0.26\textwidth,clip]{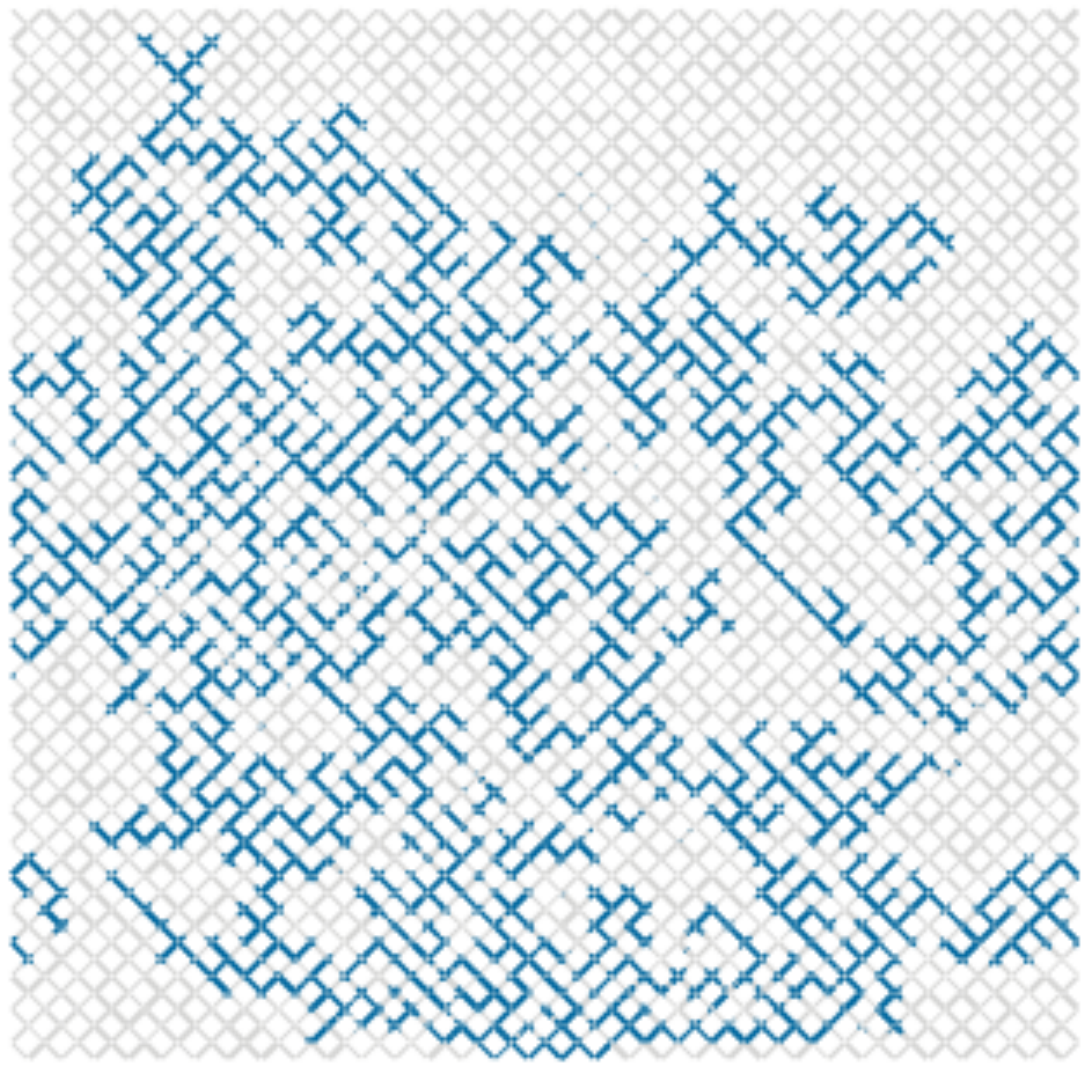}\hfill}
  \centerline{\hspace{0.06\textwidth}\hfill
    \includegraphics[width=0.26\textwidth,clip]{fig_Arrow_up.pdf}\hfill
    \includegraphics[width=0.26\textwidth,clip]{fig_Arrow_up.pdf}\hfill
    \includegraphics[width=0.26\textwidth,clip]{fig_Arrow_up.pdf}\hfill}
\caption{\label{figObtr}Development of transient flow patterns during
  the drainage simulations in a disordered square network of $64\times
  64$ links. The network is initially filled with wetting fluid (gray)
  and the non-wetting fluid (blue) is injected through four inlet
  nodes at the bottom edge of the network, shown by the arrow, with a
  constant flow rate. The top edge of the network is kept open which
  works as the outlet. Periodic boundary conditions are applied in the
  horizontal direction, therefore the left and right edges are
  connected together. Only the capillary number Ca and the viscosity
  ratio $M$ are altered during these simulations and all other flow
  parameters and the algorithms for interface dynamics are exactly the
  same. The flow patterns show the different regimes of transient
  two-phase flow, namely (a) the viscous fingering (b) the stable
  displacement and (c) the capillary fingering. We like to point out
  that, though we adopted ``democratic" rules for interface
  algorithms, it is the solution of the flow equations that lead the
  fluids to generate these different flow patterns.}
\end{figure}

\subsection{\label{secTrn} Drainage displacements}
As described in the introduction, when a non-wetting fluid is invaded
into a porous medium filled with a wetting fluid, it generates
different types of invading flow patterns depending on the capillary
number and viscosity ratio. A less viscous fluid displacing a high
viscous fluid creates viscous fingering patterns at a high Ca
\cite{cw85, mfj85} and capillary fingering patterns at low Ca
\cite{lz85}. The capillary and viscous fingering patterns resembles
with the invasion percolation \cite{ww83} and diffusion limited
aggregation (DLA) models \cite{ws81} respectively. Alternatively, when
a high viscous fluid displaces a low viscous fluid, a stable
displacement front is observed. In order to verify whether our
pore-network model can reproduce these different flow patterns during
the drainage, we run different simulations for different values of the
capillary number ${\rm Ca}$ and viscosity ratio $M$. In our
simulations, we set $\gamma\cos\theta = 0.03$ N/m. The results are
highlighted in figure \ref{figObtr}. Here $M=\mu_{\rm n}/\mu_{\rm w}$
where $\mu_{\rm n}$ and $\mu_{\rm w}$ are the viscosities of the
invading blue fluid and the wetting gray fluid. In the top row the low
viscous non-wetting fluid displaces the more viscous fluid where one
can observe the development of the viscous fingering pattern. In the
second row $M=10^{2}$, and a more viscous blue fluid is displacing the
less viscous gray fluid and a compact and stable displacement front is
observed \cite{ltz88}. Flow patterns in the third row correspond to
capillary fingering which are generated with the simulations at a very
low capillary number ${\rm Ca} = 10^{-5}$. These fingering patterns
are more fractal than the viscous fingering and depend strongly on the
system disorder. We like to point out that, these three different
transient regimes are generated only by altering the values of the
flow rate $Q$ and the viscosities of the two fluids to set the values
of Ca and M, and no modification in the interface dynamics algorithms
were made between different simulations.

\subsection{\label{secStd} Steady state}
\begin{figure}
  \offinterlineskip
  \centerline{\hfill
    \includegraphics[align=c,width=0.16\textwidth,clip]{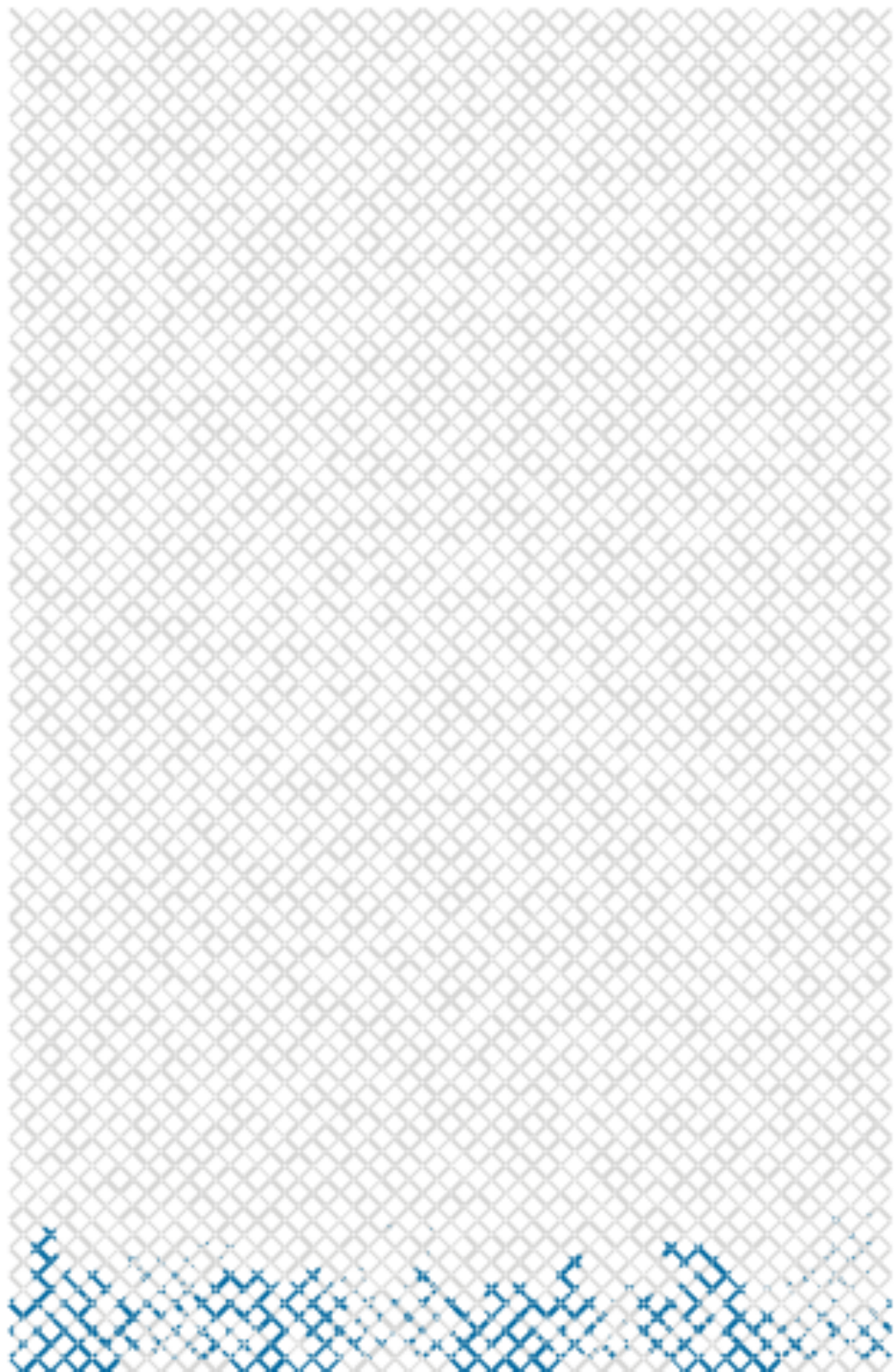}\hfill
    \includegraphics[align=c,width=0.16\textwidth,clip]{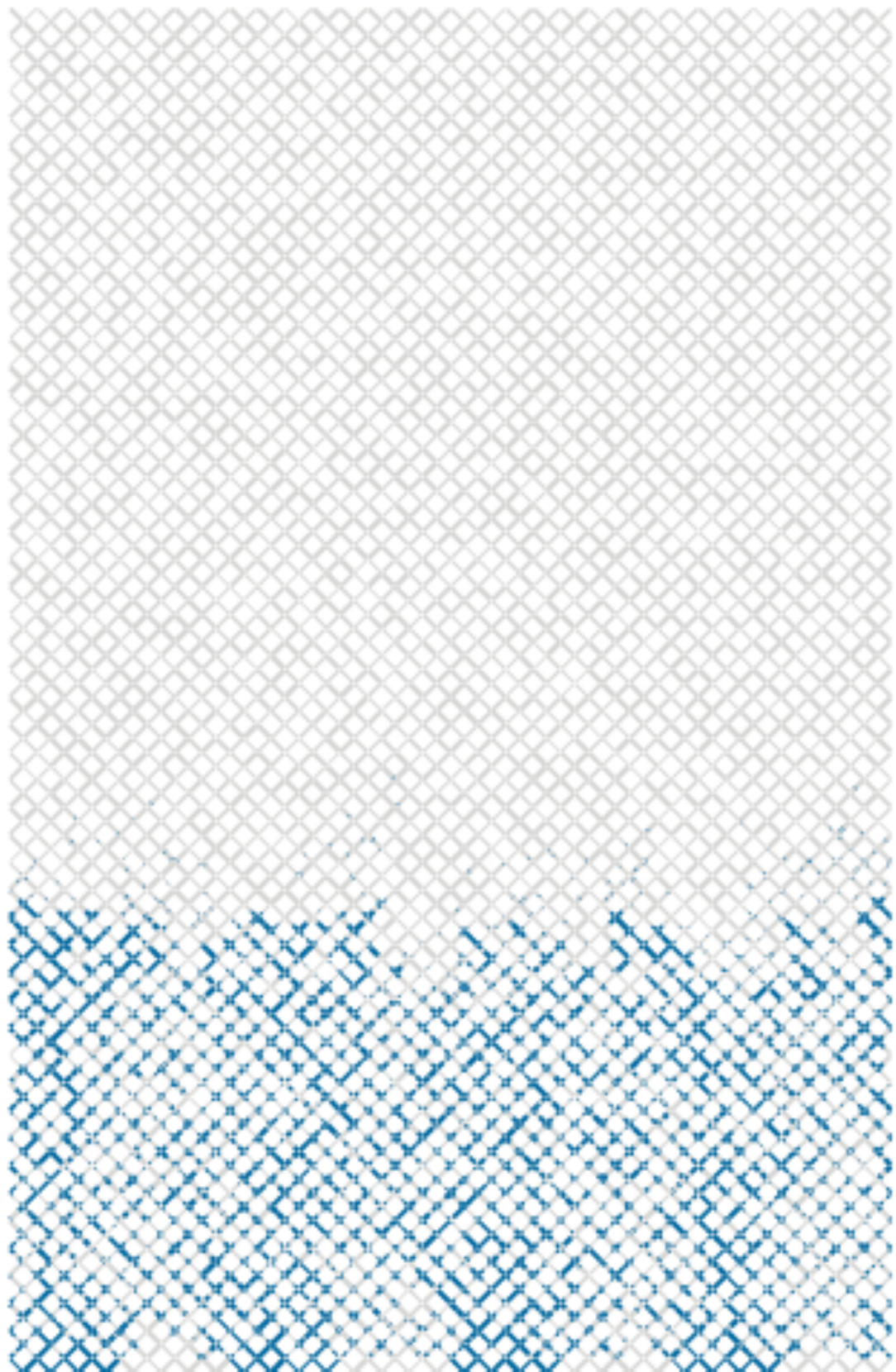}\hfill
    \includegraphics[align=c,width=0.16\textwidth,clip]{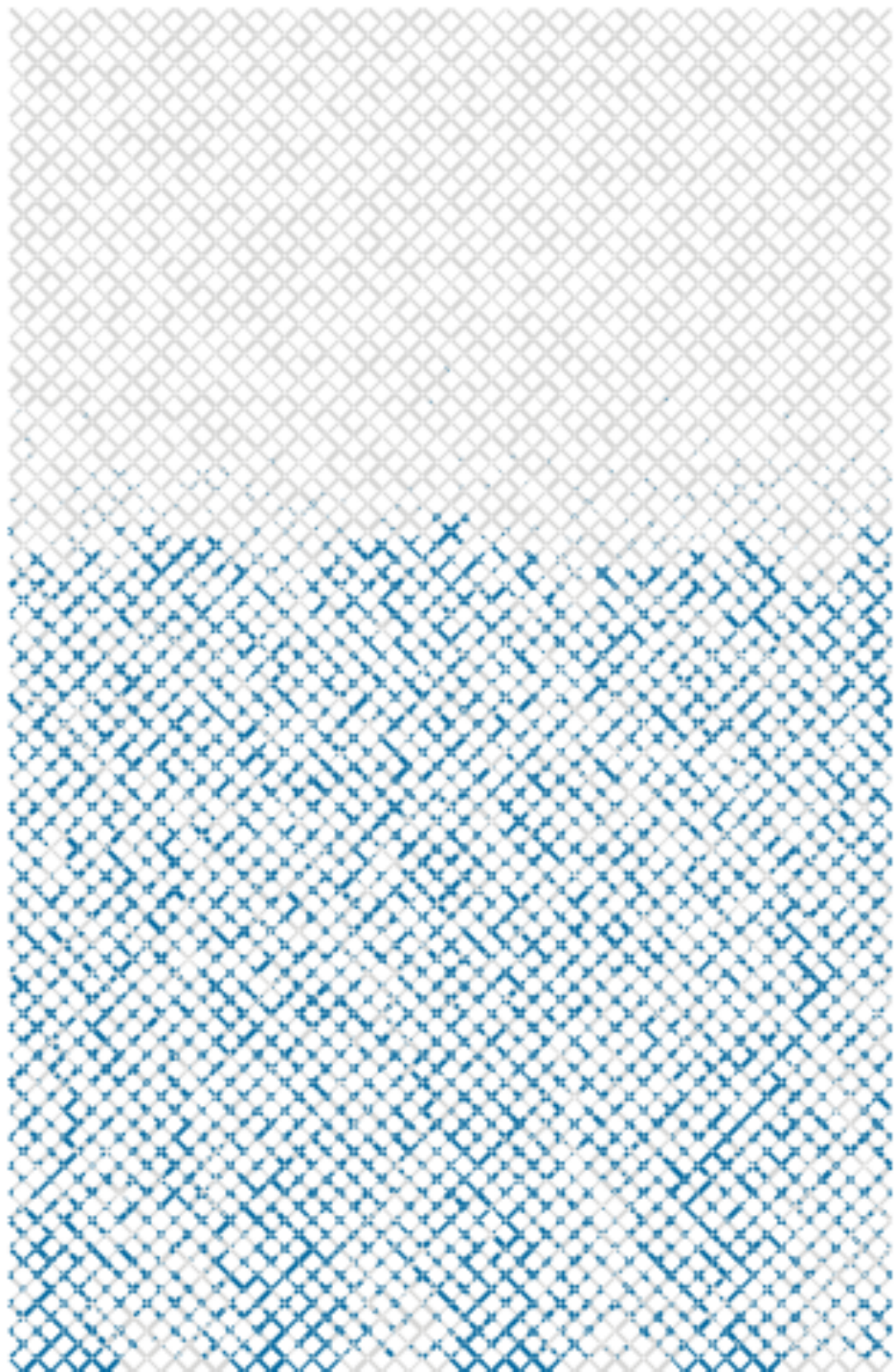}\hfill
    \includegraphics[align=c,width=0.16\textwidth,clip]{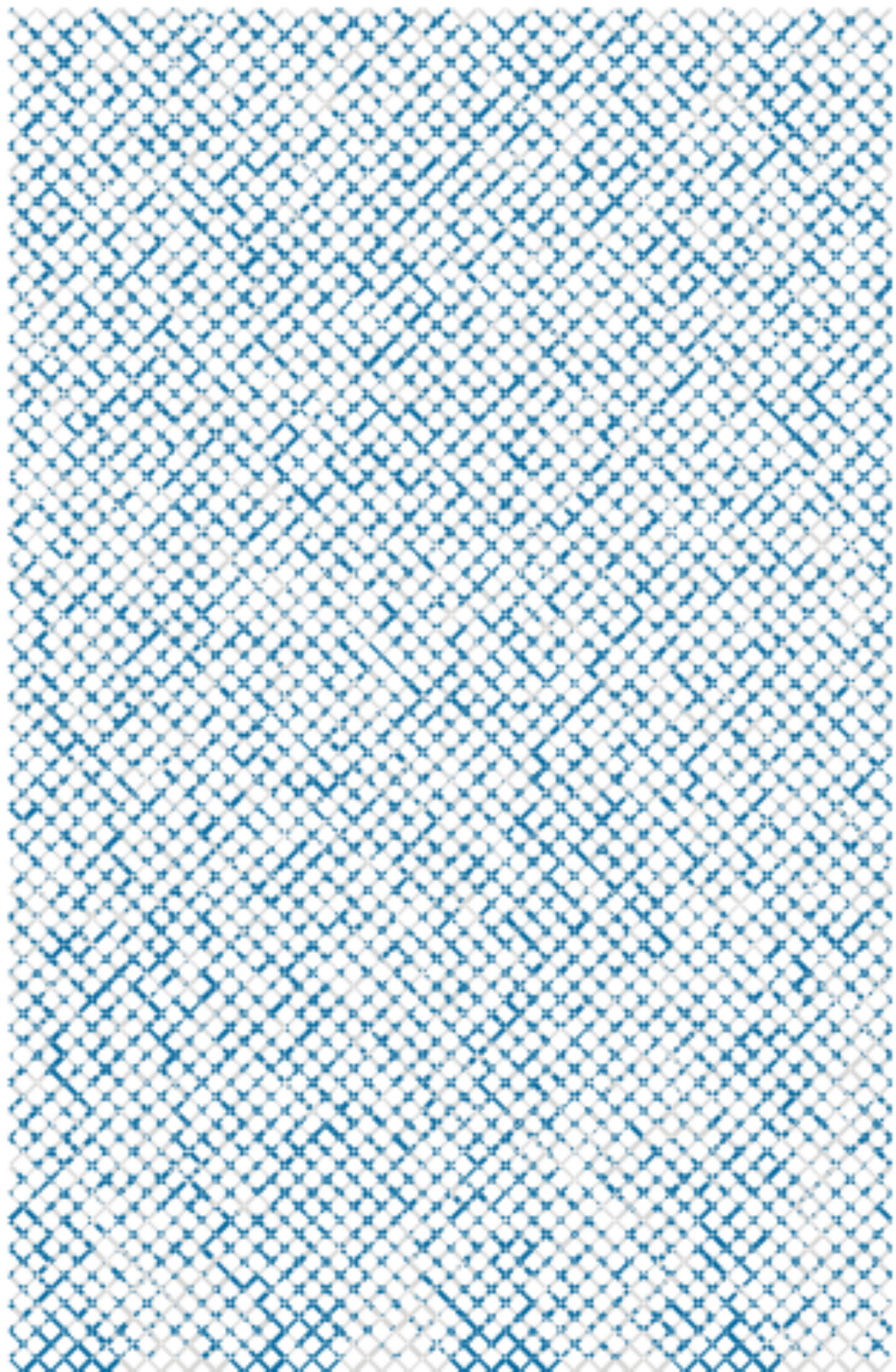}\hfill
    \includegraphics[align=c,width=0.20\textwidth,clip]{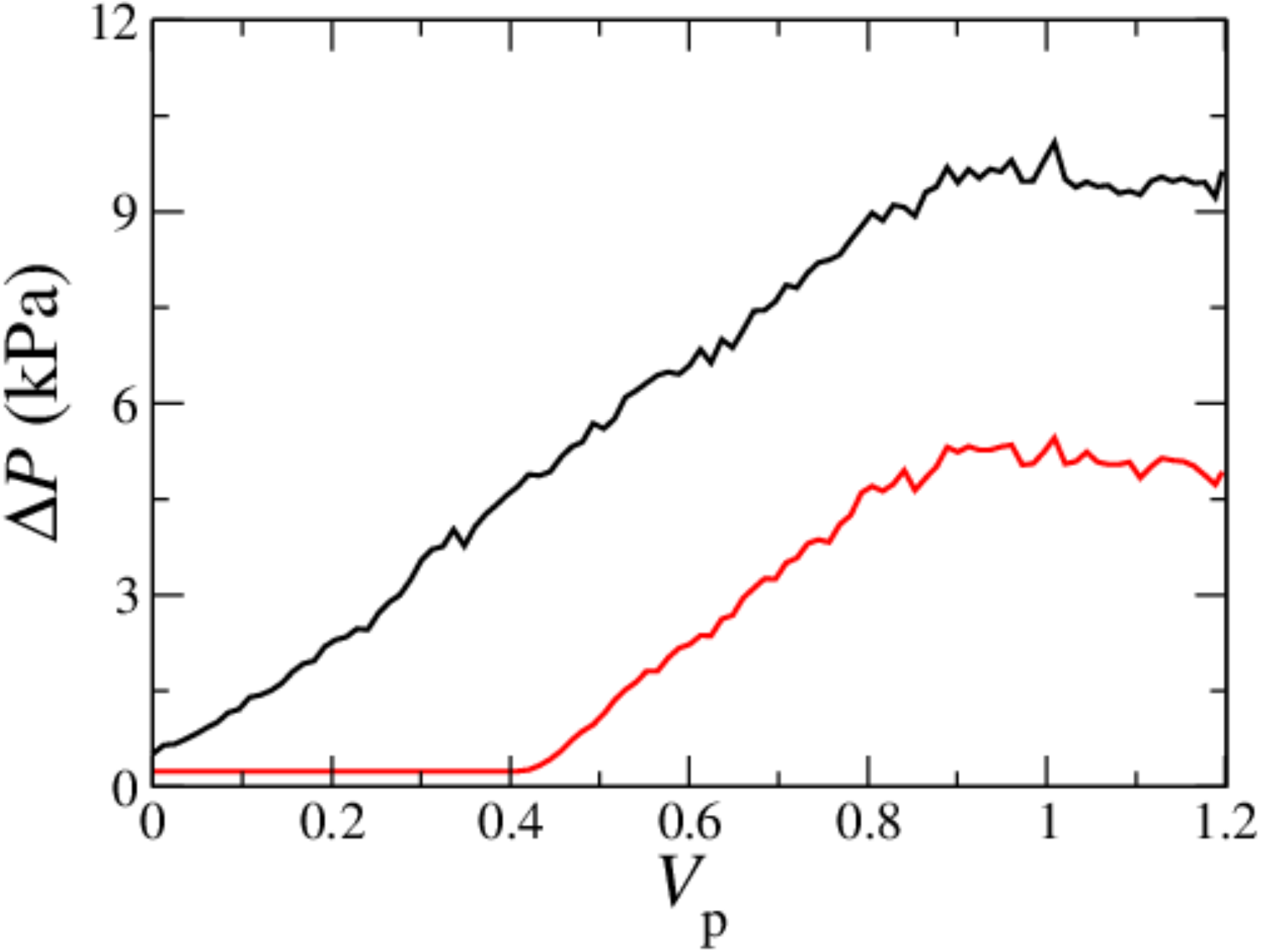}\hfill}
  \centerline{\hfill
    \includegraphics[align=c,width=0.16\textwidth,clip]{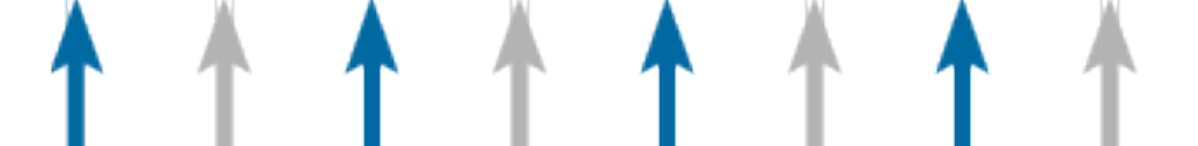}\hfill
    \includegraphics[align=c,width=0.16\textwidth,clip]{fig_Syringe4.pdf}\hfill
    \includegraphics[align=c,width=0.16\textwidth,clip]{fig_Syringe4.pdf}\hfill
    \includegraphics[align=c,width=0.16\textwidth,clip]{fig_Syringe4.pdf}\hfill
    \hspace{0.20\textwidth}\hfill}
  \medskip
  \centerline{\hfill
    \includegraphics[align=c,width=0.16\textwidth,clip]{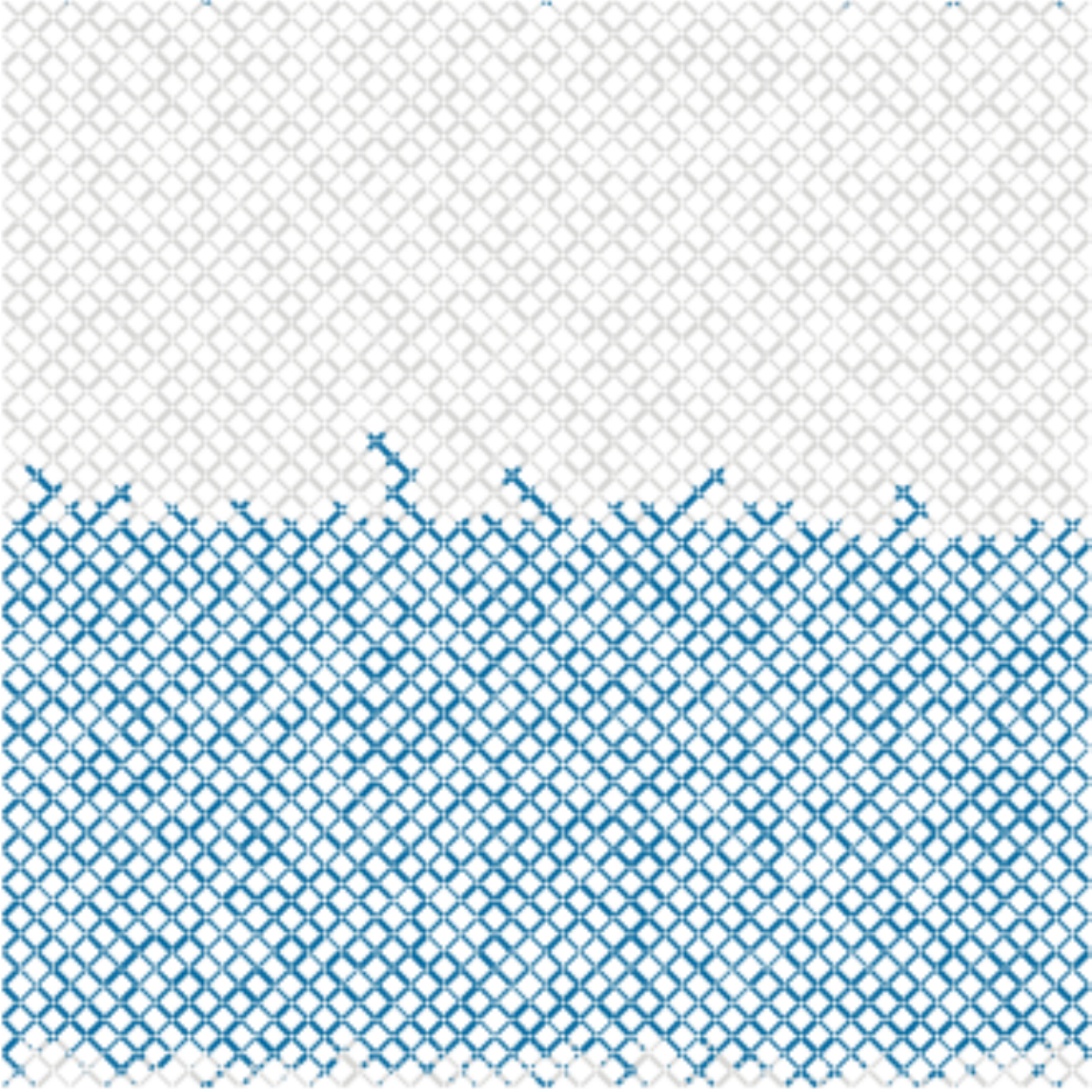}\hfill
    \includegraphics[align=c,width=0.16\textwidth,clip]{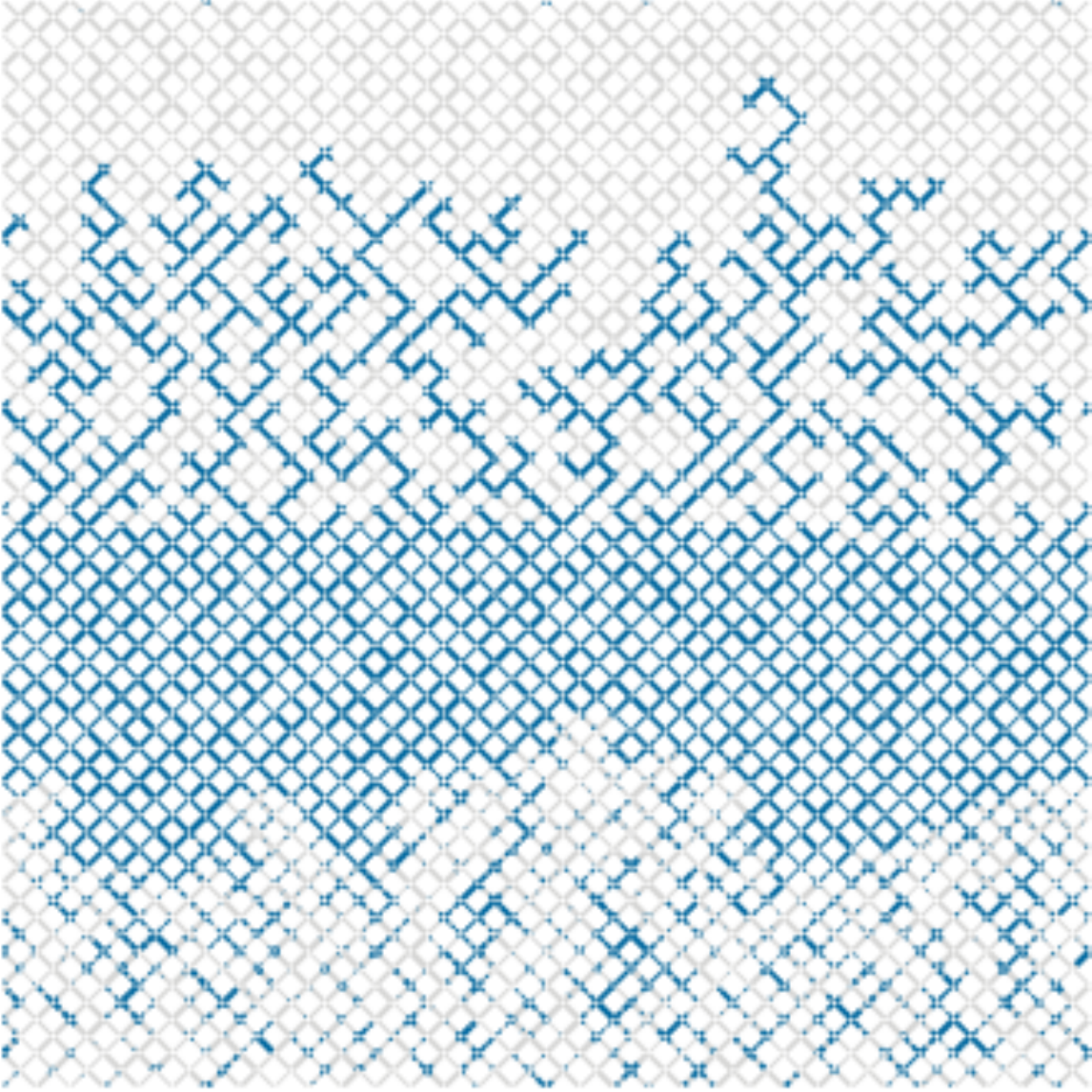}\hfill
    \includegraphics[align=c,width=0.16\textwidth,clip]{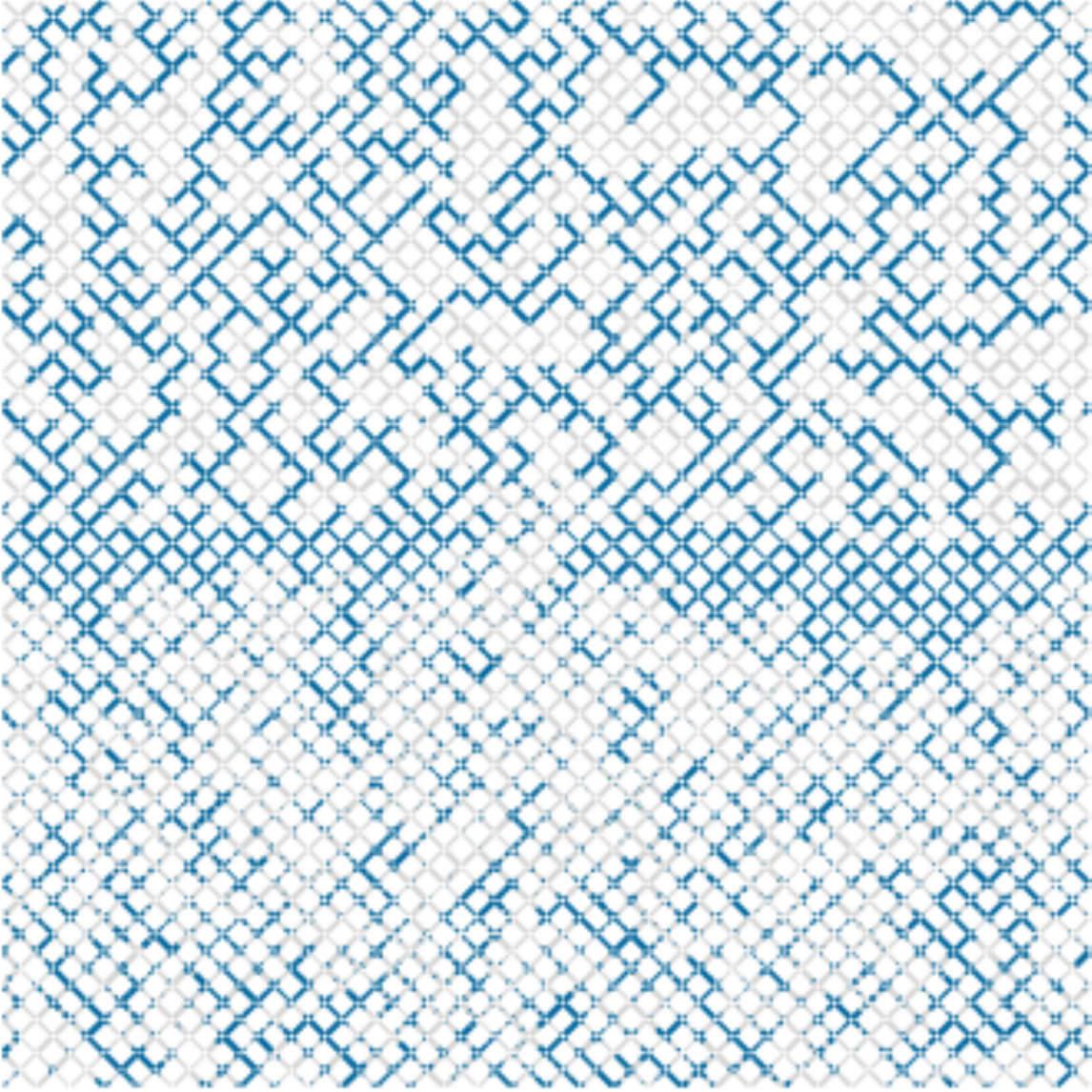}\hfill
    \includegraphics[align=c,width=0.16\textwidth,clip]{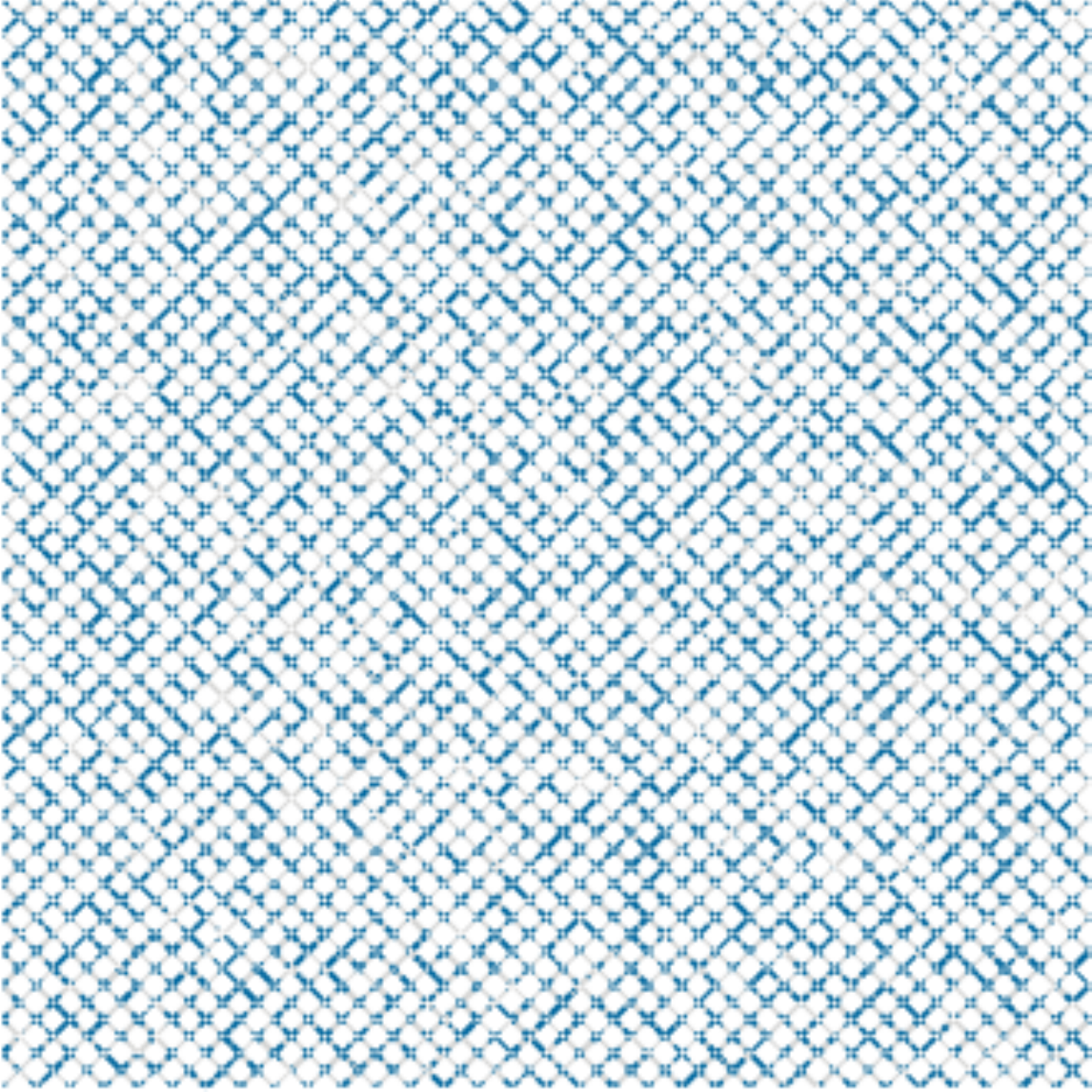}\hfill
    \includegraphics[align=c,width=0.20\textwidth,clip]{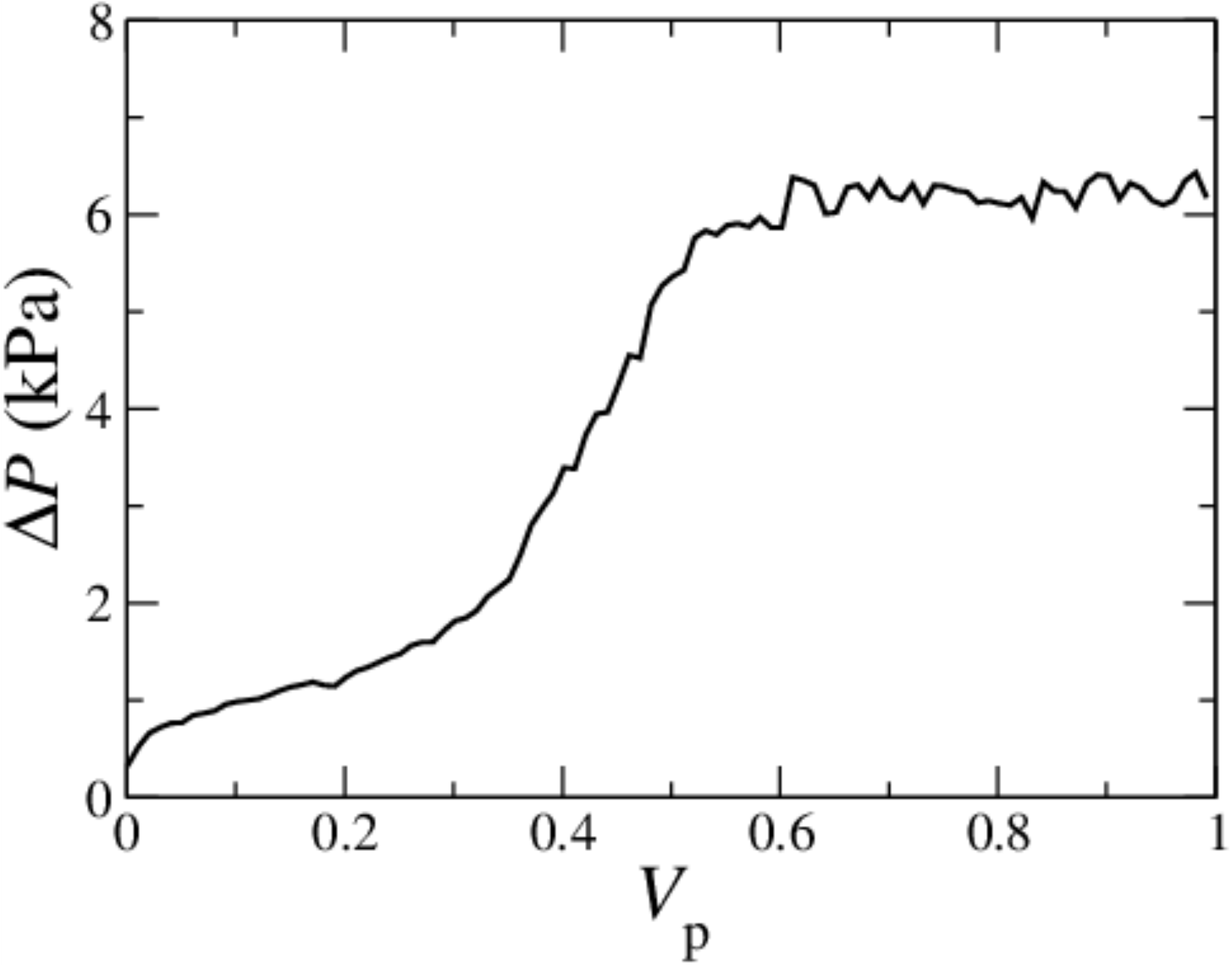}\hfill}
  \medskip
  \centerline{\hfill
    \includegraphics[align=c,width=0.24\textwidth,clip]{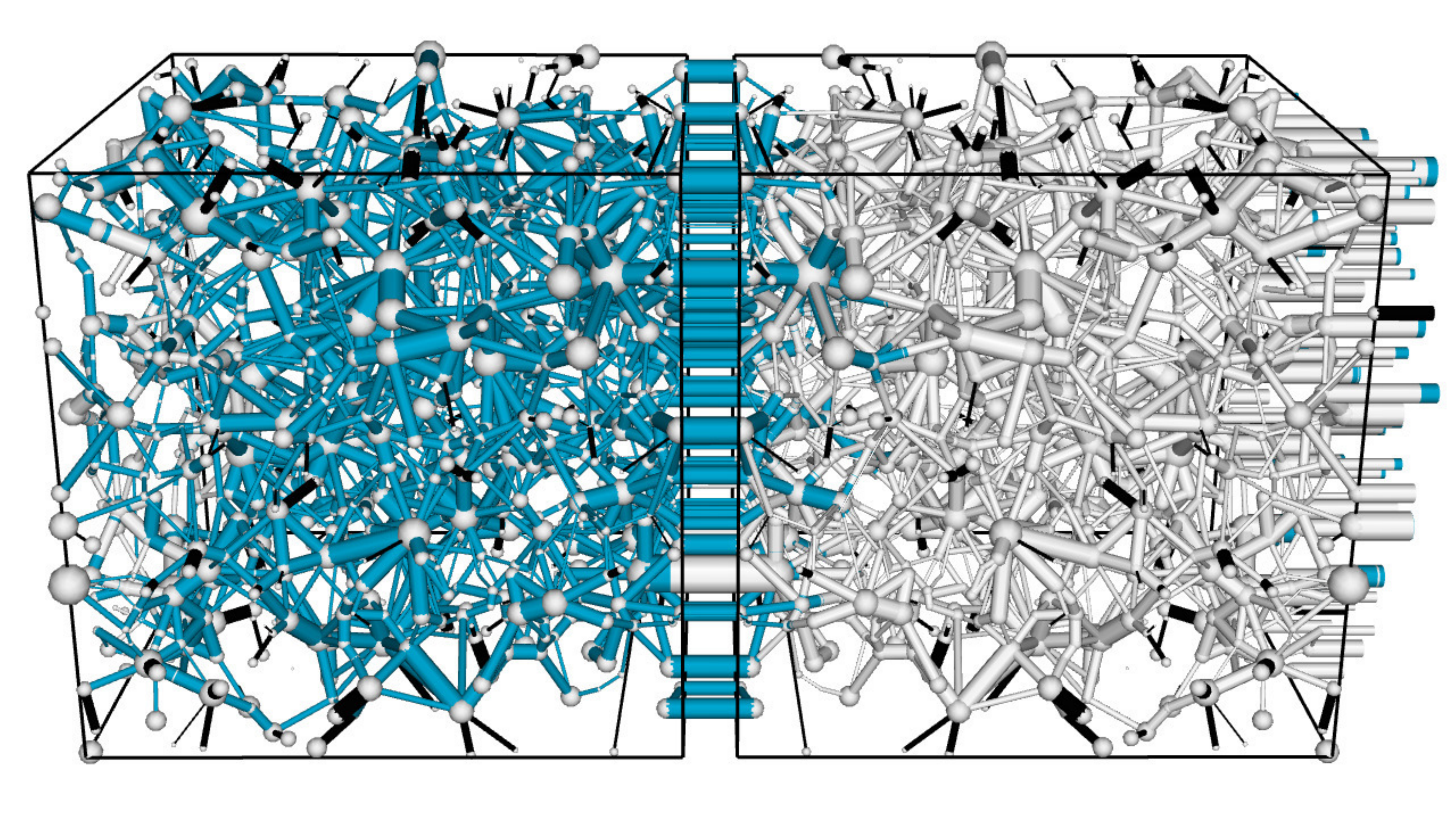}\hfill
    \includegraphics[align=c,width=0.24\textwidth,clip]{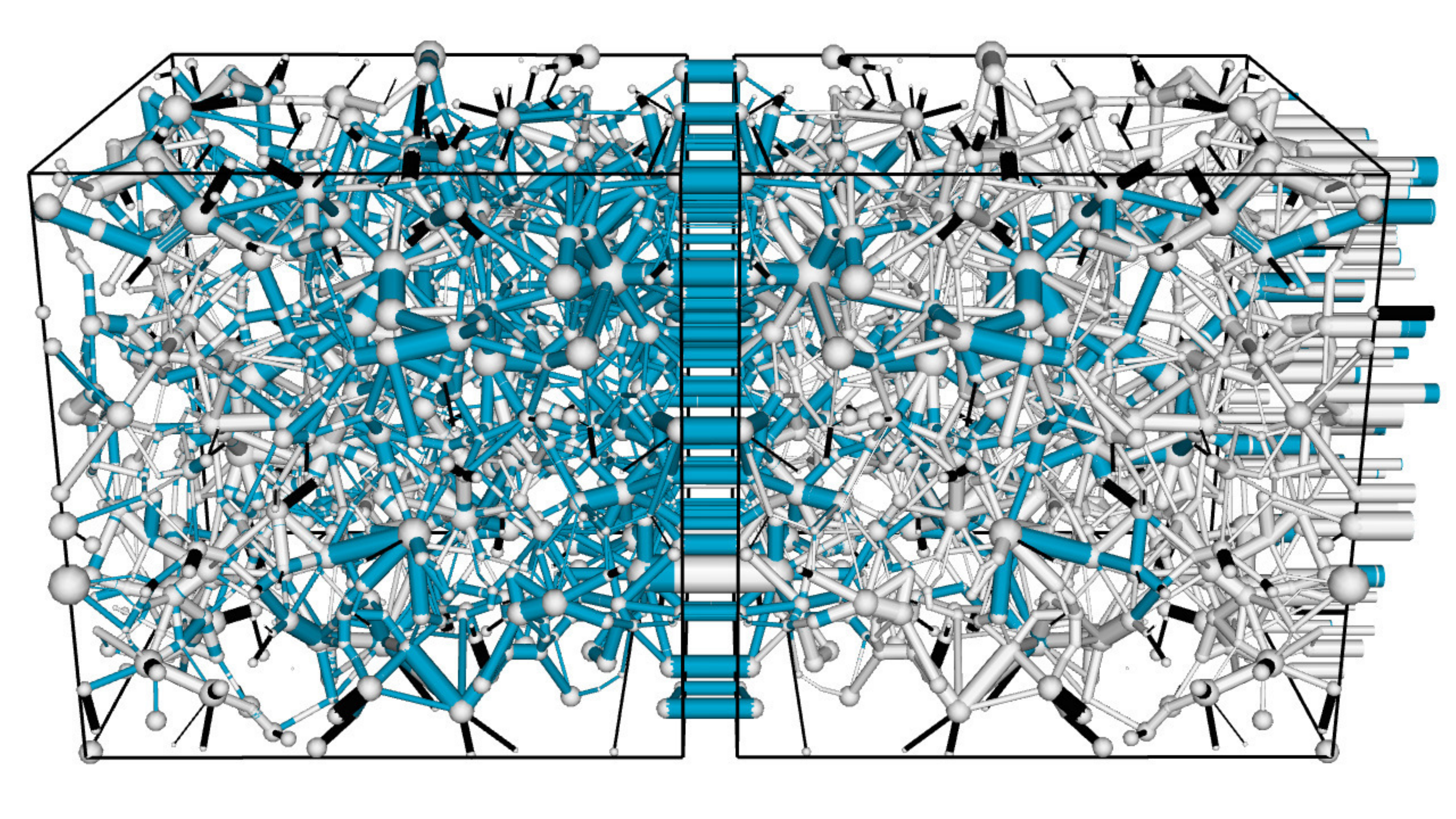}\hfill
    \includegraphics[align=c,width=0.24\textwidth,clip]{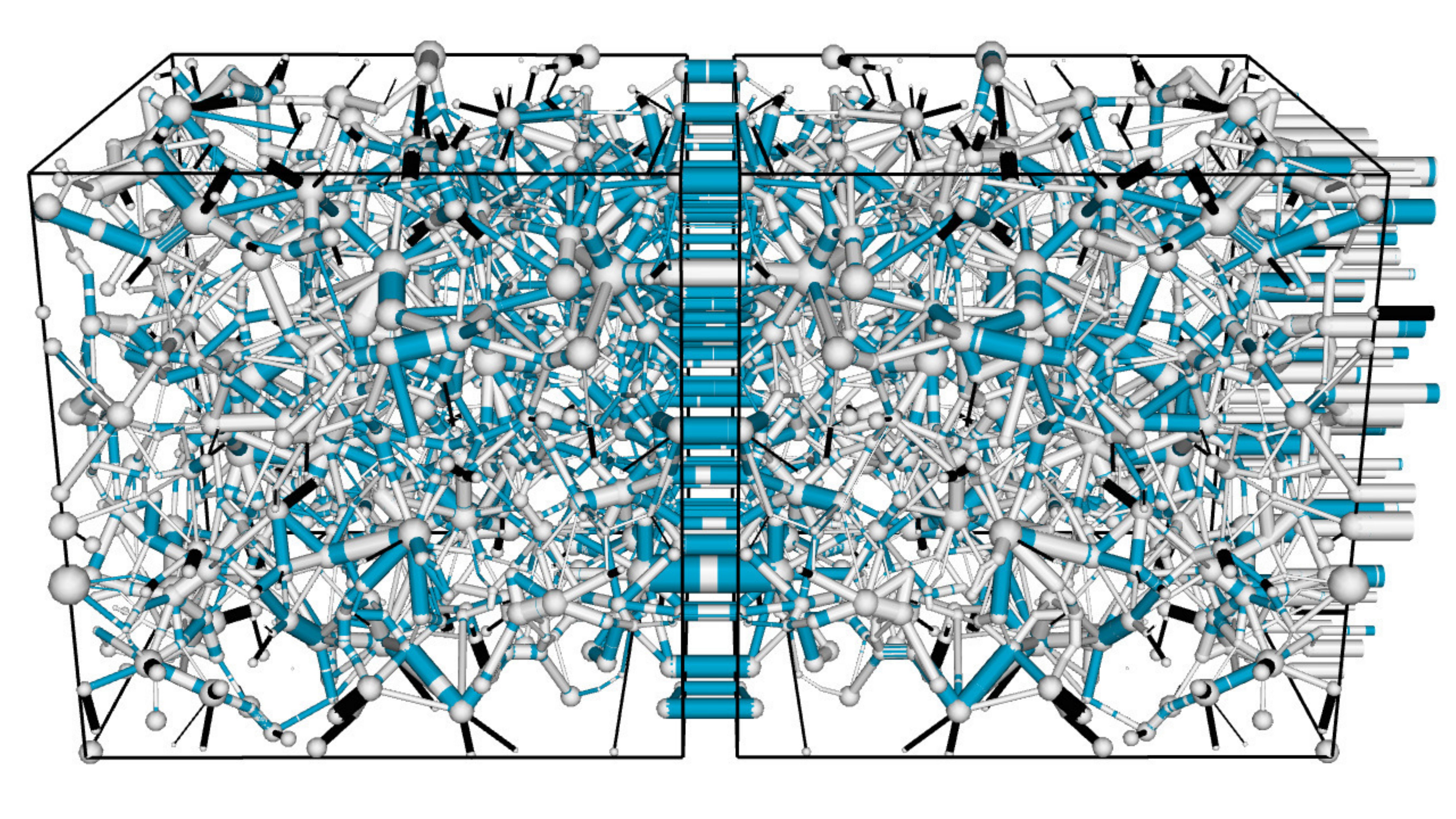}\hfill\hfill
    \includegraphics[align=c,width=0.21\textwidth,clip]{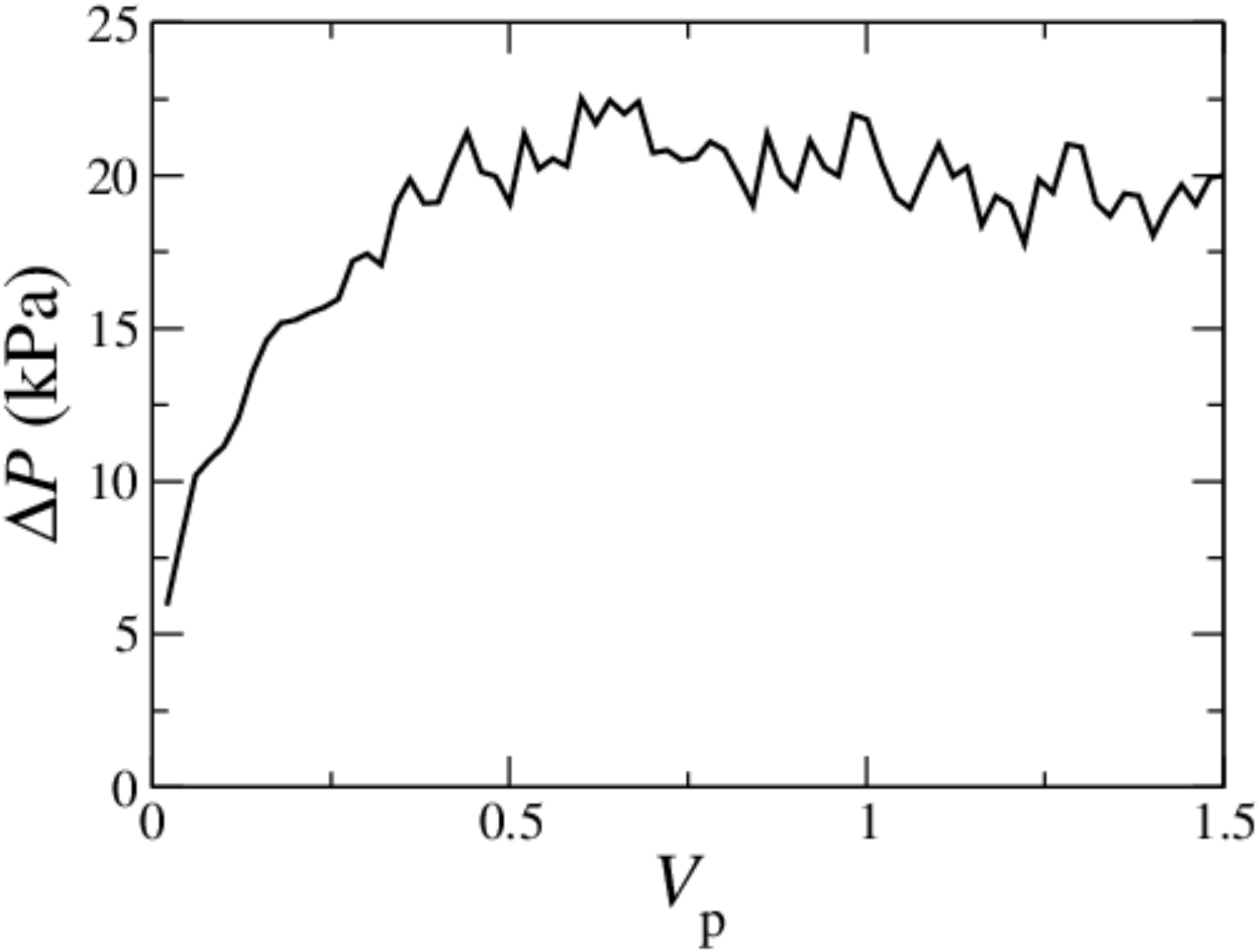}\hfill\hfill}
  \caption{\label{figStdst}Evolution to the steady state under
    different boundary conditions. All the simulations are performed
    at constant flow rate with ${\rm Ca}=0.001$ and $M=1$. The wetting
    and non-wetting fluids are colored by gray and blue. In the first
    row, simulations with open boundary conditions with a 2D square
    network of $64\times 100$ links are shown, where two fluids are
    injected through alternate injection points at the bottom inlet
    row, as shown by the arrows of two different colors. The
    fractional flow is $F_{\rm n}=0.5$ here which is controlled by
    defining flow rates at the inlet links that act as boundary
    conditions while solving the flow equations. The top outlet row is
    kept open. The second row shows the steady-state simulation with
    periodic boundary conditions with a $64\times 64$
    links. Saturation is a control parameter here and $S_{\rm w}=0.5$
    in these simulations. In the third row of figures, we show
    steady-state simulations with reconstructed 3D networks from Berea
    sandstone where the overall flow is the direction from left to
    right. At the right, we plot the global pressure drops of the
    respective systems as a function of the pore volumes of fluids
    that passed through any cross section where the red and black
    plots for the first system show the average pressure drops at the
    middle and at the inlet of the system with respect to the outlet
    row.}
\end{figure}

Steady-state flow can be simulated by implementing either open or
periodic boundary conditions. In figure \ref{figStdst}, we show the
evolution of the system to reach the steady state at a constant flow
rate with different boundary conditions. The top row shows simulations
with open boundary condition, where wetting (gray) and non-wetting
(blue) fluids are injected through alternate injection points at the
bottom edge of a two-dimensional pore network consisting $64\times
100$ links. The radii ($r_j$) of the links are drawn from a uniform
random number distribution in the range $0.1l$ to $0.4l$, where $l_j$
is the length of the links. The top edge of the network is kept open,
through which fluids leave the system. The two side edges are
connected with periodic boundary here but can also be kept closed if
necessary. We can control the flow rates of the individual fluids
which sets the fractional flow. Here the simulations are shown for
$F_{\rm w}=0.5$. As there will be traces of injection of the alternate
fluids near the inlet edge, one consider a system that is long enough
in the direction of the overall flow so that a region of spatially
homogeneous steady-state flow can be achieved away from the inlets. At
the right, we plot the average global pressure drops as a function of
the injected pore-volumes which show the evolution to the steady state
when $\Delta P$ fluctuate around an average value. The second row of
figure \ref{figStdst} shows simulations with periodic boundary
conditions in a disordered square lattice of $64\times 64$ links. Here
the system is closed and the control parameter is the fluid
saturation. The simulations are shown for $S_{\rm w} = 0.5$. In the
bottom row, we show the simulation with a three dimensional pore
network reconstructed from a real sample of Berea sandstone for
$S_{\rm w}=0.5$.

In the following, we present two examples of simulation results with
this dynamic pore network model to show excellent agreement with the
steady-state properties of two-phase flow. First, we will present the
effective rheological properties where the total flow rate shows
non-linear dependence on the pressure drop in the capillary dominated
regime \cite{tlkrfm09, rcs11, sh12, sbdk17}. Next, we measure seepage
velocities and will verify the relations between them
\cite{hsbkgv18}. We will also describe in detail how to measure
different quantities of flow, such as the flow rates and the seepage
velocities of the different fluid components.

\subsubsection{Effective rheology and crossover from linear to
  non-linear flow regime} 
\begin{figure}
  \centerline{\hfill
    \includegraphics[width=0.4\textwidth,clip]{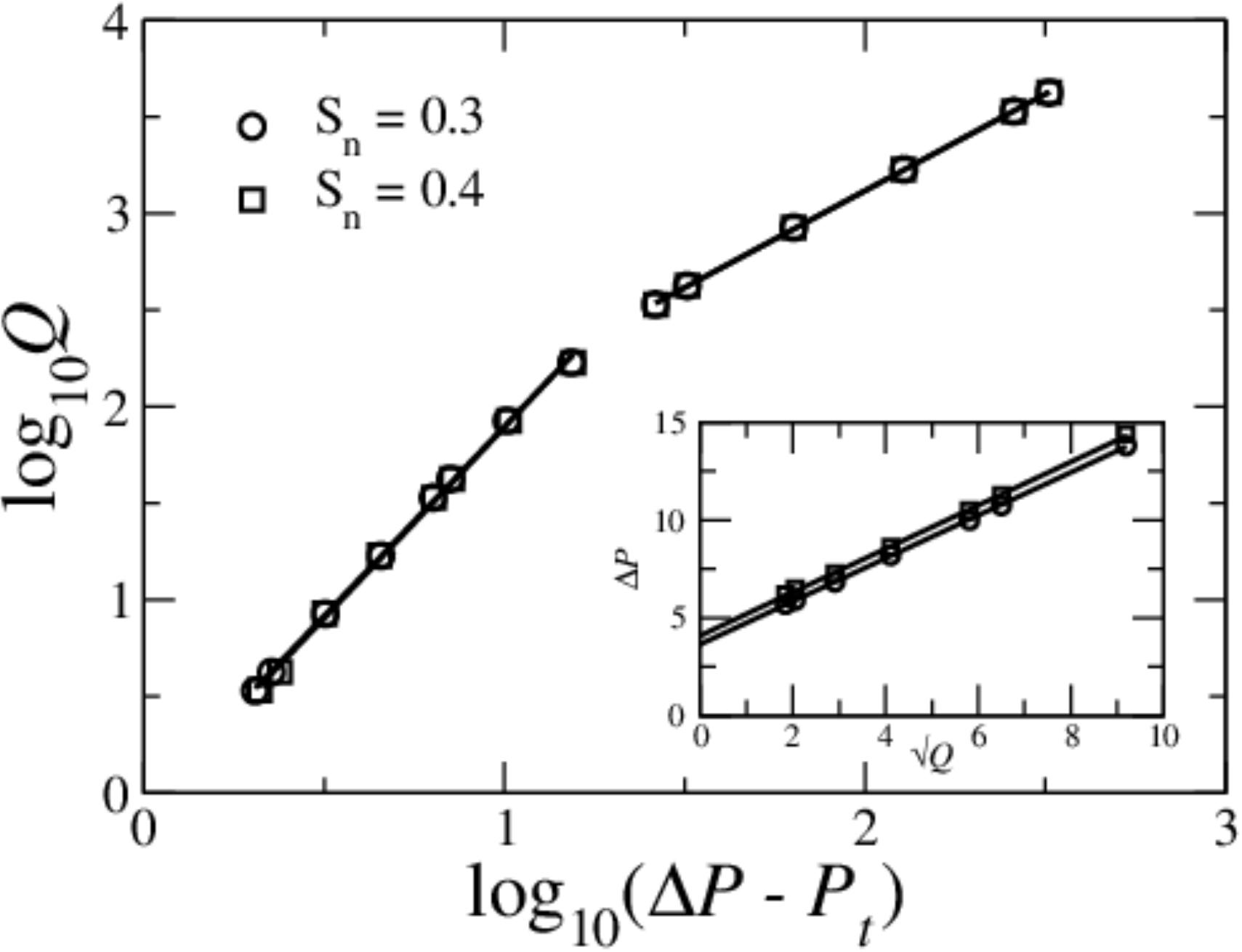}\hfill\hfill
    \includegraphics[width=0.4\textwidth,clip]{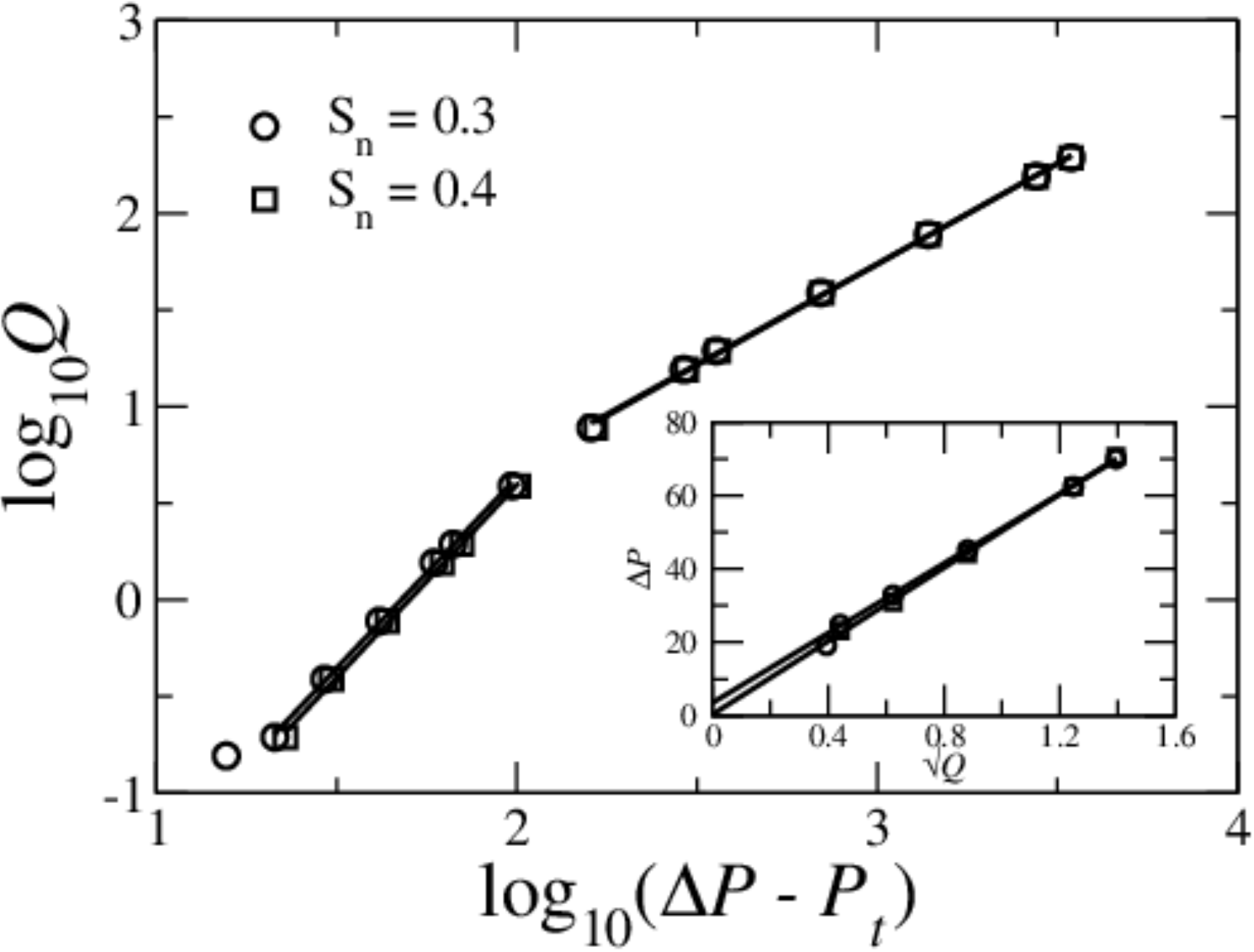}\hfill}
\caption{\label{figQp2}Variation of the total volumetric flow rate $Q$
  (${\rm mm}^3/{\rm s}$) with the overall pressure drop $\Delta P$
  (kPa) in the steady state for square and Berea networks. In the
  inset, $\Delta P$ is plotted against $\sqrt Q$ for the low Ca regime
  where the intercepts at the $y$-axis correspond to the values of
  threshold pressures ($P_t$). For 2D, we find $P_t=3.65$kPa and
  $4.09$kPa for $S_{\rm n}=0.3$ and $0.4$ respectively. For the Berea
  network in 3D, we find $P_t=3.54$kPa and $0.32$kPa for $S_{\rm
    n}=0.3$ and $0.4$ respectively. Using these values, we plot
  $\log_{10}Q$ versus $\log_{10}(\Delta P -P_t)$ which shows two
  distinct regimes at low and high pressures. For the linear regime,
  the slopes are obtained as $0.99 \pm 0.01$ and $1.00 \pm 0.01$ for
  2D and $1.03 \pm 0.01$ and $1.04 \pm 0.01$ for 3D for the two
  saturations respectively. For the quadratic regime, the slopes are
  obtained as $1.96 \pm 0.02$ and $1.98 \pm 0.03$ for 2D and $1.98 \pm
  0.03$ and $1.99 \pm 0.04$ for 3D.}
\end{figure}

Experiments have shown that the two-phase
flow in the steady state do not follow the linear Darcy relation
between the total flow rate and the pressure drop in the steady state
in capillary dominated regime \cite{tkrlmtf09, tlkrfm09, rcs11}. The
capillary pressures at the interfaces in two phase flow create
threshold barriers. The distribution of these thresholds over the
system depends on the pore sizes as well and on the interface
configurations in the steady state, which in turn depend on the flow
dynamics. These threshold pressures create a global yield threshold
pressure ($P_t$) below which there is no flow through the
system. Above $P_t$, the pores start opening while increasing the
pressure drop, due to which the total flow rate no longer vary
linearly with the pressure drop $\Delta P$ and shows non-linear
behavior. This non-linear behavior depends on the distribution of the
threshold pressure as well as on the system geometry \cite{rhs19}. For
two Newtonian fluids flowing through a porous media, it was found
experimentally \cite{tkrlmtf09, tlkrfm09, rcs11, sbdk17}, numerically
\cite{sh12, sbdk17} and with mean field calculations \cite{sh12} that,
in the regime when capillary pressures compete with the viscous
forces, that is, in the low capillary number regime, the total flow
rate $Q$ in the steady state vary quadratically with the excess
pressure drop. We can write as the following,
\begin{alignat}{2}
  \displaystyle
  Q & = 0                      &&  \quad\quad,\quad  P \le P_t \notag \\
    & \sim  (\Delta P-P_t)^2    &&  \quad\quad,\quad  P > P_t
  \label{eqnqp2}
\end{alignat}
The quadratic regime corresponds to when the individual pores keep
opening with the increasing pressure drop. Due to this, more and more
flow paths appear in the system which makes the flow rate to increase
faster than $\Delta P$. When the capillary number is high enough,
there is a transition from the quadratic regime and the flow rate
becomes linear with the pressure drop. To verify whether our model can
produce this cross-over, we performed steady-state simulations at
constant flow rate $Q$. The results are presented in figure
\ref{figQp2}. The threshold pressures $P_t$ are calculated by plotting
$\Delta P$ versus $\sqrt Q$ as shown in the insets. When the results
follow equation \ref{eqnqp2}, it will produce straight lines for the
lower values of $\Delta P$, where the intercepts of the straight lines
at $y$-axis correspond to the values of $P_t$. Using these values of
$P_t$, we plot $Q$ versus $(\Delta P-P_t)$ in the log-log scale the
disordered 2D network and the reconstructed 3D network. Results are
shown in figure \ref{figQp2} which show two distinct regimes, a
quadratic regime with slope $\approx 2$ at the low pressure drops and
a linear regime with slope $\approx 1$ at higher pressure drops. A
detailed numerical study on this non-linear effective rheological
properties using the interface algorithms presented here can be found
in \cite{sbdk17}, where the results are also compared with
experiments.

\begin{figure}
  \includegraphics[width=0.12\textwidth,clip]{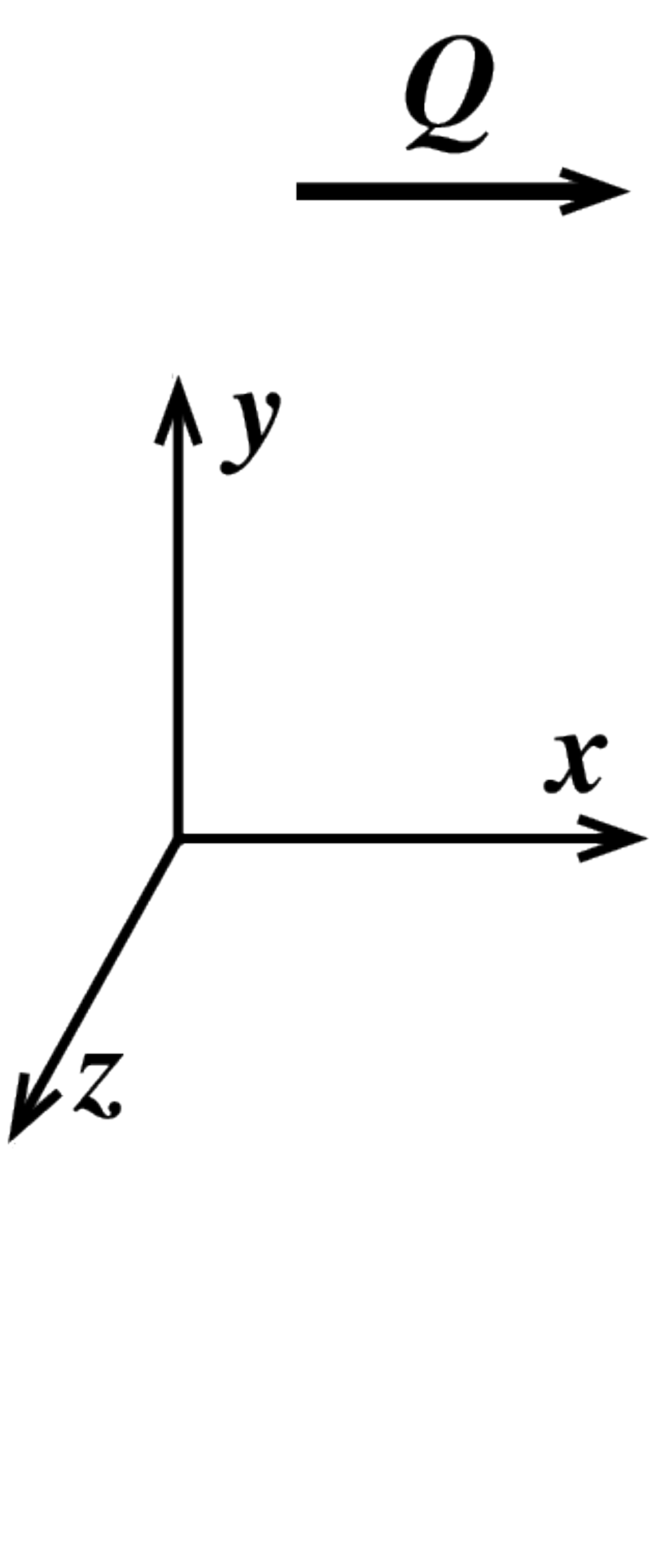}\hspace{0.09\textwidth}
  \includegraphics[width=0.24\textwidth,clip]{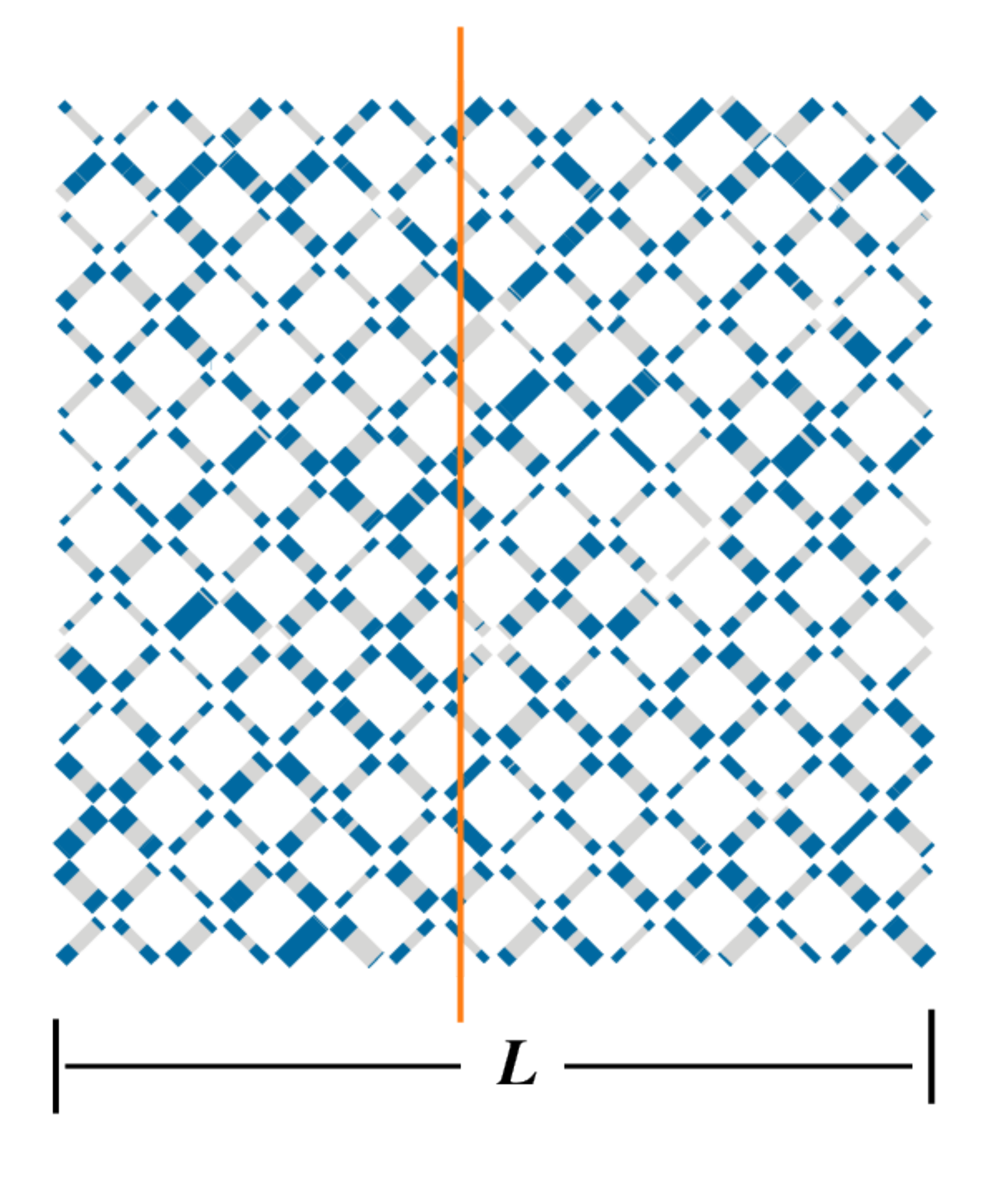}\hspace{0.09\textwidth}
  \includegraphics[width=0.46\textwidth,clip]{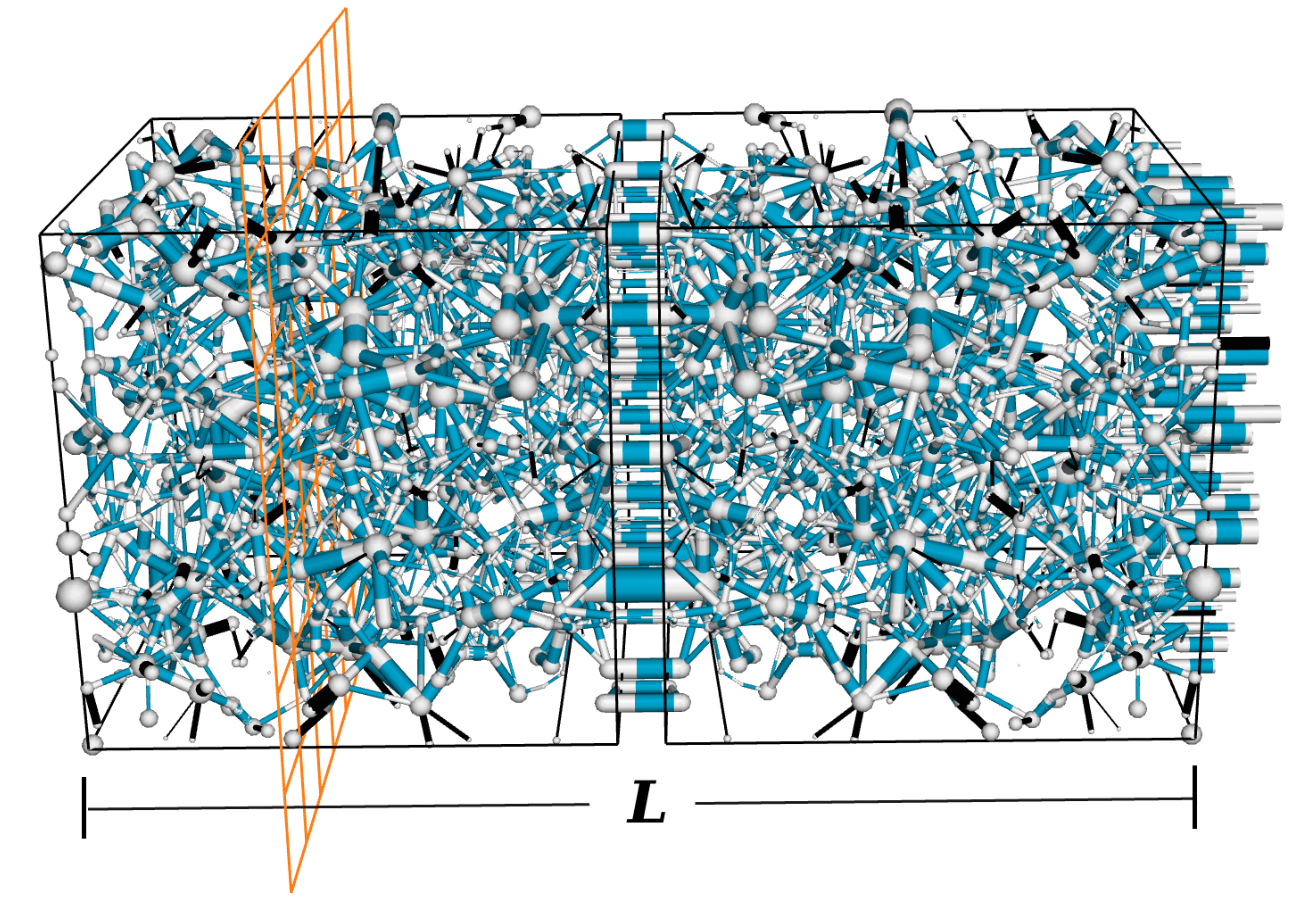}
  \offinterlineskip
  \hspace*{0.32\textwidth}\includegraphics[height=0.18\textwidth,clip]{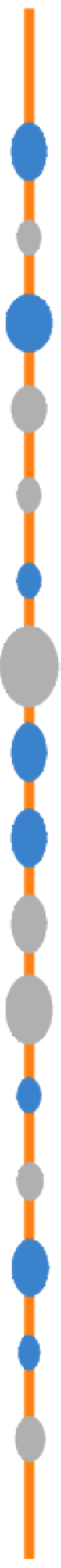}
  \hspace*{0.24\textwidth}\includegraphics[height=0.18\textwidth,clip]{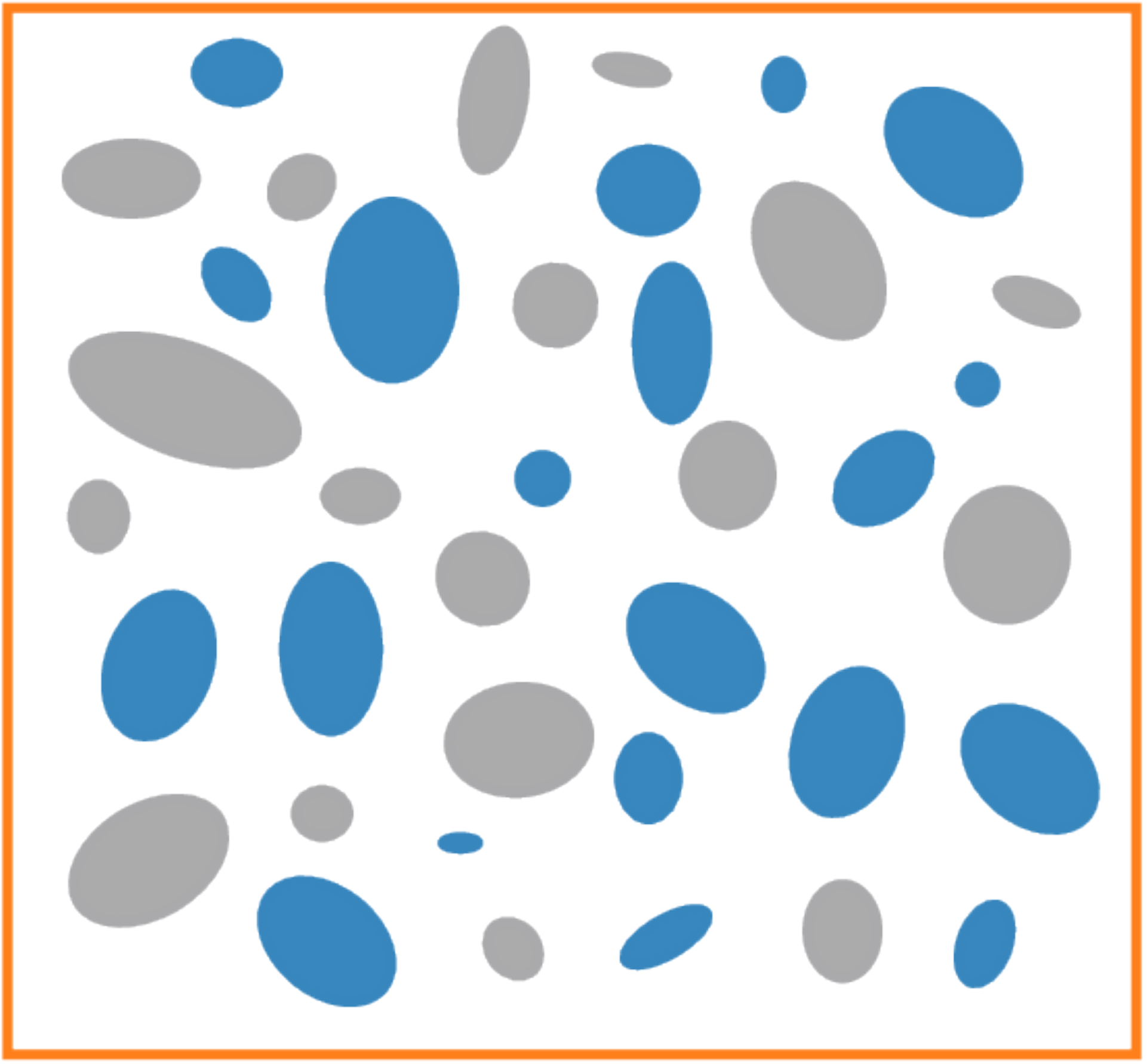}
\caption{\label{figSysArea}Description of the system to measure the
  flow rates ($Q$, $Q_{\rm w}$, $Q_{\rm n}$), the pore areas ($A$,
  $A_{\rm w}$, $A_{\rm n}$) and the seepage velocities ($v$, $v_{\rm
    w}$, $v_{\rm n}$) for the 2D and 3D networks. The global pressure
  drop $\Delta P$ is applied in the $x$ direction which is the
  direction of overall flow. A random cross section of the system
  normal to direction of overall flow is shown by the orange line for
  2D and by the orange grid plane for 3D. The normal view of the cross
  sections are shown underneath where the gray and blue patches show
  the occupation by wetting and non-wetting fluids in the cross
  section. The total gray and blue areas correspond to the wetting and
  non-wetting pore areas respectively and the sum of them correspond
  to the total pore area along this cross section. The averages of
  these areas over all the possible cross sections lead to the
  measurement of $A$, $A_{\rm w}$ and $A_{\rm n}$. In this figure, the
  gray a blue patches shown in the normal view of the cross sections
  do not reflect the actual occupations of the fluids in the above
  networks and are given as illustration purpose only.}
\end{figure}

\subsubsection{Relation between seepage velocities}
We will now measure the seepage velocities of the fluids in the steady
state with this model and will verify the relations between them. When
the system is driven under a constant pressure drop $\Delta P$, a set
of equations relating the wetting and non-wetting seepage velocities
($v_{\rm w}$, $v_{\rm n}$) to the total seepage velocity $v$ and the
fluid saturations can be derived using the Euler homogeneity property
of the total flow rate $Q$ in the steady state \cite{hsbkgv18}. These
relations necessitate a new velocity function, namely the co-moving
velocity ($v_m$), which is a characteristic of the porous medium. The
seepage velocities for the wetting and non-wetting fluids are defined
as
\begin{eqnarray}
  v_{\rm w} = \frac{Q_{\rm w}}{A_{\rm w}} & {\rm and} & v_{\rm n} = \frac{Q_{\rm n}}{A_{\rm n}}
  \label{eqnvwvndef}
\end{eqnarray}
respectively, where $Q_{\rm w}$ and $Q_{\rm n}$ are the volumetric
flow rates of the two fluids in the direction of the applied pressure
drop. Quantitatively, $Q_{\rm w}$ and $Q_{\rm n}$ are defined as the
volume of the wetting and non-wetting fluids that pass through any
cross section of the system, perpendicular to the overall flow
direction, per unit time. $A_{\rm w}$ and $A_{\rm n}$ are the wetting
and non-wetting pore areas defined as the areas occupied by the
wetting and non-wetting fluids along any orthogonal cross section
through the system. This is illustrated in figure \ref{figSysArea}
where $\Delta P$ is applied in the positive $x$ direction. The length
of the systems in this direction is $L$. A cross section normal to the
$x$ direction is shown by an orange straight line for the 2D network
and by an orange grid plane for the 3D network. Orthogonal views of
these cross sections are shown underneath where the gray and blue
patches show the pore areas occupied by the wetting and non-wetting
fluids respectively. The sum of the areas of individual colors
correspond to the wetting and non-wetting pore areas $A_{\rm w}$ and
$A_{\rm n}$ along this cross section. For a homogeneous porous medium
the average values of $A_{\rm w}$ and $A_{\rm n}$ remain same for any
orthogonal cross section of the system in the steady state. Here we
measure $A_{\rm w}$ and $A_{\rm n}$ by averaging the pore areas over
all possible cross sections along $x$. The total pore area $A$ related
to all the fluids is therefore given by $A=A_{\rm w}+A_{\rm n}=\phi
A_s$ where $\phi$ is the porosity and $A_s$ is the average
cross-sectional area of the total system including the pore space and
the solid. With this, we can express the fluid saturations $S_{\rm w}$
and $S_{\rm n}$ in terms of the pore areas by $S_{\rm w,n}=V_{\rm
  w,n}/V_p=(A_{\rm w,n}L)/(AL)=A_{\rm w,n}/A$. The total flow rate $Q$
of the two fluids is the sum of the wetting and non-wetting flow rates
given by $Q = Q_{\rm w}+Q_{\rm n} = A_{\rm w}v_{\rm w}+A_{\rm n}v_{\rm
  n}$. Correspondingly, the total seepage velocity $v$ associated with
the total flow rate $Q$ is defined by,
\begin{equation}
  v = \frac{Q}{A}
  \label{eqnv}
\end{equation}
and we can find,
\begin{equation}
  v = S_{\rm w}v_{\rm w} + S_{\rm n}v_{\rm n}
  \label{eqnswvw}
\end{equation}
by using the relations mentioned above.

In our network simulations we have the information about the local
flow rates $q_i$ and the interface positions for each link $i$ at any
time step. However, calculating the average flow rates and the pore
areas along any orthogonal cross section of the network from those
quantities may not be straight forward, specially in case of an
irregular network. For a regular network considered in 2D, all the
links are of the same length $l_j=1\,{\rm mm}$ and they are oriented
along the same angle ($45^\circ$) with respect to the overall flow
direction. The sum of the local flow rates through each row of links
normal to the flow direction is therefore the same for any row and the
flow rates can therefore be measured by summing over all the links of
the network and then dividing by the number of rows. This given by,
\begin{equation}
  Q = \frac{1}{N_L}\sum_jq_j \;{\rm,} \;\;\; Q_{\rm w} = \frac{1}{N_L}\sum_js_{{\rm w},j}q_j \;\;\;{\rm and}\;\;\; Q_{\rm w} = \frac{1}{N_L}\sum_j(1-s_{{\rm w},j})q_j
  \label{eqnQ2D}
\end{equation}
where $N_L$ is the number of rows along $L$, i. e. $64$
here. Similarly, we can calculate the cross-sectional areas as,
\begin{equation}
  A = \frac{1}{N_L}\sum_ja_j \;{\rm,} \;\;\; A_{\rm w} = \frac{1}{N_L}\sum_js_{{\rm w},j}a_j \;\;\;{\rm and}\;\;\; A_{\rm n} = \frac{1}{N_L}\sum_j(1-s_{{\rm w},j})a_j
  \label{eqnA2D}
\end{equation}
where $a_j$ is the cross-sectional area of the link $j$, projected
into the plane normal to the flow direction. The wetting saturations
of the links $s_{{\rm w},j}$ are provided by the interface
positions. Here the individual terms corresponding to the wetting and
non-wetting phases are multiplied with the corresponding link
saturations as the probability that a cross section through a link
will pass through the wetting or non-wetting phase is proportional to
the link saturation of that phase.

\begin{figure}
  \centerline{\hfill
    \includegraphics[width=0.45\textwidth,clip]{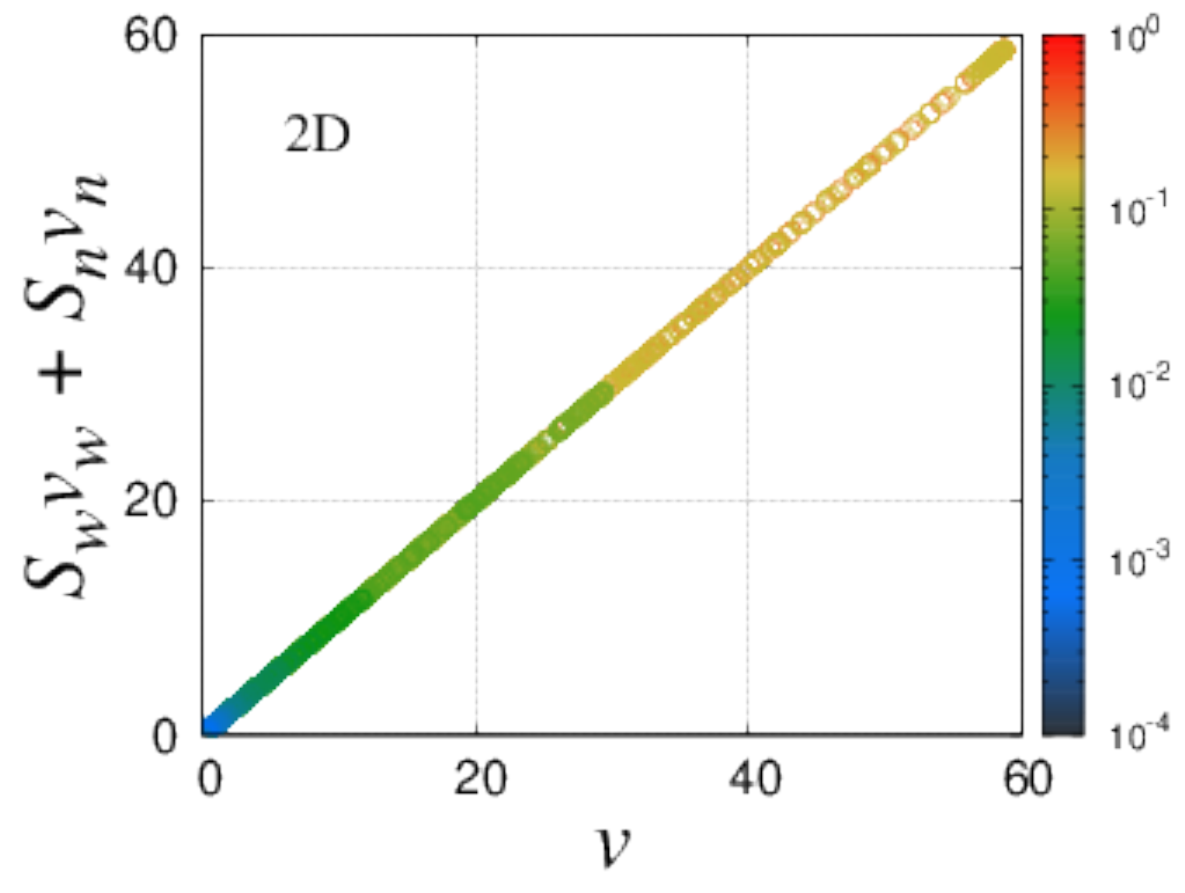}\hfill\hfill
    \includegraphics[width=0.45\textwidth,clip]{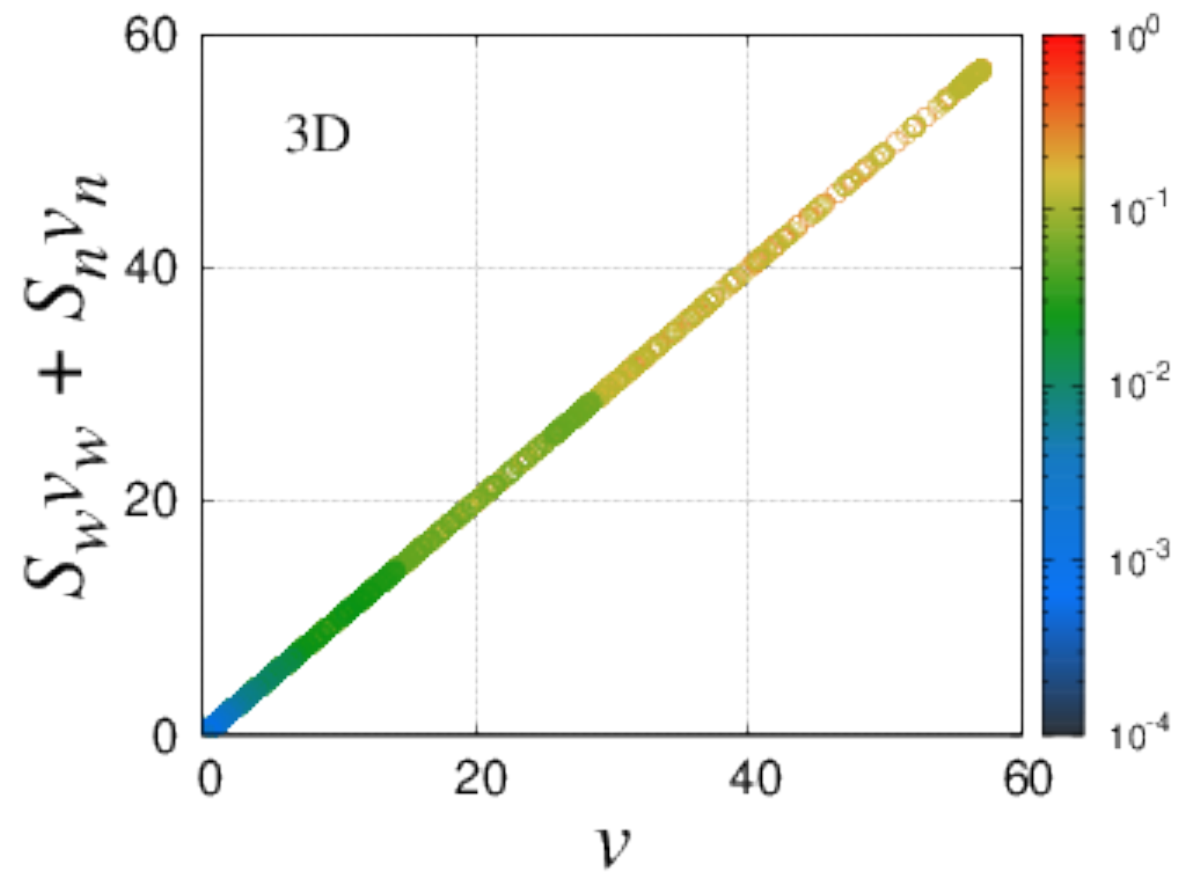}\hfill}
\caption{\label{figSwvw}Plot of $S_{\rm w}v_{\rm w}+S_{\rm n}v_{\rm
    n}$ against the total seepage velocity $v$ for 2D and 3D. The flow
  rates and pore areas to calculate $v_{\rm w}$, $v_{\rm n}$ and $v$
  are measured using the equations \ref{eqnQ2D} and \ref{eqnA2D} for
  2D and using \ref{eqnQ3D} and \ref{eqnA3D} for 3D. The velocities
  are in the unit of ${\rm mm/s}$. The colors represent the capillary
  numbers which are in the range of $0.0007 - 0.2985$ for 2D and
  $0.001 - 0.271$ for 3D. The measurements show the exact match of
  equation \ref{eqnswvw} for both 2D and 3D.}
\end{figure}

For an irregular network that is considered here in 3D, the links are
of different lengths and oriented in different directions. In that
case, measurement of the flow rates and the areas by summing over all
the links and dividing by the number of rows using the equations
\ref{eqnA2D} leads to wrong results. In this case, we measure these
quantities in the following way. Let us consider an orthogonal
cross-sectional plane at a random position through which we like to
measure the flow rates (figure \ref{figSysArea}). The probability that
any link $j$ will pass through this plane will be proportional to
$l_{x,j}/L$ where $l_{x,j}$ is the length of the link $j$ in the $x$
direction, the direction of the overall flow. After the link is
selected, the probability that the plane will pass through the wetting
fluid inside the link will be proportional to the local wetting
saturation $s_{{\rm w},j}$ of the link. Considering these
probabilistic terms, the total flow rates $Q$, $Q_{\rm w}$ and $Q_{\rm
  n}$ through a random orthogonal cross section can therefore be
calculated from the sum of the local flow rates over the links which
pass through this cross section,
\begin{equation}
  Q = \frac{1}{L}\sum_jl_{x,j}q_j \;{\rm,} \;\;\; Q_{\rm w} = \frac{1}{L}\sum_jl_{x,j}s_{{\rm w},j}q_j \;\;\;{\rm and}\;\;\; Q_{\rm n} = \frac{1}{L}\sum_jl_{x,j}(1-s_{{\rm w},j})q_j
  \label{eqnQ3D}
\end{equation}
where $L$ is the length of the network in the $x$ direction. The areas
can be measured in the similar way given by,
\begin{equation}
  A = \frac{1}{L}\sum_jl_{x,j}a_j \;{\rm,} \;\;\; A_{\rm w} = \frac{1}{L}\sum_jl_{x,j}s_{{\rm w},j}a_j \;\;\;{\rm and}\;\;\; A_{\rm n} = \frac{1}{L}\sum_jl_{x,j}(1-s_{{\rm w},j})a_j.
  \label{eqnA3D}
\end{equation}
For the regular 2D network, $l_{x,j}$ are the same for all the links
($=l$) and we recover equations \ref{eqnQ2D} and \ref{eqnA2D} by using
$N_L=L/l$. After measuring the flow rates and the pore areas, the
seepage velocities $v$, $v_{\rm w}$ and $v_{\rm n}$ are calculated
using the equations \ref{eqnv} and \ref{eqnvwvn}. Results are averaged
over time in the steady state.

The calculation of $v$, $v_{\rm w}$ and $v_{\rm n}$ from the
measurements of the flow rates and the pore areas should satisfy
equation \ref{eqnswvw}. In figure \ref{figSwvw}, we plot $S_{\rm
  w}v_{\rm w}+S_{\rm n}v_{\rm n}$ against the total seepage velocity
$v$ where we used the equations \ref{eqnQ2D} and \ref{eqnA2D} for 2D
and the equations \ref{eqnQ3D} and \ref{eqnA3D} for 3D. Figure
\ref{figSwvw} shows an exact match of equation \ref{eqnswvw} for the
whole range of parameters.

\begin{figure}
  \centerline{\hfill
    \includegraphics[width=0.45\textwidth,clip]{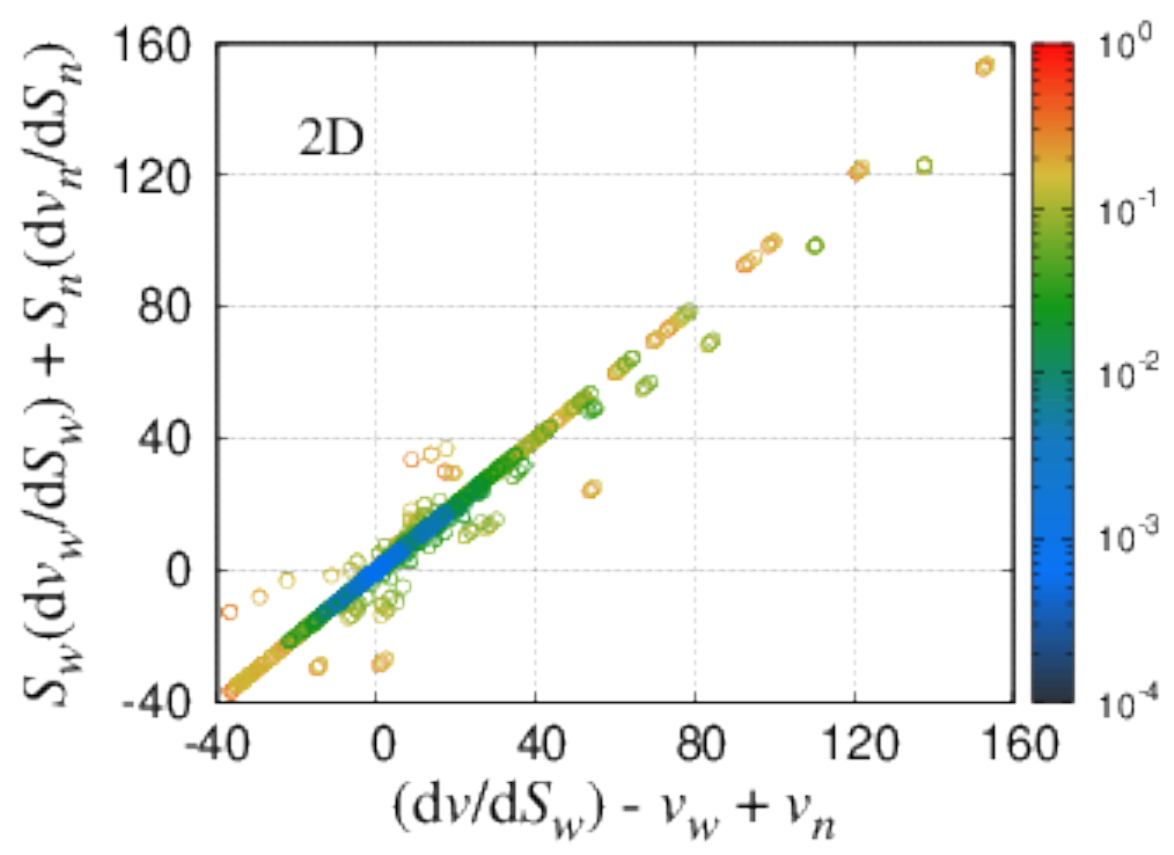}\hfill\hfill
    \includegraphics[width=0.45\textwidth,clip]{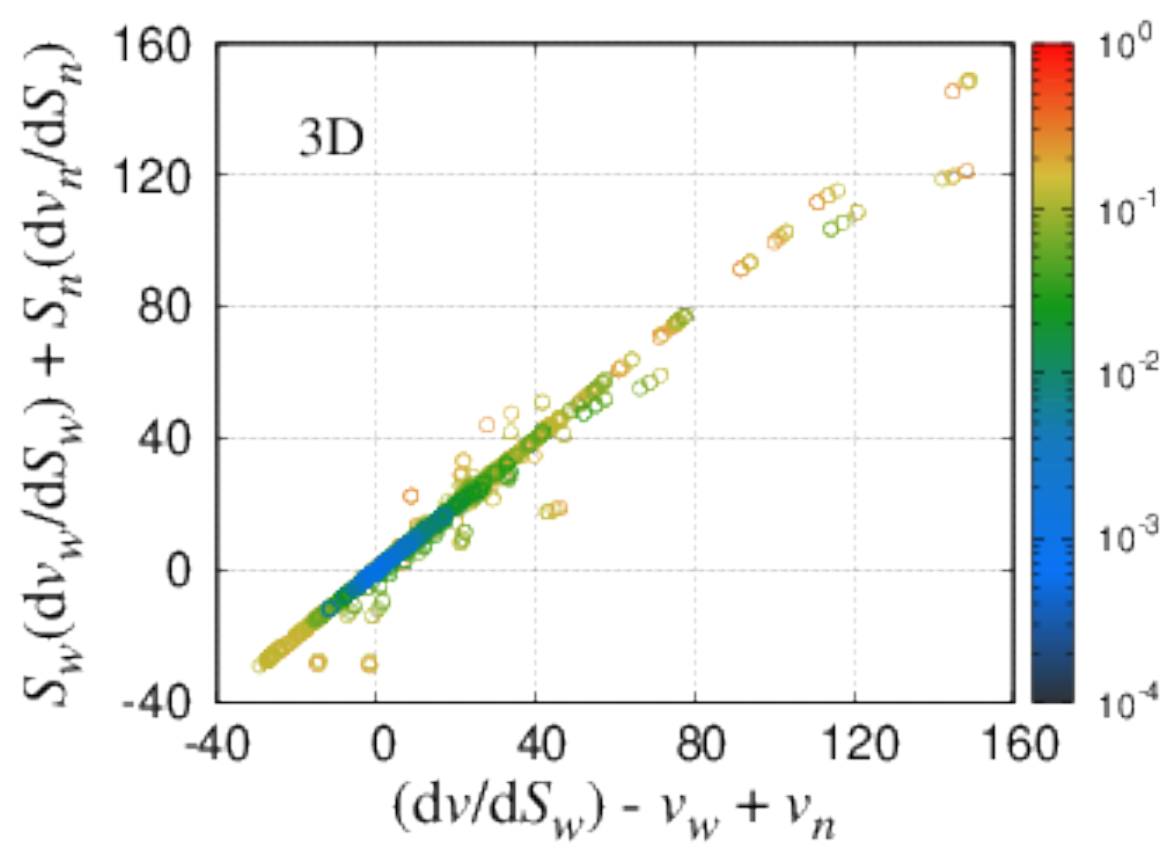}\hfill}
\caption{\label{figVm3637} Plot of the co-moving velocity $v_m$
  calculated using equation \ref{eqnvm37}, $v_m=S_{\rm w}(\mathrm
  dv_{\rm w}/\mathrm dS_{\rm w})+S_{\rm n}(\mathrm dv_{\rm n}/\mathrm
  dS_{\rm n})$ versus the same using equation \ref{eqnvm36},
  $v_m=(\mathrm dv/\mathrm dS_{\rm w})-v_{\rm w}+v_{\rm n}$ for the
  square and Berea networks. The velocities are in the unit of ${\rm
    mm/s}$. The data shows good a agreement with the equations
  \ref{eqnvm37} and \ref{eqnvm36} for the whole range of capillary
  numbers.}
\end{figure}

The total flow rate $Q$ in the steady state is a homogeneous function
of order one of the pore areas $A_{\rm w}$ an $A_{\rm n}$, that is, if
we scale the three areas by $A\to\lambda A_{\rm w}$, $A_{\rm
  n}\to\lambda A_{\rm n}$ and $A_s\to\lambda A_s$ by keeping the
porosity $\phi$ constant, the volumetric flow rate $Q$ scales as,
$Q(\lambda A_{\rm w}, \lambda A_{\rm n})=\lambda Q(A_{\rm w}, A_{\rm
  n})$. This property of $Q$ leads to a new set of equations between
the seepage velocities. Complete derivations of the equations can be
found in reference \cite{hsbkgv18} and here we will present them in
brief and will use them to validate our model. Taking the derivative
of the homogeneity equation of $Q$ with respect to $\lambda$ and then
setting $\lambda = 1$ we get,
\begin{equation}
  \displaystyle
  v = S_{\rm w}\left(\frac{\partial Q}{\partial A_{\rm w}}\right)_{A_{\rm n}} +  S_{\rm n}\left(\frac{\partial Q}{\partial A_{\rm n}}\right)_{A_{\rm w}}.
  \label{eqndQ}
\end{equation}
These two partial derivatives in the above equation have the units of
velocity and correspondingly they define two thermodynamic velocities
$\hat{v}_{\rm w}$ and $\hat{v}_{\rm n}$ given by,
\begin{eqnarray}
  \displaystyle
  \hat{v}_{\rm w} = \left(\frac{\partial Q}{\partial A_{\rm w}}\right)_{A_{\rm n}} & {\rm and} & \hat{v}_{\rm n} = \left(\frac{\partial Q}{\partial A_{\rm n}}\right)_{A_{\rm w}}.
  \label{eqnvhat}
\end{eqnarray}
With these definitions equation \ref{eqndQ} becomes,
\begin{equation}
  \displaystyle
  v = S_{\rm w}\hat{v}_{\rm w} + S_{\rm n}\hat{v}_{\rm n} {\rm ,}
  \label{eqnswvwhat}
\end{equation}
which has the similar form of equation \ref{eqnswvw}. However, this
does {\it not} imply that the thermodynamic velocities $\hat{v}_{\rm
  w}$ and $\hat{v}_{\rm n}$ are the same as the seepage velocities
$v_{\rm w}$ and $v_{\rm n}$ that we measure. These two types of
velocities can be related by a new velocity function is $v_m$ given
by,
\begin{eqnarray}
  \displaystyle
  \hat{v}_{\rm w} = v_{\rm w} + S_{\rm n}v_m & {\rm and} & \hat{v}_{\rm n} = v_{\rm n} - S_{\rm w}v_m
  \label{eqnvhv}
\end{eqnarray}
which fulfill both the equations \ref{eqnswvw} and
\ref{eqnswvwhat}. This velocity function $v_m$ is a function of the
saturation $S_{\rm w}$ and called as the co-moving velocity, which is
a property of the pore-network. With this definition of $v_m$, we can
derive two equations that are related to the variation of saturation,
\begin{equation}
  \displaystyle
  \frac{\mathrm dv}{\mathrm dS_{\rm w}} = v_{\rm w} - v_{\rm n} + v_m
  \label{eqnvm36}
\end{equation}
and
\begin{equation}
  \displaystyle
  S_{\rm w}\frac{\mathrm dv_{\rm w}}{\mathrm dS_{\rm w}} + (1-S_{\rm w}) \frac{\mathrm dv_{\rm n}}{\mathrm dS_{\rm n}} = v_m .
  \label{eqnvm37}
\end{equation}
In order to verify whether our pore-network model with the set of
interface algorithms described here do satisfy these equations, we
perform a large number of simulations with a wide range of parameters
for the 2D square network and the 3D Berea network. Five viscosity
ratios, $M=0.5$, $1$, $2$, $5$ and $10$ are considered where the
wetting viscosity is chosen as $\mu_{\rm w} = 0.1\,{\rm Pa\,s}$. The
non-wetting viscosity $\mu_{\rm n}$ is then chosen accordingly in
order to set the value of $M$. For each value of $M$, three different
values of the surface tension, $\gamma = 0.02$, $0.03$ and $0.04\;{\rm
  N/m}$, are considered. For each set of $M$ and $\gamma$, we
considered a set of pressure drops $|\Delta P/L|=0.16$, $0.20$,
$0.40$, $0.80$, $1.0$ and $2.0\;{\rm MPa/m}$ for 2D and $10$, $20$,
$40$, $80$ and $160\;{\rm MPa/m}$ for 3D. These values of pressure
drops are chosen in order to get capillary numbers ${\rm Ca}$ in a
range around $10^{-3}$ to $10^{-1}$. Specifically, we find ${\rm Ca}$
in the range of $0.0007-0.2985$ for 2D and $0.001-0.271$ for 3D. For
any set of parameters, saturations are varied in the range of $0$ to
$1$ in the steps of $0.05$ which correspond to $21$ saturation
values. This led to a total of $1890$ independent simulations for 2D
and $1575$ simulations for 3D in steady state.

\begin{figure}
  \centerline{\hfill
    \includegraphics[width=0.45\textwidth,clip]{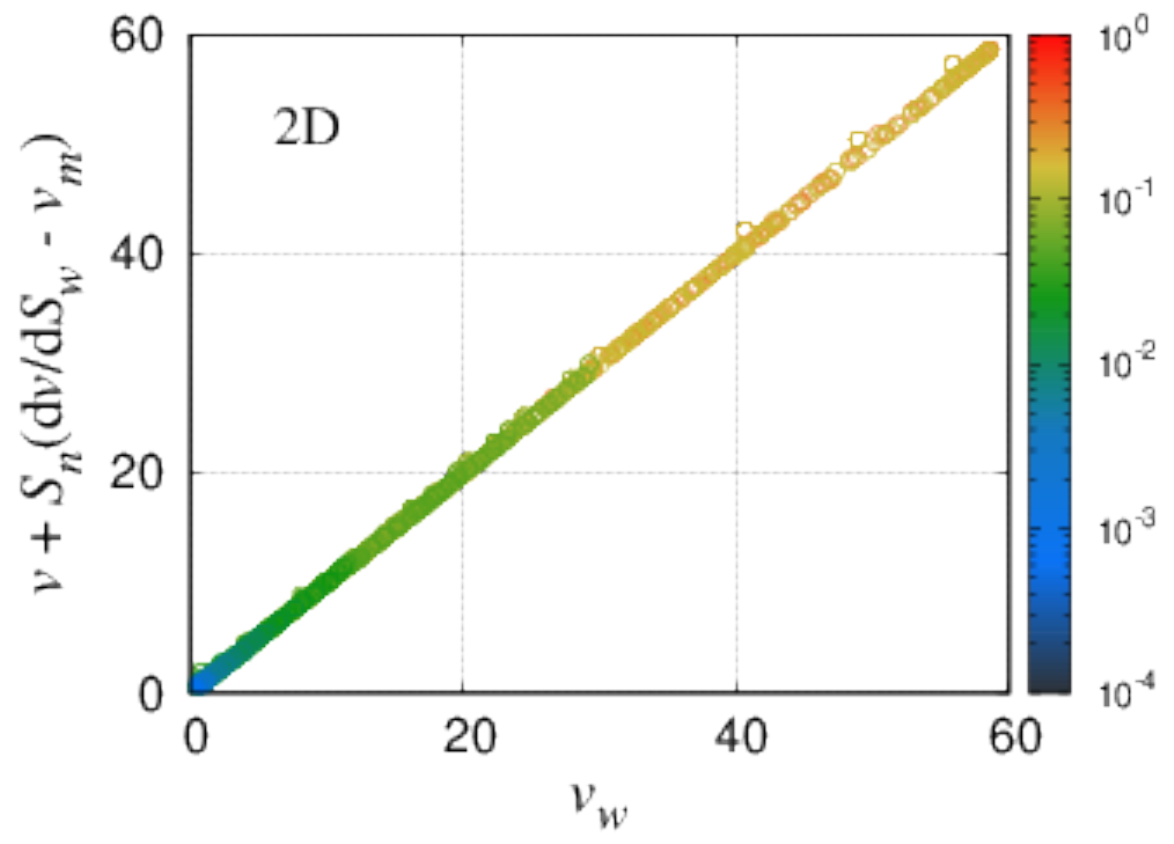}\hfill
    \includegraphics[width=0.45\textwidth,clip]{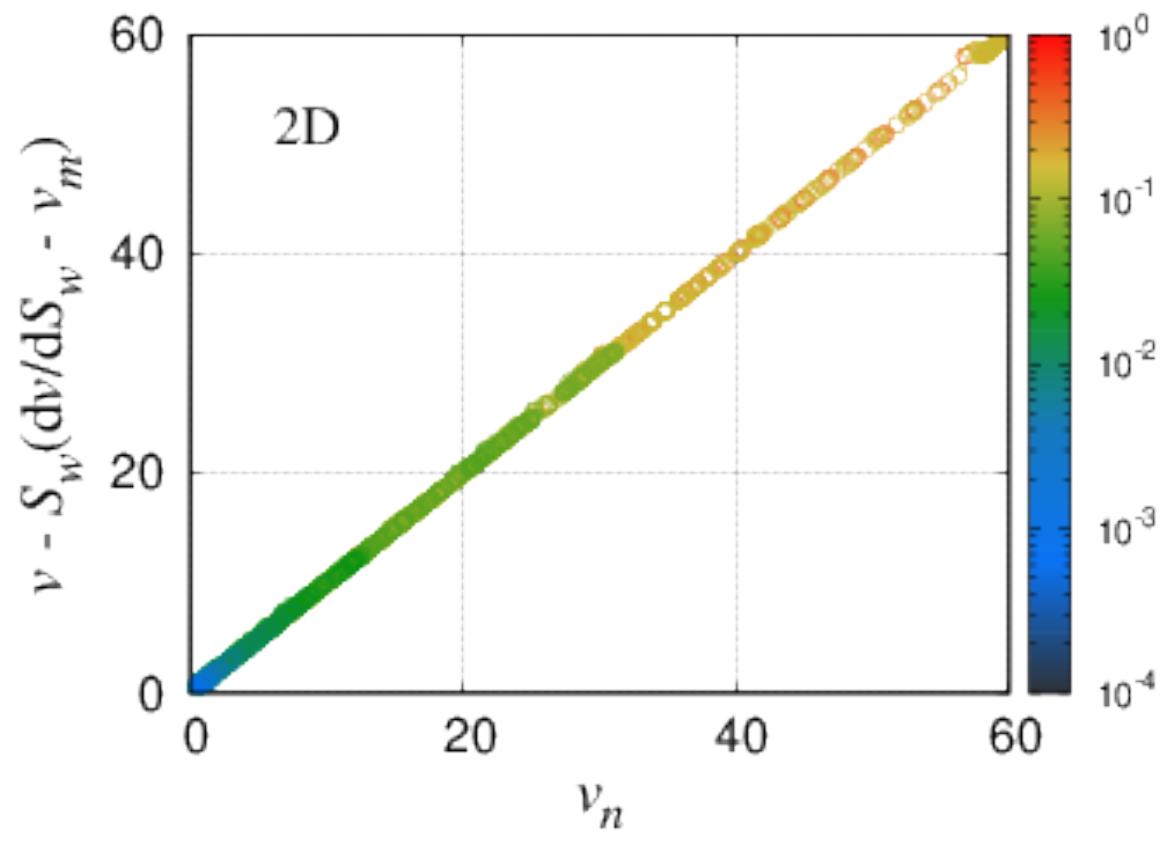}\hfill}
  \centerline{\hfill
    \includegraphics[width=0.45\textwidth,clip]{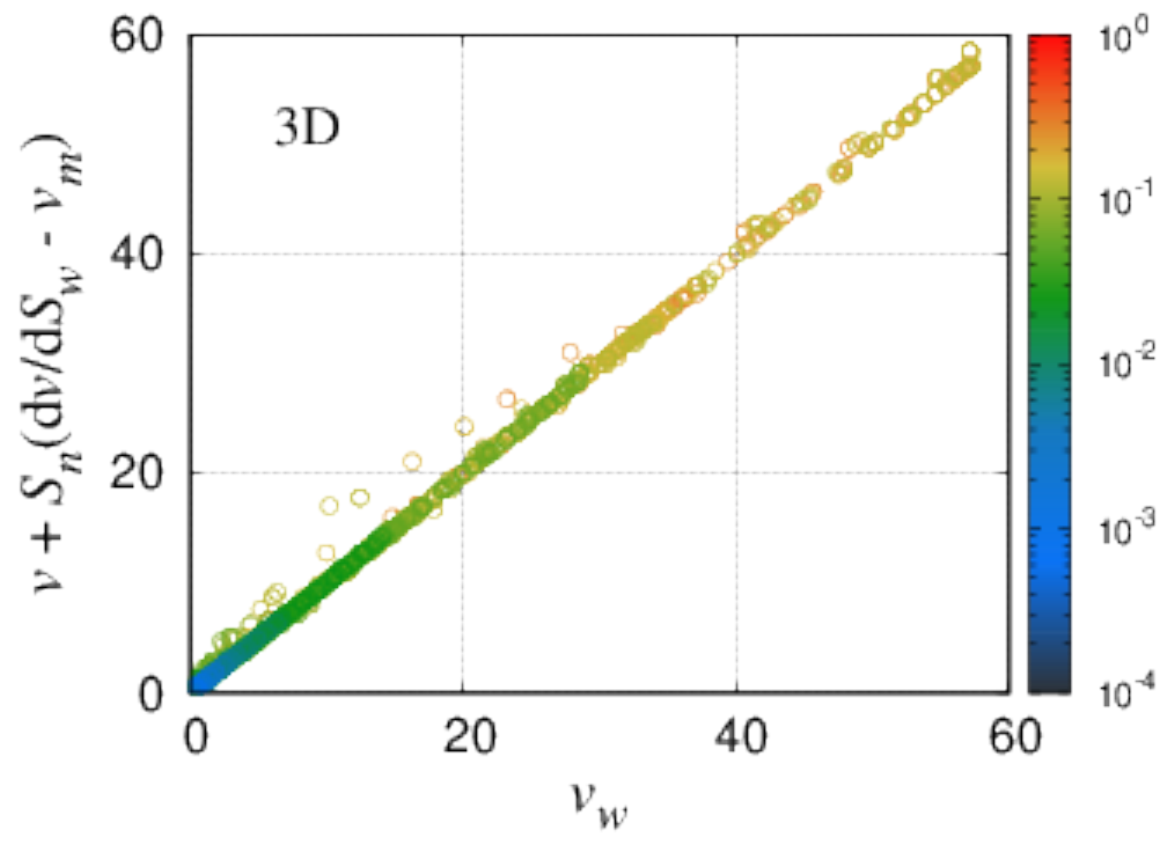}\hfill
    \includegraphics[width=0.45\textwidth,clip]{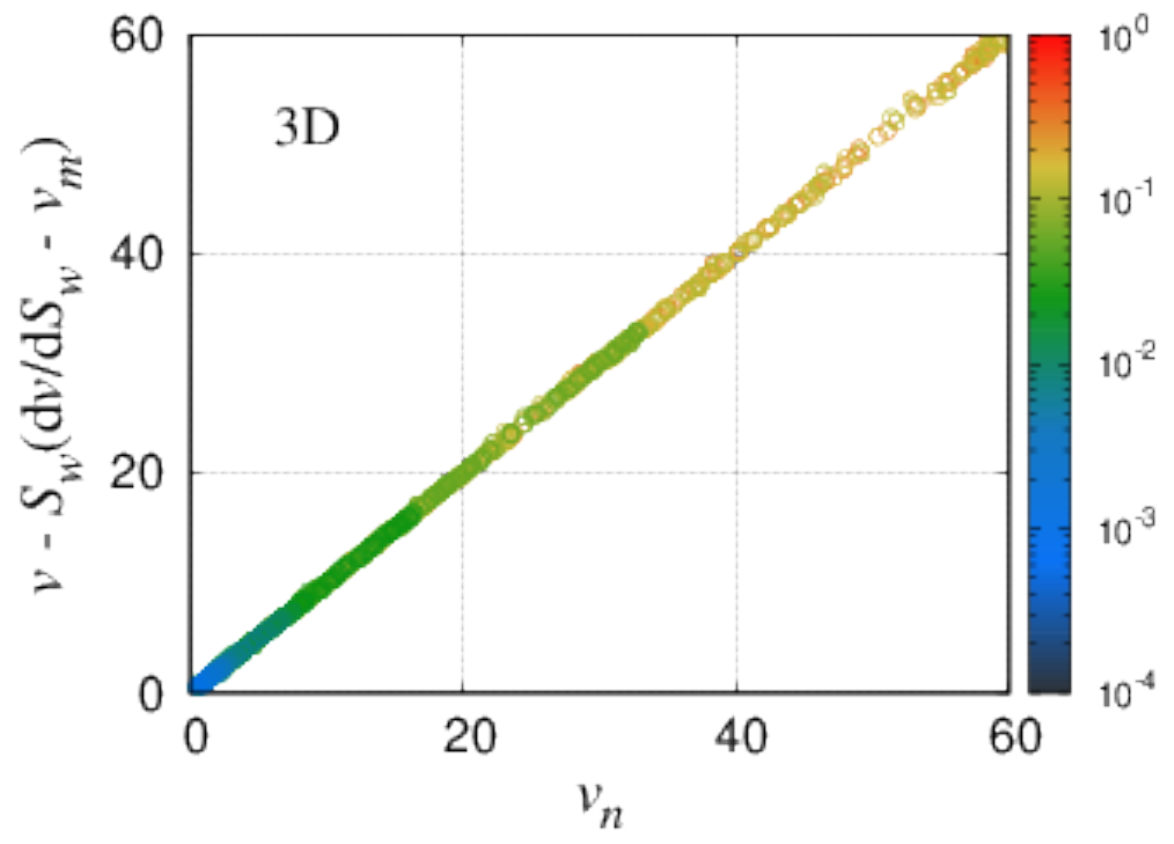}\hfill}
  \caption{\label{figVwvn}Plot of the wetting and non-wetting seepage
    velocities (mm/sec) calculated using equations \ref{eqnvwvn}
    against the measured values of $v_{\rm w}$ and $v_{\rm n}$ for the
    square (top row) and Berea (bottom row) network. The derivatives
    are calculated using the central difference techniques. The color
    scale shows the capillary numbers.}
\end{figure}

The seepage velocities are measured for all the simulations and the
derivatives with respect to the saturations are measured with central
difference technique. We calculate the values of $v_m=(\mathrm
dv/\mathrm dS_{\rm w})-v_{\rm w}+v_{\rm n}$ (equation \ref{eqnvm36})
and $v_m=S_{\rm w}(\mathrm dv_{\rm w}/\mathrm dS_{\rm w})+S_{\rm
  n}(\mathrm dv_{\rm n}/\mathrm dS_{\rm n})$ (equation \ref{eqnvm37})
and plot in figure \ref{figVm3637}. The results show good agreement
with the equation \ref{eqnvm36} and \ref{eqnvm37} for both square and
Berea networks for the whole range of the capillary numbers as
indicated by the color scale. The few data points that are outside the
straight line mostly correspond to the simulations near $S_{\rm w}=0$
or $1$ where the system undergoes from two-phase to single phase
regime that creates a jump in the derivatives.

Equations \ref{eqnswvw} and \ref{eqnvm37} transforms the wetting and
non-wetting velocities ($v_{\rm w}$, $v_{\rm n}$) to ($v$, $v_m$)
while varying the saturation. By inverting the velocity
transformation, it is possible to find equations to transform the
total velocity and the co-moving velocity ($v$, $v_m$) to ($v_{\rm
  w}$, $v_{\rm n}$), which are given by,
\begin{eqnarray}
  \displaystyle
  v_{\rm w} = v + S_{\rm n}\left(\frac{\mathrm dv}{\mathrm dS_{\rm w}}-v_m\right) & {\rm and} & v_{\rm n} = v - S_{\rm w}\left(\frac{\mathrm dv}{\mathrm dS_{\rm w}}-v_m\right).
  \label{eqnvwvn}
\end{eqnarray}
With these equations we can verify the measured values of wetting and
non-wetting seepage velocities $v_{\rm w}$ and $v_{\rm n}$ against the
ones calculated from these equation. In figure \ref{figVwvn}, we plot
$v+S_{\rm n}(\mathrm dv/\mathrm dS_{\rm w} - v_m)$ and $v-S_{\rm
  w}(\mathrm dv/\mathrm dS_{\rm w} - v_m)$ against the measured value
of $v_{\rm w}$ and $v_{\rm n}$ respectively. While calculating $v_{\rm
  w}$ and $v_m$ using above equations, we used the values of $v_m$
that are obtained from equation \ref{eqnvm37}. For the whole range of
the capillary numbers, good match with equations \ref{eqnvwvn} can be
observed for both the square and the Berea networks.

\begin{figure}
  \centerline{\hfill
    \includegraphics[width=0.48\textwidth,clip]{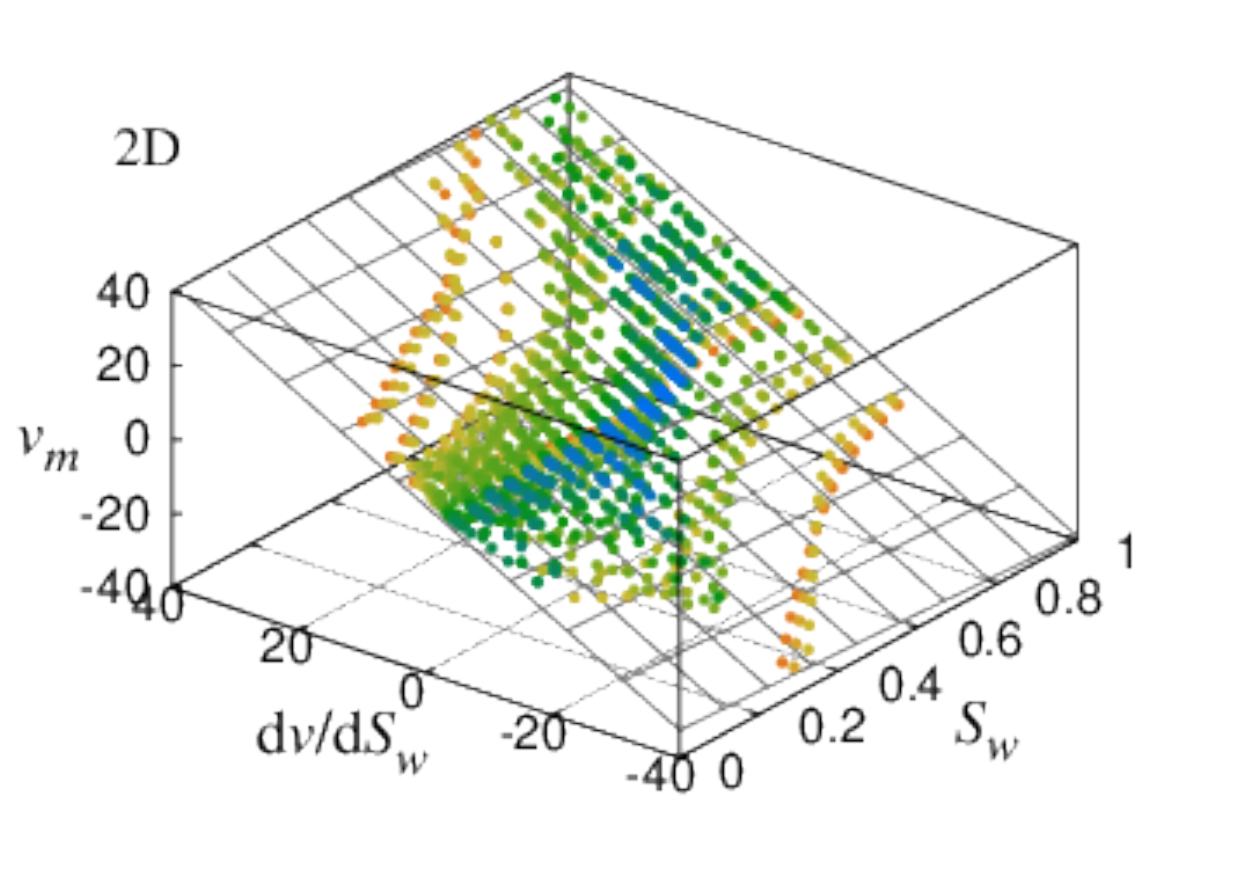}\hfill
    \includegraphics[width=0.48\textwidth,clip]{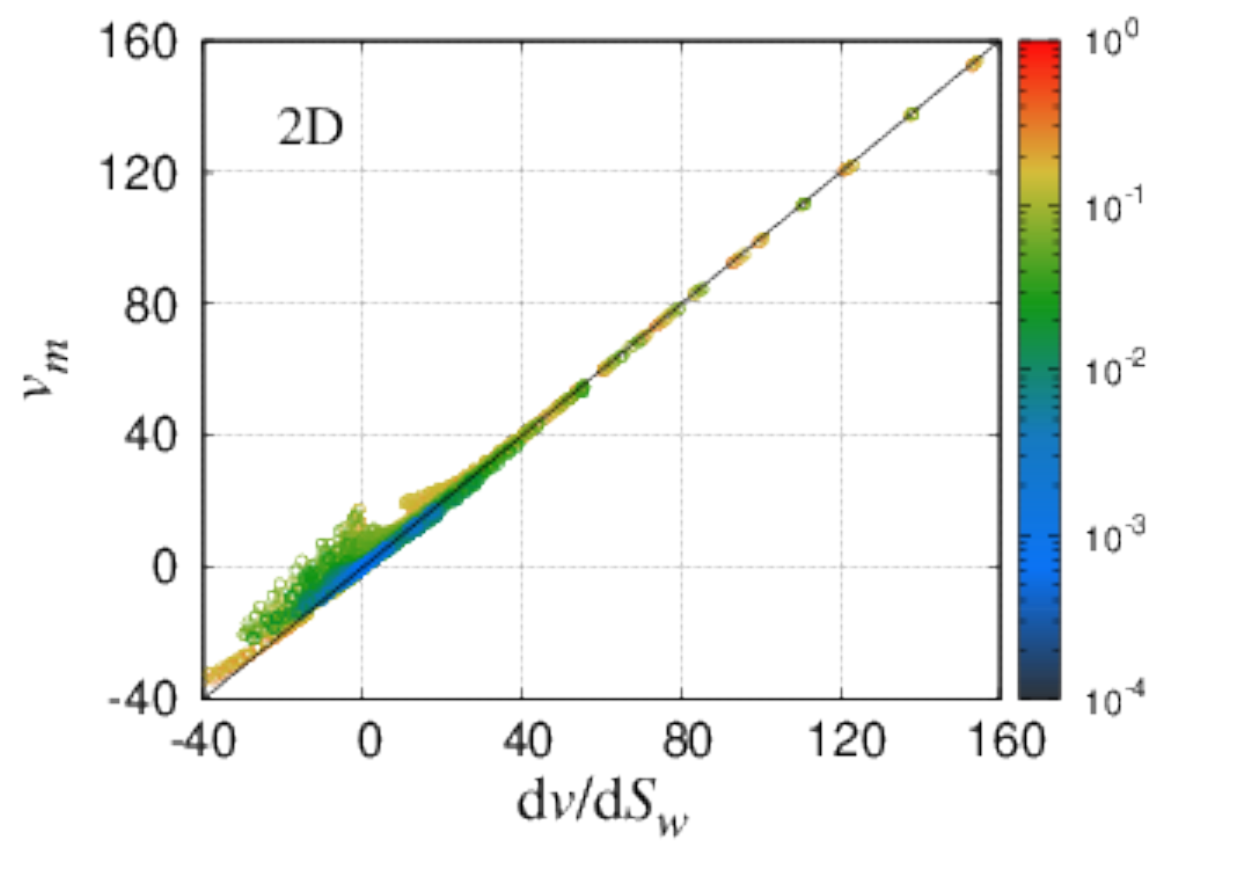}\hfill}
  \centerline{\hfill
    \includegraphics[width=0.48\textwidth,clip]{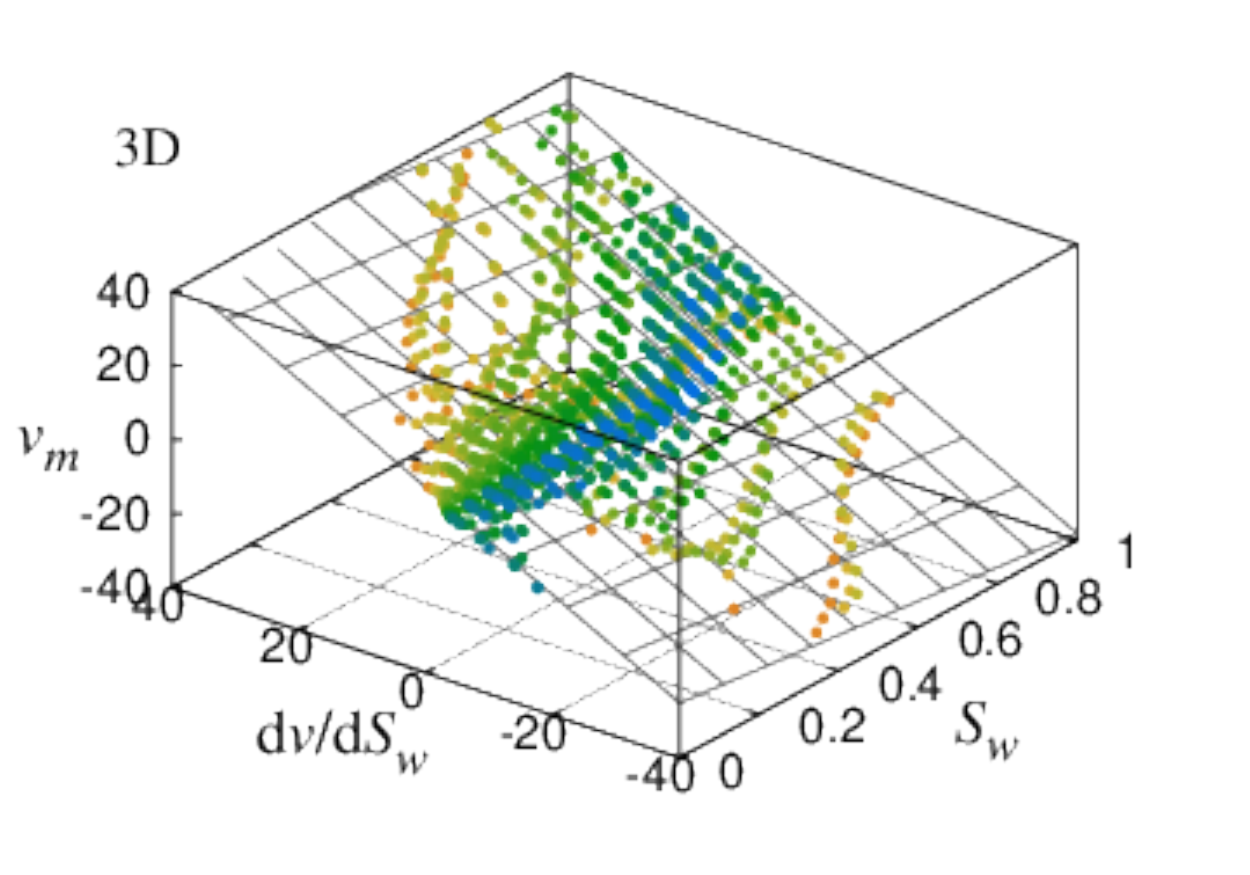}\hfill
    \includegraphics[width=0.48\textwidth,clip]{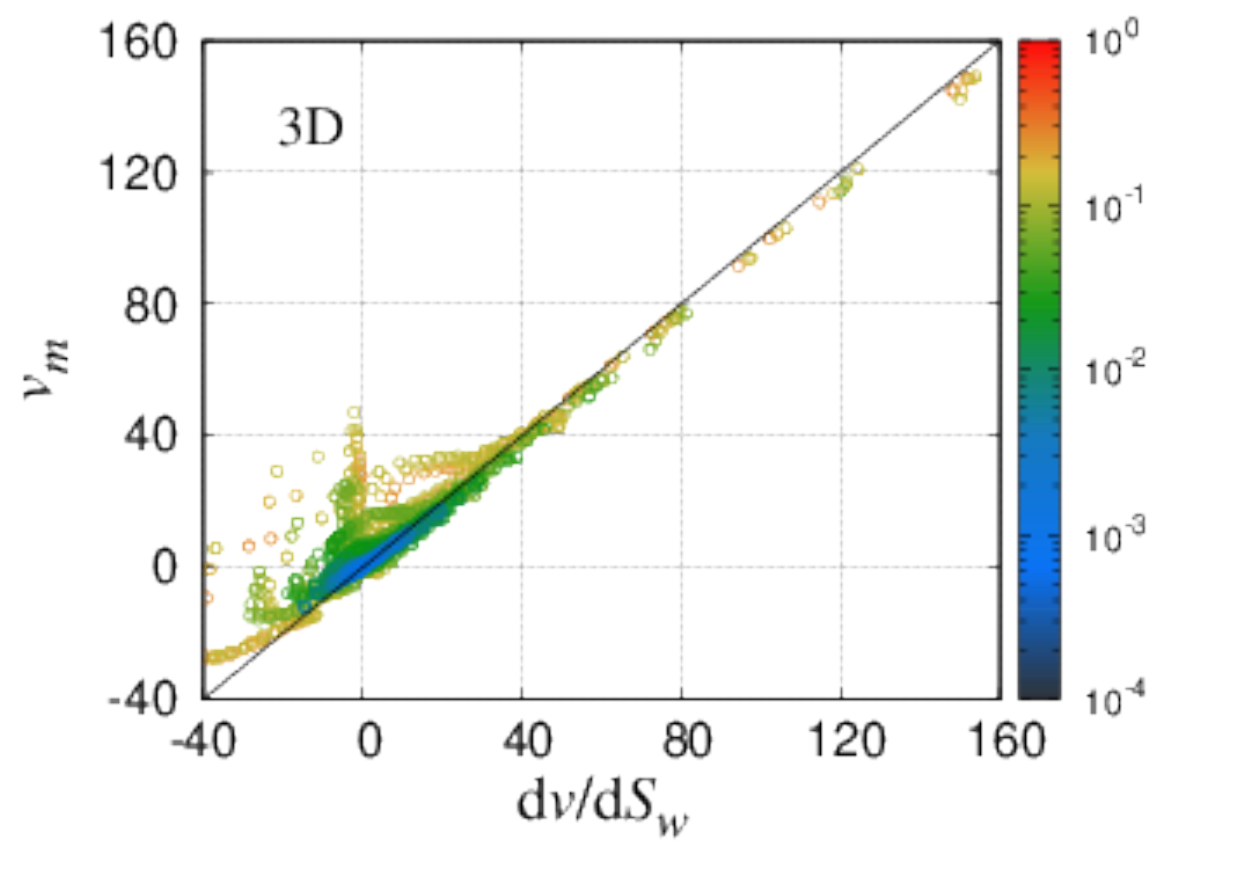}\hfill}
  \caption{\label{figVmdvm} Plot of the co-moving velocity $v_m$ as a
    function of $S_{\rm w}$ and $\mathrm dv/\mathrm dS_{\rm w}$ for
    the square (2D) and Berea (3D) networks are shown in the left. The
    data shows a rough planner behavior. The projection of the plots
    on the $y-z$ plane is shown in the right, where the values of
    $v_m$ roughly follow the $y=z$ straight line.}
  %Col: 0.0 0.45 0.70
\end{figure}

Finally, we plot the co-moving velocity $v_m$ as a function of both
$S_{\rm w}$ and $\mathrm dv/\mathrm dS_{\rm w}$ in figure
\ref{figVmdvm}. The co-moving velocity is a property of the porous
material and a function of the saturation $S_{\rm w}$, total seepage
velocity $v$ and the variation of $v$ while changing the
saturation. It is not enough to specify only the $S_{\rm w}$ and $v$
to determine $v_m$, as $\mathrm dv/\mathrm dS_{\rm w}$ depends on how
the external parameters are controlled while varying the $S_{\rm
  w}$. Here $v_m$ was calculated using equation \ref{eqnvm36} in
figure \ref{figVmdvm}. The data for $v_m$ roughly shows a planer form
given by,
\begin{equation}
v_m = aS_{\rm w} + b\frac{\mathrm dv}{\mathrm dS_{\rm w}} + c.
  \label{eqnvmplane}
\end{equation}
By fitting all the data points for the whole set of simulations, we
find $a=-6.36\pm 0.25$, $b=0.94\pm 0.01$ and $c=5.00\pm 0.13$ for the
square network and $a=-12.94\pm 0.62$, $b=0.88\pm 0.01$ and
$c=10.10\pm 0.32$ for the Berea network. The planes using these
parameters are shown in the respective figures with grid lines.
Interestingly, the values of $b$ are close to $1$ which leads the data
points to fall around the $y=x$ straight line while plotting $v_m$
against $\mathrm dv/\mathrm d S_{\rm w}$ as shown the figure
\ref{figVmdvm}.

\section{\label{secCon} Summary}
We presented a detailed description of a set of algorithms for
transporting fluids in a dynamic pore-network model of two-phase flow
in porous media. The displacements of the fluids in this model are
monitored by updating the positions of all the interfaces with
time. The basic concept of the algorithms are simple, at every time
step all the fluids arriving at a node from the incoming links are
displaced to all the outgoing links, and the volumes of the fluids are
distributed according to the ratio of the fluxes of the outgoing
links. Our aim in this article is to present these algorithms with all
the technical details so that it is possible for the reader to
reproduce this model. We have illustrated that this pore-network model
and the interface algorithms are applicable for both the regular and
irregular network topologies as well as for both two and three
dimensional pore networks. Moreover, by reproducing some of the
fundamental results of two-phase flow, we have also shown that the
model can be used to simulate both the transient and steady-state
flow. We have shown different drainage flow patterns that can be
generated with this model when a fluid displaces the other in a porous
medium. In steady state, the model successfully reproduces the linear
to non-linear transition in the effective rheological properties as
well as the relations between the seepage velocities.

In a recent paper by Zhao et al.\ \cite{vhj19} ten different groups
with different approaches to modeling two-phase flow in porous media
were invited to reproduce fluid injection in a circular Hele-Shaw cell
at different capillary numbers and wetting properties, ranging from
drainage to strong imbibition, i.e., imbibition where film flow
dominates the process. The conclusion of that work was, whereas all
the different approaches were able to reproduce the drainage processes
well, none succeeded in reproducing strong imbibition. Film flow is an
important mechanism during imbibition and in a first attempt, we
expanded our model in \cite{toh12} to include films. However, this
work has so far not been followed up by us.

The time integration procedure is time consuming. In \cite{sshbkv16},
Savani et al. proposed a Monte Carlo algorithm to replace the time
integration. This approach promises a large increase in efficiency of
the model. However, the method needs to be tested at low capillary
numbers. Moreover, it has so far only been implemented for regular
lattices and needs to be generalized to realistic pore networks.

\section*{Acknowledgments}
We thank our co-workers, past and present, who have been involved in
developing this model through its various stages, Eyvind Aker,
G.\ George Batrouni, M.\ Gr{\o}va, Henning A.\ Knudsen, Knut J{\o}rgen
M{\aa}l{\o}y, P{\aa}l Eric {\O}ren, Thomas Ramstad, Subhadeep Roy,
Isha Savani and Glen T{\o}r{\aa}. This work was partly supported by
the Research Council of Norway through its Centres of Excellence
funding scheme, project number 262644. SS was supported by the
National Natural Science Foundation of China under grant number
11750110430.

--------------------------------
\end{document}